\documentclass[twocolumn,hyperpdf,amsmath,amssymb,aps,prd,10pt,superscriptaddress,nofootinbib,noeprint,preprintnumbers,floatfix]{revtex4-1}

\usepackage{graphicx, color}

\usepackage[letterspace=-10]{microtype} 

\usepackage{bm, amsmath, amsfonts, amssymb,xfrac}
\usepackage{multirow, tabularx, dcolumn}
\usepackage{mathtools, leftidx, braket, slashed, cancel, bigdelim}
\usepackage{blkarray}
\usepackage[figures]{rotating}

\usepackage[utf8]{inputenc} 
\usepackage{hyperref}
\definecolor{jlab_red}{RGB}{192,39,45}
\definecolor{jlab_orange}{RGB}{249,102,0}
\definecolor{jlab_blue}{RGB}{47,122,121}
\definecolor{jlab_green}{RGB}{65,125,10}
\definecolor{jlab_gray}{gray}{0.6}

\hypersetup{%
pdftitle = {Radiative decay of the resonant $K^*$ and the $\gamma K \!\to\! K \pi$ amplitude from lattice QCD},
pdfsubject = {},
pdfkeywords = {},
pdfauthor = {Hadron Spectrum Collaboration},
colorlinks = {true},
filecolor = {black},
linkcolor = {jlab_blue},
menucolor = {black},
citecolor = {jlab_green},
urlcolor = {jlab_green},
}{}

\begin{document}

\preprint{JLAB-THY-22-3709}
\title{ Radiative decay of the resonant $K^*$ and the $\gamma K \!\to\! K \pi$ amplitude from lattice QCD}
\author{Archana Radhakrishnan}
\email{aradhakrishnan01@wm.edu}
\affiliation{Department of Physics, College of William and Mary, Williamsburg, VA 23187, USA}
\affiliation{\lsstyle Thomas Jefferson National Accelerator Facility, 12000 Jefferson Avenue, Newport News, VA 23606, USA}
\author{Jozef~J.~Dudek}
\email{dudek@jlab.org}
\affiliation{Department of Physics, College of William and Mary, Williamsburg, VA 23187, USA}
\affiliation{\lsstyle Thomas Jefferson National Accelerator Facility, 12000 Jefferson Avenue, Newport News, VA 23606, USA}
\author{Robert~G.~Edwards}
\email{edwards@jlab.org}
\affiliation{\lsstyle Thomas Jefferson National Accelerator Facility, 12000 Jefferson Avenue, Newport News, VA 23606, USA}
\collaboration{for the Hadron Spectrum Collaboration}
\date{\today}
\begin{abstract}
We present the first calculation in lattice QCD of the process $\gamma K \to K\pi$ in which the narrow $K^*$ vector resonance appears. Using a lattice on which the pion has a mass of 284 MeV, we determine the transition amplitude at 128 points in the $(Q^2, E_{K\!\pi})$ plane, and find suitable resonant scattering descriptions. We demonstrate the need to account for $S$--wave $K\pi$ elastic scattering when converting the finite-volume matrix elements computed in lattice QCD to the physically relevant infinite-volume matrix elements, even when we are primarily interested in the $P$--wave amplitude. Analytically continuing parameterizations of the $\gamma K \to K\pi$ amplitude to the $K^*$ resonance pole, we obtain the $K^{*+} \to K^+ \gamma$ transition form-factor, and compare the $Q^2=0$ value to the corresponding value extracted from the experimental partial-decay width.
\end{abstract}
\maketitle

\section{Introduction}
  \label{Intro}

The $K^*$ is a well established experimental resonance, typically observed as a narrow peak in any process leading to a $K\pi$ final state. As well as its sole hadronic decay mode to $K\pi$, the $K^*$ also has a much weaker decay mode to $K \gamma$, and the rate for this has been determined experimentally by considering the process $\gamma K \to K \pi$. This is achieved using a kaon beam on nuclear targets, where the very small momentum transfer region is dominated by nearly-real photon exchange, in what is known as the Primakoff process. Experiments were performed with both charged and neutral ($K^0_L$) kaon beams in the 1970s and 1980s~\cite{Carithers:1975cg, Berg:1980cp, Chandlee:1983hf, Carlsmith:1985ep}, yielding estimates for the partial decay widths (averaged by the PDG) of
\begin{align*}
\Gamma(K^{*\pm} \to K^\pm \gamma) &= 50(5) \, \mathrm{keV} \, ,\nonumber \\
\Gamma(K^{*0} \to K^0 \gamma)     &= 116(10) \, \mathrm{keV} \, .
\end{align*}
At the time of the experiments, the most sophisticated theoretical tool available to describe these numbers was the constituent quark model, and these rates were interpreted in terms of constituent-quark magnetic moments, leading to estimates of the degree of $SU(3)$ flavor symmetry breaking. 

Today we would seek to understand these numbers within non-perturbative Quantum Chromodynamics, either directly, as we will do in this paper, using lattice QCD, or indirectly using chiral effective field theory. Dax \emph{et al.} have emphasized the connection of these radiative processes to the chiral anomaly which is a leading effect in such an effective field theory approach~\cite{Dax:2020dzg}.

In order to study this process rigorously, one should utilize theoretical techniques which respect, to the highest degree possible, the general constraints of scattering theory. A particularly important constraint in the current case is that of \emph{unitarity} -- the strong rescattering of $K\pi$ through the $K^*$ resonance is limited by unitarity, and correct application leads to a restriction on the possible behavior of the $K\gamma$ production amplitude. Similarly, a rigorous description of the $K^*$ resonance comes through associating it with a pole singularity at a complex value of the scattering energy, where the mass and width are related to the pole position, and where the residue of the pole yields couplings to $K\pi$ and $K\gamma$. 

The relevant $\gamma K \to K \pi$ scattering process can be studied in first-principles lattice QCD in a two--stage process: In the first stage, the discrete spectrum of states with the quantum numbers of $K\pi$ is extracted from two-point correlation functions that are computed in the finite spatial volume of the periodic lattice, and these energies are used to constrain the elastic $K\pi \to K\pi$ scattering process in one or more partial waves~\cite{Briceno:2017max}. In the second stage, three-point correlation functions are computed in which the source operators interpolate $K\pi$ finite-volume states, the sink operators interpolate the kaon, and the electromagnetic current is inserted between the two. From these correlation functions, \emph{finite-volume matrix-elements} are extracted, but these must undergo correction by a factor which is sensitive to the volume of the lattice and to the $K\pi$ elastic scattering amplitude~\cite{Lellouch:2000pv,Briceno:2014uqa,Briceno:2015csa}. While the experimental determinations discussed earlier are for real photons, within the lattice QCD calculation it is natural to consider the transition process also as a function of photon virtuality, $Q^2$.

The only previous applications of this finite-volume formalism to a similar process are to $\gamma \pi \to \pi \pi$~\cite{Briceno:2015dca, Briceno:2016kkp, Alexandrou:2018jbt}, where the contribution of the $\rho$ resonance was determined. An important difference between extraction of $\gamma K \to K\pi$ and $\gamma \pi \to \pi \pi$ in lattice QCD is the presence of \emph{both parities} of $K\pi$ in moving-frame irreducible representations~\cite{Leskovec:2012gb}. In this case, even though the \emph{vector} $K\pi$ channel is of interest to us, the \emph{scalar} wave also plays a role -- while the transition ${\gamma K \to (K\pi)_{\ell=0}}$ is forbidden by parity and angular momentum conservation, the \emph{normalization of the finite-volume states} depends upon the elastic $K \pi$ $S$-wave scattering amplitude. In this paper we will, for the first time, account for this effect in an explicit calculation, showing that application of the finite-volume formalism leads to consistent infinite-volume amplitudes.

In order to sample a wide region of the $E^\star_{K\!\pi}$ dependence of the transition process, from threshold, through the $K^*$ resonance, and into the high-energy tail, we make use of optimized operators to access correlation functions for $K \pi$ finite-volume states \emph{beyond the ground-state} in each irrep~\cite{Shultz:2015pfa}. We use up to the second excited state in some irreps, observing no significant increase in statistical uncertainty relative to the ground states. To sample a large number of points in photon virtuality $Q^2$ we consider many momenta for the kaon operator, the $K\pi$ operator and the current insertion. We obtain a data set of 128 unique $(Q^2, E^\star_{K\!\pi})$ points at 18 values of $E^\star_{K\!\pi}$ which we describe with parameterizations of the transition amplitude.

In this first calculation we compute the transition $\gamma K \to K\pi$ in the ${I=\tfrac{1}{2}, I_z = + \tfrac{1}{2}}$ isospin state, which is related to the definite charge final states by appropriate isospin Clebsch-Gordan coefficients,
\begin{equation*}
\Big| K\!\pi \big(\tfrac{1}{2}, +\tfrac{1}{2} \big) \Big\rangle = -\tfrac{1}{\sqrt{3}} \big| K^+ \pi^0 \big\rangle - \sqrt{\tfrac{2}{3}} \big| K^0 \pi^+ \big\rangle .\, 
\end{equation*}
These factors are such that if one considers the final state $K^0$ to be a superposition of the decay eigenstates ($K^0_S, K^0_L$), we would expect equal decay rates of $K^{*+}$ to ${K^+\pi^0, K^0_S \pi^+}$, and $K^0_L \pi^+$.

The parameters of the anisotropic $2+1$ flavor Clover lattice used in this calculation are given in Table~\ref{lattice}, and details of the lattice action can be found in Refs.~\cite{Edwards:2008ja, HadronSpectrum:2008xlg}. Using the $\Omega$-baryon mass to set the scale, the pion here has a mass of 284 MeV, and the lattice has a spatial extent of $\sim 2.7$ fm. 
This calculation takes advantage of the prior determination on this lattice of the $K\pi$ spectrum and elastic $K\pi$ scattering amplitudes as reported in Ref.~\cite{Wilson:2019wfr}. Herein we use slightly fewer configurations than in that calculation, with the same\footnote{There is a typographic error in Table I of Ref.~\cite{Wilson:2019wfr} which suggests that 162 vectors were used, when in fact the actual number was 160 vectors, as in the current calculation.} number of distillation vectors~\cite{HadronSpectrum:2009krc}, but the extracted discrete energy levels are found to be statistically compatible with those presented therein. The fact that we are using only a single lattice volume means that we are able to use single-elimination jackknife to propagate uncertainty through the entire calculation using the original ensemble of lattice configurations.

\begin{table}[h]
\renewcommand{\arraystretch}{1.18}
\begin{tabular}{r|l}
$(L/a_s)^3 \times (T/a_t)$ & $24^3\times 256$ \\
$N_\mathrm{cfgs}$          & 348 \\
$N_\mathrm{vecs}$          & 160 \\
$a_t m_\pi$                & 0.04735(22) \\
$a_t m_K$                  & 0.08659(14) \\
$\xi = a_s/a_t$			   & 3.455(6)			 
\end{tabular}
\caption{Parameters of the lattice configurations used in this calculation. See Ref.~\cite{Wilson:2019wfr} for more details. } \label{lattice}
\end{table}


\section{Infinite-volume transition amplitude}

Initial production of a $(K\pi)_{\ell=1}$ system when a photon is absorbed by a kaon is unavoidably followed by strong rescattering of the $K \pi$ system, subject to the constraint of unitarity. Considering the time-reversed $K \pi \to \gamma K$ process, a solution\footnote{
This is not a \emph{unique} solution to the unitarity constraint. Another commonly made choice is to make use of the Omn\`es function, which implements some additional properties of analyticity using a dispersion relation applied to the elastic scattering phase-shift~\cite{Omnes:1958hv}.} to the unitarity constraint for the transition amplitude takes the form\footnote{In this paper all variables with a star superscript are evaluated in the center-of-momentum frame.},
\begin{align}\label{H}
\mathcal{H}&^\mu_{\lambda_{\!K\!\pi}}( \mathbf{p}_K, \mathbf{p}_{K\! \pi}; Q^2, E^\star_{K\!\pi} ) \nonumber \\
&\equiv \big\langle K; \mathbf{p}_K  \big| j^\mu(0) \big| K\pi(\ell\!=\!1, \lambda_{K\!\pi}); E^\star_{K\!\pi}, \mathbf{p}_{K\!\pi} \big\rangle \nonumber \\
&= \mathcal{A}^\mu_{\lambda_{\!K\!\pi}} \big( \mathbf{p}_K, \mathbf{p}_{K\!\pi}; Q^2, E^\star_{K\!\pi} \big) \cdot \frac{1}{k^\star_{K\!\pi}} \cdot \mathcal{M}^{\ell=1}(E^\star_{K\!\pi})\, .
\end{align}
The function $\mathcal{A}$ parameterizes the production process, and is expected to be a relatively featureless function of $E^\star_{K\!\pi}$, having neither the unitarity branch cut, nor any possible resonance pole singularities, both of which live in the elastic scattering amplitude, $\mathcal{M}^{\ell=1}$. The factor of $1/k^\star_{K\!\pi}$, featuring the momentum of the kaon (or the pion) in the center-of-momentum frame, is included to cancel the unwanted final-state $P$--wave threshold factor in $\mathcal{M}$ describing $K\pi \to K \pi$, when we are considering $K\pi \to K \gamma$.

The production amplitude depends upon the helicity of the $K \pi$ state and the direction of the current\footnote{Or the helicity of the photon if the amplitude is projected as $ \mathcal{H}_{\lambda_{\gamma}, \lambda_{K\!\pi}} = \epsilon^*_\mu( \mathbf{q}, \lambda_\gamma )  \mathcal{H}^\mu_{\lambda_{K\!\pi}}$.}. The constraints arising from Poincare symmetry can be satisfied by introducing a \emph{kinematic factor}  multiplying a \emph{transition form-factor},
\begin{align} \label{A}
 \mathcal{A}&^\mu_{\lambda_{\!K\!\pi}}  \big( \mathbf{p}_K, \mathbf{p}_{K\!\pi}; Q^2, E^\star_{K\!\pi} \big) \nonumber \\
 &= \tfrac{2}{m_K} \epsilon^{\mu\nu\rho\sigma} (\mathbf{p}_K)_\nu \,(\mathbf{p}_{K\!\pi})_\rho \, \epsilon_\sigma( \mathbf{p}_{K\!\pi}, \lambda_{K\!\pi} ) \,\cdot F(Q^2, E^\star_{K\!\pi}) \,.
\end{align}
Alternative forms for the kinematic factor are presented in Ref.~\cite{Briceno:2016kkp} where they are shown to be equivalent to this one. For physical scattering with spacelike, or modestly timelike photon virtualities, $F(Q^2, E^\star_{K\!\pi})$ should be a real function without nearby singularities.

In the case we are to consider of $K \pi$ in a $P$-wave, we expect there to be a relatively narrow $K^*$ resonance that will manifest as a pole singularity in $\mathcal{M}$ at a complex value of $E^\star_{K\!\pi} = m_R - i\Gamma_R/2$, and this pole will also be present in $\mathcal{H}$. The residue of this pole in $\mathcal{H}$ can be used to provide a rigorous definition of the resonance transition form-factor, ${K^* \to K \gamma}$. The scattering amplitude near the pole can be written,
\begin{equation*}
 \mathcal{M}(E^\star_{K\!\pi}) = 16\pi \frac{ (c_R)^2}{\big(m_R - i\Gamma_R/2 \big)^2 - E^{\star2}_{K\!\pi}} + \ldots  \, ,
\end{equation*} 
and it follows that the singular part of the transition amplitude, $\mathcal{H}^\mu$, takes the form, 
\begin{equation*}
\Big( K^\mu \, F\, \sqrt{16\pi} \, \hat{c}_R \Big) 
\cdot \frac{1}{\big(m_R - i\Gamma_R/2 \big)^2 - E^{\star2}_{K\!\pi}}
\cdot \Big( \sqrt{16\pi} \, \hat{c}_R \, k^{\star}_{K\!\pi} \Big) \, ,
\end{equation*}
where the reduced couplings $\hat{c}_R \equiv c_R / k^{\star}_{K\!\pi R}$, and where $K^\mu$ is the kinematic factor in Eqn.~\ref{A}. The leftmost object is interpreted as the ${K^* \to K \gamma}$ vertex, and we define a \emph{resonance transition form-factor},
\begin{equation} \label{fRdefn}
f_R(Q^2) \equiv F \big(Q^2, m_R \!-\! i \tfrac{1}{2}\Gamma_R \big) \cdot \sqrt{16\pi} \, \hat{c}_R \, ,
\end{equation}
which will be a complex-valued function of the photon virtuality. To obtain this function we must analytically continue $F(Q^2, E^\star_{K\!\pi})$ into the complex energy plane, but since this function does not have the unitarity cut and will be parameterized by finite-order polynomials in ${s=(E^\star_{K\!\pi})^2}$, this continuation will be trivial. 

Relationships between the quantities introduced in this section, and the partial decay width $\Gamma(K^* \to K\gamma)$ and a cross-section $\sigma(\gamma K \to K \pi)$ will be presented in Section~\ref{results}.

\begin{table*}
\renewcommand{\arraystretch}{1.3}
\begin{tabular}{r | c|c|c|c|c|c|c|c|c|c|c}
$\mathbf{p}_{K\!\pi} \, \Lambda$ & $[000] A_1^+$ & $[000] T_1^-$ & $[100] A_1$ & $[100] E_2$ & $[110] A_1$ & $[110] B_1$ & $[110] B_1$ & $[111] A_1$ & $[111] E_2$ & $[200] A_1$ & $[100] E_2$\\ \hline
$\ell \le 2$                        & $0$        & $1$           & $0,1,2$     & $1,2$       & $0,1,2$     & $1,2$       & $1,2$       & $0,1,2$     & $1,2$       & $0,1,2$     & $1,2$
\end{tabular}
\caption{$K\pi$ partial-waves of angular momentum $\ell \le 2$ subduced into irreps in a finite cubic volume -- see Refs.~\cite{Leskovec:2012gb,Thomas:2011rh} for more details.}
\label{subd}
\end{table*}

\section{Transition process in finite volume}
With an infinite-volume transition amplitude decomposition in hand, we move to consider how the transition process will look in the finite spatial volume of lattice QCD, where the spectrum of $K\pi$ states is not continuous.

The relationship between the discrete spectrum of states in a finite-volume and the infinite-volume scattering amplitudes is now regularly used in lattice QCD (see Ref.~\cite{Briceno:2017max} and references therein), and takes the form of a quantization condition that can be expressed as
\begin{equation} \label{qc}
0 = \det \Big[  F^{-1}(E^\star_{K\!\pi}, \mathbf{p}_{K\!\pi}; L) + \mathcal{M}(E^\star_{K\!\pi}) \Big] \, ,
\end{equation}
where $F$ and $\mathcal{M}$ are matrices which in the current case of elastic $K\pi$ scattering are in the space of possible partial waves $\ell = 0, 1, 2 \ldots$. $F$ is a dense matrix of known kinematic functions sensitive to the $L \!\times\! L \!\times\! L$ volume, while $\mathcal{M}$ is a diagonal matrix of the (a priori unknown) elastic scattering amplitudes,
\begin{equation*}
\mathcal{M}_{\ell, \ell'} = \delta_{\ell, \ell'} \cdot 16\pi \tfrac{1}{\rho(E^\star_{K\!\pi})} e^{i \delta_\ell( E^\star_{K\!\pi})} \sin \delta_\ell( E^\star_{K\!\pi}),  \,
\end{equation*}
where the phase-space, $\rho = \tfrac{2 k^{\star}_{K\!\pi}}{E^\star_{K\!\pi}}$. The discrete solutions of Eqn.~\ref{qc} in any given volume, $E^\star_n(L)$, are the energy levels expected in a lattice QCD calculation.

The breaking of rotational symmetry by the cubic nature of the lattice boundary means that solutions are sought in irreducible representations, or \emph{irreps}, of the relevant reduced symmetry group. In the case of $K\pi$ scattering, where the scattering hadrons have unequal masses, the subductions of partial waves into these irreps are given in Table~\ref{subd}.

The impact of the finite-volume in $1 + j \to 2$ processes can be subsumed into an effective finite-volume normalization for the discrete hadron-hadron states having energies $E^\star_n(L)$. When there are multiple partial-waves subduced into a particular irrep (or similarly multiple coupled hadron-hadron scattering channels), the finite-volume eigenstates are sensitive to all non-negligible scattering amplitudes. In the case of elastic $K\pi$ scattering in $\mathbf{p}_{K\!\pi} \!\neq\! [000], A_1$ irreps, the state normalizations are sensitive to both $S$--wave and $P$--wave scattering amplitudes\footnote{ The $D$-wave amplitude is estimated, using energy levels in irreps where $\ell=2$ is the leading partial wave, to be negligibly small in the energy region we consider~\cite{Wilson:2019wfr}. }. 

Reproducing the basic argument presented in Ref.~\cite{Briceno:2021xlc}, which itself is a summary and reformulation of the work presented in Refs.~\cite{Lellouch:2000pv,Briceno:2014uqa,Briceno:2015csa}, we start with the relationship between the finite-volume current matrix element, and the infinite-volume transition matrix element defined in the previous section,
\begin{equation} \label{defn}
 \Big|
  \prescript{}{L\!}{ \big\langle} K \big| j^\mu(0) \big| K\pi \big\rangle_L 
    \Big| 
 = \frac{1}{L^3} \frac{1}{\sqrt{2E_K}} \frac{1}{\sqrt{2E_n}} \Big( \mathcal{H}^\mu \cdot \widetilde{R}_n \cdot \mathcal{H}^\mu \Big)^{1/2} \, ,
\end{equation}
where the residue of the finite-volume $K\pi$ propagator is
\begin{align*}
\widetilde{R}&_n(\mathbf{p}_{K\!\pi}, L) \\
&\equiv 2 E_n \cdot \!\lim_{E\to E_n}\! \big( E - E_n \big) \Big( F^{-1}(E^\star, \mathbf{p}_{K\!\pi}; L) + \mathcal{M}(E^\star) \Big)^{-1} \,.
\end{align*}
For $K\pi$ scattering at low energies, the finite-volume object $F$ is a dense matrix in a space $(\ell\!=\!0, \ell\!=\!1)$, while $\mathcal{M}$ is a diagonal matrix in that space\footnote{Except in those irreps listed in Table~\ref{subd} which do not feature $\ell=0$, where these objects are just scalars if we neglect the impact of higher angular momentum partial waves.}. Using an eigen-decomposition of $F + \mathcal{M}^{-1} = \sum_i \mu_i\, \mathbf{w}_i \mathbf{w}_i^\intercal$, we can express $\widetilde{R}_n$ in terms of the slope in energy of that eigenvalue which has a zero-crossing and hence gives rise to a solution of the quantization condition. The resulting form,
\begin{equation*}
 \widetilde{R}_n = \left( - \frac{2E^\star_n}{\mu_0^{\star\prime}}  \right) \mathcal{M}^{-1} \mathbf{w}_0\, \mathbf{w}_0^\intercal \mathcal{M}^{-1} \, ,
\end{equation*}
factorizes and when used in Eqn.~\ref{defn} reduces the equation to a linear relationship. The presence of an explicit factor of $\mathcal{M}^{-1}$ acts to cancel the rapid energy-dependence of $\mathcal{M}$ in Eqn.~\ref{H}. Since the amplitude for $\gamma K \to (K\pi)_{\ell=0}$ is zero by angular momentum and parity, and hence cannot contribute to the finite-volume matrix element, we obtain the result
\begin{align*}
\Big|& \prescript{}{L\!}{ \big\langle} K \big| j^\mu(0) \big| K\pi \big\rangle_L   \Big| \\
  &= \frac{1}{L^3}  \frac{1}{\sqrt{2E_K}} \frac{1}{\sqrt{2E_n}}
  \; \sqrt{-\frac{2E^\star_n}{\mu^{\star\prime}_0}} \, \big| \mathbf{w}_0^{\ell\!=\!1} \big| 
  \;   \frac{1}{k_{K\!\pi}^\star} \mathcal{A}^\mu_{\lambda_{K\!\pi}}  \, .
\end{align*}
The finite-volume correction factor in this expression appears repeatedly, so we will give it the symbol
\begin{equation} \label{rtilde}
   \tilde{r}^\Lambda_n(L) =   \sqrt{-\frac{2E^\star_n}{\mu^{\star\prime}_0}} \, \big| \mathbf{w}_0^{\ell\!=\!1} \big| \frac{1}{k_{K\!\pi}^\star} \, ,
\end{equation}
where the dependence on the $K\pi$ irrep, $\Lambda$, and the volume, $L$, is implicit in the finite-volume energy, the eigenvalue slope, and the $P$--wave component of the eigenvector for each state.  

If we define a ``finite-volume form-factor'' via
\begin{align} \label{FVff}
  \prescript{}{L\!}{ \big\langle} K \big| j^\mu(0) \big|& K\pi; E^\star_n \big\rangle_{\!L} \nonumber \\
  &= \frac{1}{L^3} \frac{1}{\sqrt{2 E_K}} \frac{1}{\sqrt{2 E_n}} \cdot K^\mu \cdot F_L(Q^2, E^\star_n ) \, ,
\end{align}
where $K^\mu$ is the same kinematic factor which appears in Eqn.~\ref{A}, then the relationship between finite-volume and infinite-volume form-factors is simply
\begin{equation} \label{FVcorr}
 F(Q^2, E^\star_{K\!\pi} \!=\! E^\star_n) = \frac{1}{\tilde{r}_n(L)} F_L(Q^2, E^\star_n) \, .
\end{equation}

The job of the lattice calculation is to determine finite-volume form-factors, $F_L$, at multiple discrete $(Q^2,E^\star_n)$ points by analyzing the time-dependence of appropriate three-point correlation functions. We can then correct for the finite-volume effect using $\tilde{r}$, which can be computed once the $K \pi$ scattering amplitudes in \mbox{$S$--wave} and \mbox{$P$--wave} are known. These scattering amplitudes can be constrained through the quantization condition in Eqn.~\ref{qc} using the discrete energy spectrum obtained in variational analysis of two-point correlation functions.

\section{Finite-volume spectrum and $K\pi$ elastic scattering amplitudes} \label{Kpi}

Our approach to determining the lattice QCD spectrum in a range of irreps has been described in detail elsewhere~\cite{Briceno:2017max, Dudek:2012xn, Dudek:2012gj, Wilson:2019wfr}, but in short, we make use of a large basis of both ``single-meson-like'' and ``meson-meson-like'' operators to compute a matrix of correlations functions, which is then analysed variationally by solving a generalized eigenvalue problem. The eigenvalues of this problem as a function of timeslice are fitted to obtain the discrete energy values, while the eigenvectors provide the weights in a linear superposition of the original operators which optimally overlap with each state in the spectrum. We will later make use of these optimized operators to access three-point correlation functions featuring excited states~\cite{Shultz:2015pfa}.

\begin{figure*}
\includegraphics[width=1.0 \textwidth]{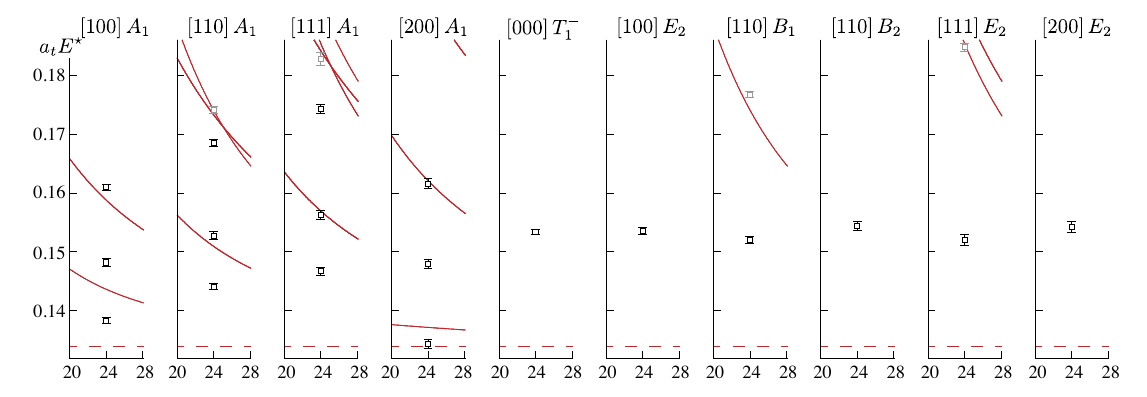}
\caption{Finite-volume spectra by irrep extracted from variational analysis of matrices of correlation functions. Solid red curves indicate non-interacting $K\pi$ energy levels and the dashed red line the $K\pi$ threshold. The first inelastic threshold, into $K\pi\pi$, is at $a_t E^\star = 0.181$. Energy levels indicated by gray points are not used in the subsequent transition matrix-element analysis. }
\label{spec}
\end{figure*}

\begin{figure*}
\vspace{1cm}
\includegraphics[width=1.0\columnwidth]{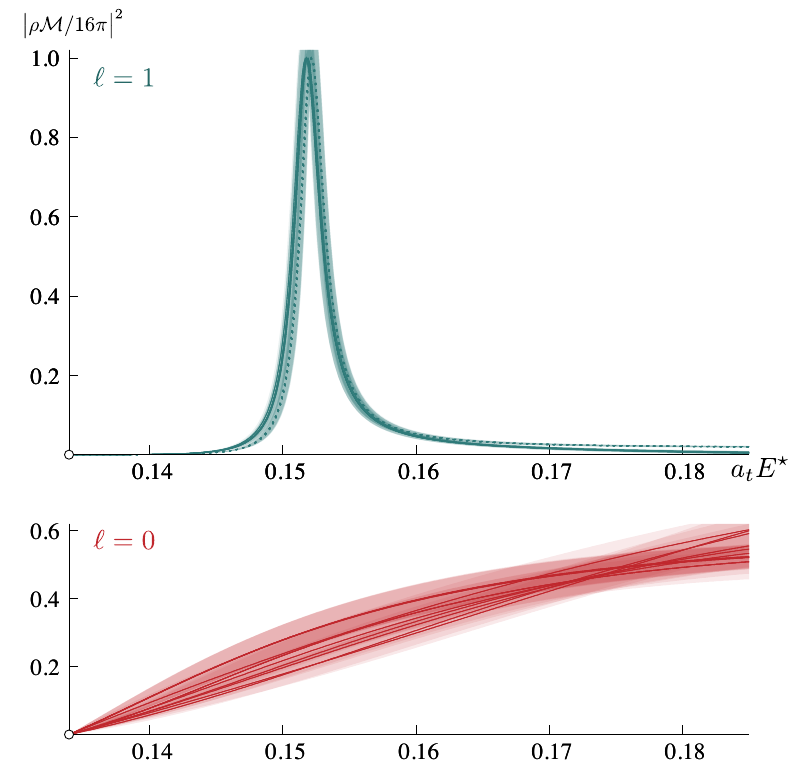}
\includegraphics[width=0.75\columnwidth]{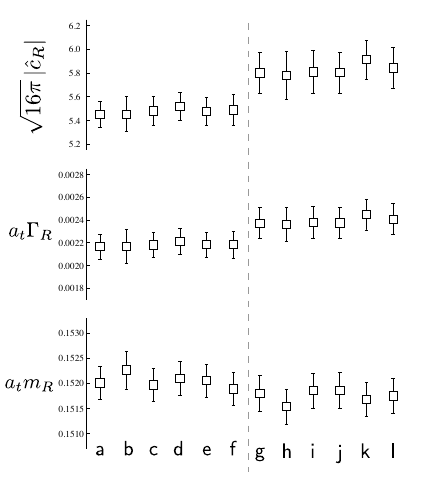}
\caption{Left panel: Elastic $K\pi$ scattering amplitudes in $P$--wave and $S$--wave determined using the finite-volume spectra in Figure~\ref{spec}. Twelve different amplitude parameterizations are used, generating the visible spread in the lower panel, and the more modest spread in the upper panel. In the upper panel, parameterizations using a $P$--wave Breit-Wigner are indicated by the dashed lines. Right panel: $K^*$ pole parameters obtained by analytic continuation into the complex plane of the twelve $P$--wave amplitude parameterizations.}
\label{amps}
\end{figure*}

The current system of elastic $K\pi$ scattering\footnote{The system is elastic up to $a_t E^\star = 0.181$ where $K\pi\pi$ opens, followed immediately by $K\eta$.} was previously considered in Ref.~\cite{Wilson:2019wfr} using a slightly larger number of configurations. The spectra extracted in this calculation, shown in Figure~\ref{spec}, are statistically compatible with the spectra in Ref.\cite{Wilson:2019wfr}. The irreps in the six rightmost columns, which receive no contribution from \mbox{$S$--wave} scattering, have a clear isolated state near to $a_t E^\star = 0.152$ which signals the presence of a narrow resonance in \mbox{$P$--wave}. Using all these irrep spectra as constraint, plus the $[000]\, A_1^+$ irrep, solving the quantization condition using parameterizations of $S$--wave and $P$--wave elastic scattering leads to the amplitudes shown in Figure~\ref{amps}.

As was previously reported in Ref.~\cite{Wilson:2019wfr}, there is a well-determined narrow $K^*$ resonance in $P$--wave, while the $S$--wave shows a slow growth from threshold, that may or may not be due to a broad $\kappa$ resonance. Our focus in this paper is on the $K^*$ resonance, and in order to explore the sensitivity to amplitude parameterization we make use of twelve different choices: six amplitudes having a Breit-Wigner description of the $P$--wave, and various descriptions of the $S$--wave ($\mathrm{BW}_\mathsf{a \ldots f}$), and six amplitudes in which the $P$--wave is described by a $K$--matrix featuring a single pole plus a polynomial ($\mathrm{KM}_\mathsf{g \ldots l}$). More complete descriptions of the amplitudes are given in Appendix~\ref{app::Mparams}.


We observe in Figure~\ref{amps} that the $P$--wave amplitude is largely insensitive to the details of the parameterization, with the only notable feature being the more rapid fall off at energies well above the resonance for the more flexible $K$--matrix forms relative to the Breit-Wigner form. Also shown in Figure~\ref{amps} are the $P$--wave resonance pole parameters for the twelve variations, and we note that compared to the $K$--matrix forms, the Breit-Wigner forms have a resonance mass that is slightly higher, and widths and reduced couplings that are systematically slightly smaller. In subsequent analysis we will default to amplitude $\mathrm{KM}_\mathsf{g}$, which has resonance pole parameters,
\begin{align*}
	a_t\, m_R & = 0.1518(4) \, ,\\         
	a_t \, \Gamma_R &= 0.0024(1) \, ,\\    
	\sqrt{16\pi}\, \hat{c}_R & = 5.80(17) - i \, 0.19(3)   \, ,
\end{align*}
and consider the others as systematic variations.

\section{Three-point correlation functions}

In order to access finite-volume transition matrix-elements we compute three-point correlation functions having an optimized kaon operator\footnote{Constructed from the linear superposition of ``single-hadron-like'' operators that gave the ground-state in variational analysis of two-point correlation functions.} at a definite momentum $\mathbf{p}_K$  at a fixed timeslice $\Delta t/a_t = 32 $, an optimized $K\pi$ operator at a definite momentum $\mathbf{p}_{K\!\pi}$ in an irrep $\Lambda$\footnote{Taken from the analysis described in the previous section.} at a fixed timeslice $0$, and an insertion of the spatially-directed electromagnetic current, projected to definite momentum $\mathbf{q} = \mathbf{p}_K \!-\! \mathbf{p}_{K\!\pi}  $, at each timeslice $t/a_t$ between $0$ and $32$. \\

The optimized operators are normalized such that
\begin{equation*}
  \big\langle n^\prime \big | \Omega^\dag_n(0) \big| 0 \big\rangle = \sqrt{2E_n} \, \delta_{n, n^\prime} + \ldots \, ,
\end{equation*}
and it follows that the three-point correlation functions have a spectral decomposition,
\begin{align} \label{threept}
 \big\langle &0 \big| 
 \Omega^{\vphantom{\dag}}_K(\mathbf{p}_K, \Delta t) \, j(\mathbf{q}, t) \, \Omega^\dag_{K\!\pi}(\mathbf{p}_{K\!\pi}, 0)
 \big| 0 \big\rangle \nonumber \\
  &= L^3  \sqrt{2E_K} \sqrt{2E_n} \, e^{- E_K ( \Delta t - t) } \, e^{-E_n t} \cdot\!  \prescript{}{L\!}{ \big\langle} K \big| j(0) \big| K\pi; E^\star_n \big\rangle_{\!L} \nonumber \\
  &\;\;\; +\ldots \, ,
\end{align}
where the ellipsis represents possible suppressed contributions from source and sink states other than the ones optimally produced by the source and sink operators. Using our definition of a ``finite-volume form-factor'' in Eqn.~\ref{FVff}, we obtain
\begin{align*}
 \big\langle 0& \big| 
 \Omega^{\vphantom{\dag}}_K(\mathbf{p}_K, \Delta t) \, j(\mathbf{q}, t) \, \Omega^\dag_{K\!\pi}(\mathbf{p}_{K\!\pi}, 0)
 \big| 0 \big\rangle \nonumber \\
  &= e^{- E_K ( \Delta t - t) } \, e^{-E_n t} \cdot K \, F_L(Q^2, E^\star_n ) + \ldots \, ,
\end{align*}
and from this we can define an effective (timeslice-dependent) form-factor that can be constructed by multiplying the correlation function by the appropriate time-dependence and dividing by the kinematic factor,
\begin{align} \label{FLt}
F_L(Q^2&, E^\star_n; t) \equiv \nonumber \\
  &  e^{E_K ( \Delta t - t) } \cdot e^{E_n t} \cdot \tfrac{1}{K} \cdot \big\langle 0 \big| 
 \Omega^{\vphantom{\dag}}_K(\Delta t) \, j(t) \, \Omega^\dag_{K\!\pi}(0)
 \big| 0 \big\rangle  \, .
\end{align}
We will describe below how this object is analysed, but first we turn to the construction of the relevant three-point functions within a lattice QCD computation.

\subsection{Construction of three-point functions}

\begin{figure*}
\includegraphics[width=0.74\textwidth]{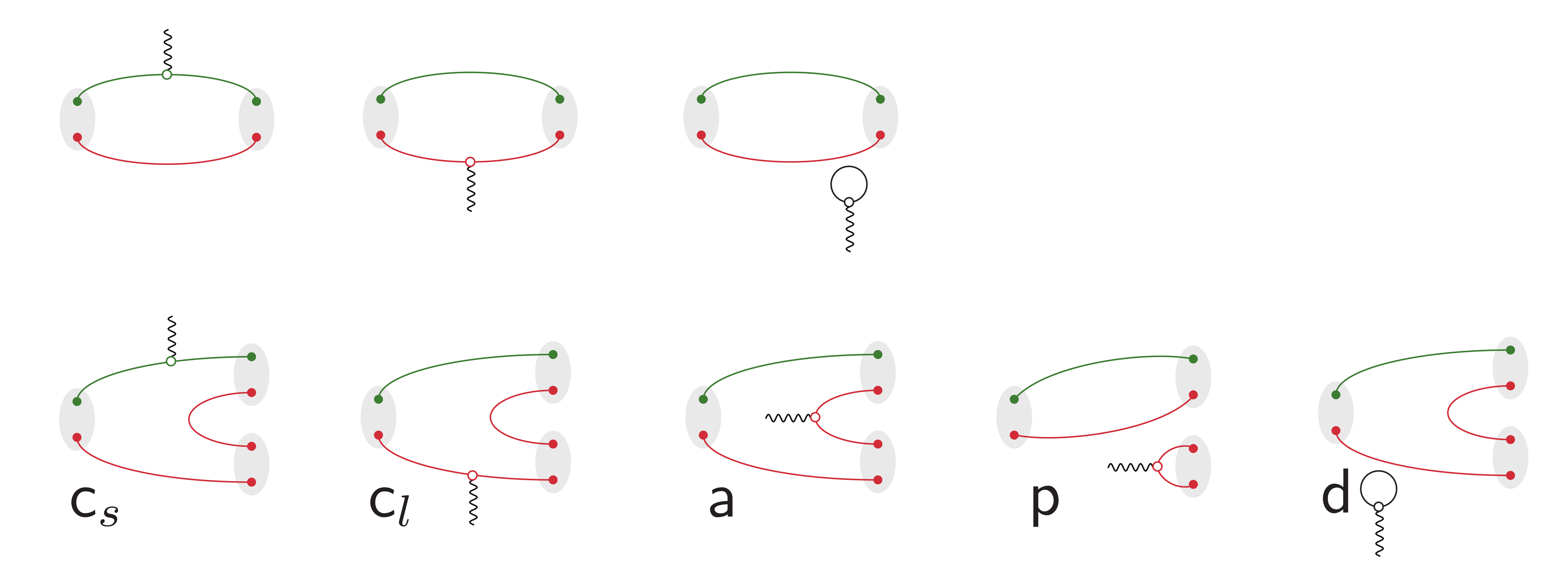}
\caption{Wick diagrams required in the computation of three-point functions needed to determine $\gamma K \to K\pi$. In each diagram timeslice $0$ is on the right and $\Delta t$ is on the left. Red lines represent light-quark propagation, and green lines strange-quark propagation.
The rightmost diagram in each row includes a completely disconnected loop in which all three quark flavors participate -- the electric charges are such that in the limit of equal quark masses these sum to zero. The weights of these diagrams in the construction of the correlation functions we use are presented in Appendix~\ref{app::wicks}. }
\label{wicks}
\end{figure*}

The three-point functions we wish to compute feature the electromagnetic current, $+\tfrac{2}{3}\bar{u} \gamma^\mu u - \tfrac{1}{3}\bar{d} \gamma^\mu d - \tfrac{1}{3}\bar{s} \gamma^\mu s$, and as such the current insertion will appear on both light-quark and strange-quark propagators. The optimized $K\pi$ operators we use at the source are, in general, linear superpositions of both ``single-meson-like'' operators and ``meson-meson-like'' operators, and as such the three-point function Wick contractions contain diagrams from both the top and bottom rows of Figure~\ref{wicks}. All these diagrams are computed without further approximation, with the exception of the rightmost entry in each row which features a completely disconnected current insertion -- these are more challenging to compute, and are expected to contribute relatively little, indeed they vanish in the limit of exact $SU(3)$ flavor symmetry. These completely disconnected diagrams are not computed. The particular combinations of diagrams that are needed are discussed in Appendix~\ref{app::wicks}.

The diagrams we compute require, in addition to light and strange perambulators describing quark propagation within the distillation framework~\cite{HadronSpectrum:2009krc}, also ``generalized perambulators'' which carry the current insertion. The ${(0\to t \to \Delta t)}$ versions of these are are described in Ref.~\cite{Shultz:2015pfa}, while the extension for the case ${(0 \to t \to 0)}$ needed for the third and fourth diagrams in the bottom row is straightforward.

The renormalized electromagnetic current can be expressed as
\begin{equation} \label{j}
j_{\mathrm{em}, \mathrm{phys}} = Z_V^l \Big( \tfrac{1}{\sqrt{2}} j_{\rho, \mathrm{lat}} + \tfrac{1}{3\sqrt{2}} j_{\omega_l, \mathrm{lat}} \Big) 
+ Z_V^s \Big( -\! \tfrac{1}{3} j_{\omega_s, \mathrm{lat}}   \Big) \, ,
\end{equation}
where isospin-basis currents are,
\begin{align*}
j_\rho \equiv \tfrac{1}{\sqrt{2}} \big( \bar{u} \Gamma u - \bar{d} \Gamma d \big), \, j_{\omega_l} \equiv \tfrac{1}{\sqrt{2}} \big( \bar{u} \Gamma u + \bar{d} \Gamma d \big), \, j_{\omega_s} \equiv \bar{s} \Gamma s \, .
\end{align*}
On the anisotropic Clover-improved lattices used here, the appropriate current includes a tree-level $\mathcal{O}(a)$ improvement term\footnote{A derivation is presented in Ref.~\cite{Shultz:2015pfa}.}, such that for a spatially-directed insertion,
\begin{equation*}
\bar{\psi}\Gamma\psi = \bar{\psi}\gamma_i \psi + \tfrac{1}{4}(1-\xi) a_t \partial_4 \big(\bar{\psi} \sigma_{4i} \psi \big) \, .
\end{equation*}

\begin{figure}
\includegraphics[width=1.06\columnwidth]{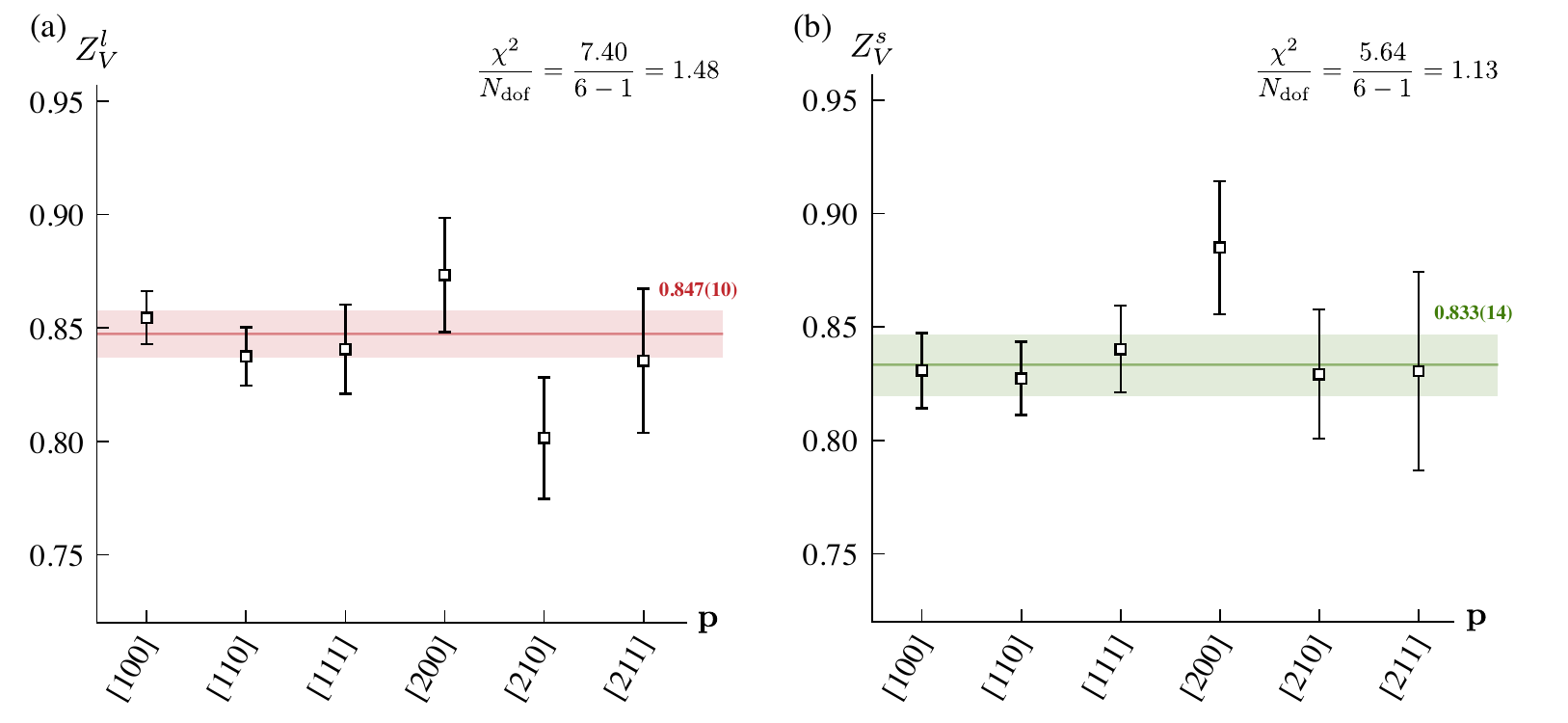}
\caption{Vector current renormalization factors for light and strange quarks determined as described in the text.
}
\label{ZV}
\end{figure}

The light-- and strange--quark current renormalization factors are determined non-perturbatively using calculations of the charged pion and kaon electromagnetic form-factors at zero virtuality, which should both be equal to the electromagnetic charge of these particles, 1. Considering
\begin{align*}
\big\langle \pi^+(\mathbf{p}) \big| j^i_{\mathrm{em}, \mathrm{phys}} \big| \pi^+(\mathbf{p})\big\rangle &= F^{\pi^+}_\mathrm{em}(Q^2\!=\!0) \cdot 2 p^i \\
&= Z_V^l \tfrac{1}{\sqrt{2}} \big\langle \pi^+(\mathbf{p}) \big| j^i_{\rho, \mathrm{lat}} \big| \pi^+(\mathbf{p})\big\rangle \, ,
\end{align*}
and setting $F^{\pi^+}_\mathrm{em}(Q^2\!=\!0) = 1$ yields
\begin{equation*}
Z_V^l = \frac{1}{\tfrac{1}{\sqrt{2}} \big\langle \pi^+(\mathbf{p}) \big| j^i_{\rho, \mathrm{lat}} \big| \pi^+(\mathbf{p})\big\rangle / 2p^i }	\, .
\end{equation*}
The matrix element in the denominator is computed for $\mathbf{p} = [100], [110], [111], [200], [210], [211]$ (averaged over rotations), and the results are shown in Figure~\ref{ZV}(a). A correlated constant fit to the six data points gives $Z_V^l = 0.847(10)$.

Considering the charged kaon,
\begin{align*}
\big\langle K^+(\mathbf{p}) \big| j^i_{\mathrm{em}, \mathrm{phys}} &\big| K^+(\mathbf{p})\big\rangle = F^{K^+}_\mathrm{em}(Q^2\!=\!0) \cdot 2 p^i \\[1.2ex]
= &Z_V^l \tfrac{1}{\sqrt{2}} \big\langle K^+(\mathbf{p}) \big| j^i_{\rho, \mathrm{lat}} \big| K^+(\mathbf{p})\big\rangle \\
+ &Z_V^l \tfrac{1}{3\sqrt{2}} \big\langle K^+(\mathbf{p}) \big| j^i_{\omega_l, \mathrm{lat}} \big| K^+(\mathbf{p})\big\rangle \\
- &Z_V^s \tfrac{1}{3} \big\langle K^+(\mathbf{p}) \big| j^i_{\omega_s, \mathrm{lat}} \big| K^+(\mathbf{p})\big\rangle \, ,
\end{align*}
and noting that the first two matrix elements are equal up to (neglected) disconnected current contributions, setting $F^{K^+}_\mathrm{em}\!(Q^2\!=\!0) = 1$ yields
\begin{equation*}
Z_V^s = \frac{ Z_V^l 2\sqrt{2}\,  \big\langle K^+(\mathbf{p}) \big| j^i_{\rho, \mathrm{lat}} \big| K^+(\mathbf{p})\big\rangle/2p^i - 3  }{\big\langle K^+(\mathbf{p}) \big| j^i_{\omega_s, \mathrm{lat}} \big| K^+(\mathbf{p})\big\rangle / 2p^i}	\, .
\end{equation*}
Computing the two relevant matrix elements and using the previously obtained value of $Z_V^l$, we obtain the results shown in Figure~\ref{ZV}(b), and an estimate of ${Z_V^s = 0.833(14)}$.

\begin{figure}
\includegraphics[width=0.96\columnwidth]{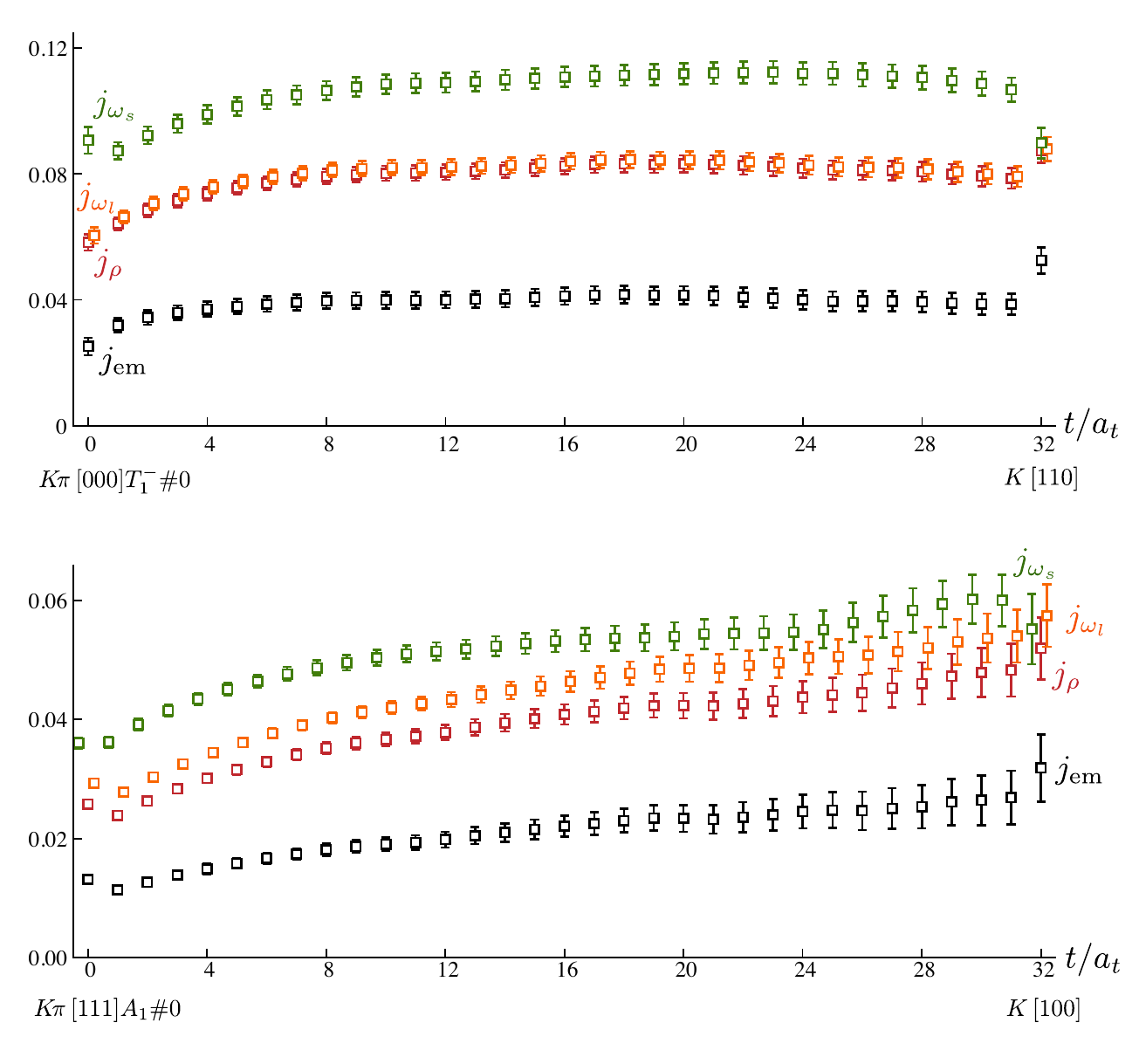}
\caption{Sample three-point functions as in Eqn.~\ref{threept} with the leading time-dependence divided out. Shown are the insertions of the isospin-basis currents, and the relevant combination defined in Eqn.~\ref{j}.
}
\label{jem}
\end{figure}

Three-point functions of the type given in Eqn.~\ref{threept} using the current in Eqn.~\ref{j} are constructed by summing weighted combinations of three-point functions computed using the isospin-basis currents. Examples are shown in Figure~\ref{jem}, where we observe comparable statistical quality for each current.

\pagebreak
\subsection{Analysis of three-point functions}

We compute a set of three-point functions based upon the following choices: 
\begin{itemize}
\item at $t=0$, an optimized operator corresponding to each black point in Figure~\ref{spec}, having any allowed lattice rotation of the specified momentum. If the irrep is more than one-dimensional, all rows are considered;
\item at all $0 \le t/a_t \le 32$ a spatial current insertion having momentum $[000], [100], [110], [111]$ or $[200]$ (and \emph{not} rotations of these specific directions). Rather than three cartesian directions for the current, the subductions of a vector for the relevant momentum are used;
\item at $\Delta t/a_t = 32$, an optimized operator for a kaon with a momentum $\le [211]$, with all allowed lattice rotations considered.
\end{itemize}

\begin{figure}[h]
\includegraphics[width=0.95\columnwidth]{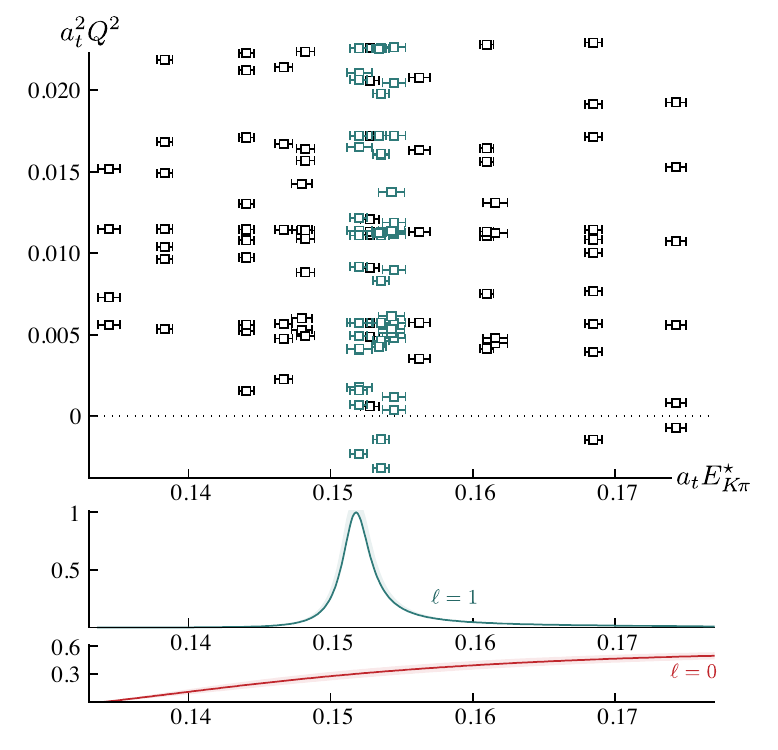}
\caption{Kinematic coverage of the $E^\star, Q^2$ plane provided by our computed set of three-point correlation functions. Points in cyan indicate levels in $K\pi$ irreps that have no contribution from $K\pi$ $S$--wave. Also shown for illustration are the $\mathrm{KM}_\mathsf{g}$ $P$--wave and $S$--wave scattering amplitudes over the same $E^\star$ range. }
\label{kin}
\end{figure}

Within these we compute all correlation functions which have a non-zero kinematic factor, $K$. This leads to over 1000 three-point correlation functions, but many of these correspond to common kinematic points, $(Q^2, E_n^\star)$. In practice we find 128 such points spread over 18 values of $E_n^\star$, distributed as shown in Figure~\ref{kin}. It is clear that the use of $A_1$ irreps (which suffer ``pollution'' from the $K\pi$ $S$--wave) allows access to a much broader region in $E^\star$, along with an increased density of points close to the resonance.

For each computed correlation function, the quantity $F_L(t)$ defined in Eqn.~\ref{FLt} is formed, which if the source and sink operators were perfectly optimized would be a constant in time, but which in practice can have curvature from source and/or sink excitation contributions. Correlated fits to the time-dependence are carried out using fit-forms: a constant, a constant plus a decaying exponential at the source, a constant plus a decaying exponential at the sink, or a constant plus both source and sink exponentials. Such fits are performed for a large number of fit-windows in time, and for each such fit, a version of the Akaike Information Criterion is formed from the $\chi^2$ and the number of degrees of freedom, $w = \exp\left[-\tfrac{1}{2}(\chi^2 - 2 N_\mathrm{dof}) \right]$. This number is treated as a probability in an average over fits along the lines presented in Ref.~\cite{Jay:2020jkz}\footnote{The average is done on the ensemble of fit values, allowing correlations to be retained.}. In most cases the difference between the value and uncertainty of the extracted constant from the fit-window with the highest probability and the value and uncertainty of the ``model average'' is rather small. The procedure is described in more detail in Appendix~\ref{app::threept}. A small number of correlation functions prove to not have even an approximate plateau region, such that timeslice fits are unreliable -- these are excluded from further analysis.

In the large number of cases where there are multiple correlation functions having the same $(Q^2, E^\star_{K\!\pi})$ but which differ in momentum directions and/or current insertion irrep, we perform a correlated average of the extracted constant values, to form a single $F_L$ value to be used later. An example of the procedure is presented in Appendix~\ref{app::threept}.

\section{Finite-volume correction}

\begin{figure*}
\includegraphics[width=0.95\textwidth]{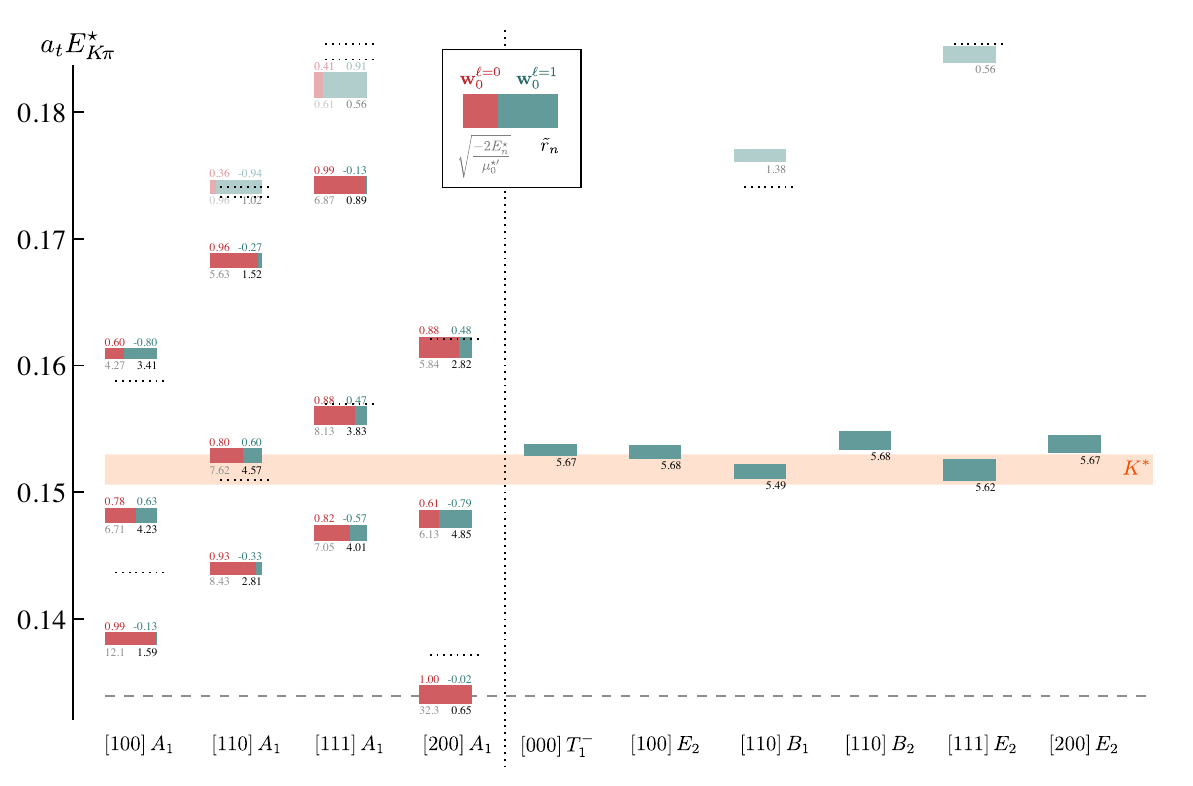}
\caption{Components in the construction of the finite-volume normalization factor, $\tilde{r}_n$, defined in Eqn.~\ref{FVcorr} for each discrete energy level using the amplitude $\mathrm{KM}_\mathsf{g}$. The red/cyan coloring indicates the relative contribution of $\ell=0, \ell=1$ to the eigenvector, normalized as $(\mathbf{w}_0^{\ell=0})^2 + (\mathbf{w}_0^{\ell=1})^2 = 1$. The vertical height of each box indicates the statistical uncertainty on the energy level. Small dashed lines show the $K\pi$ non-interacting energies for this lattice volume, and the location of the $K^*$ resonance ($m_R - \tfrac{1}{2}\Gamma_R : m_R + \tfrac{1}{2}\Gamma_R$) is indicated by the orange band. }
\label{ll}
\end{figure*}

As indicated by Eqn.~\ref{FVcorr}, the finite-volume form-factors, $F_L(Q^2, E^\star_n)$ must be scaled by a factor $1/\tilde{r}_n$ to correctly describe the transition in infinite-volume. These factors, $\tilde{r}_n$, computed using Eqn.~\ref{rtilde}, for the amplitude parameterization $\mathrm{KM}_\mathsf{g}$, are presented in Figure~\ref{ll}.

The relative contributions from $S$--wave and $P$--wave scattering in finite-volume are quantified by size of eigenvector components, shown as the red/cyan fractions of each box, recalling that only the $P$--wave component enters $\tilde{r}_n$ in Eqn.~\ref{FVcorr}. In the figure we see clearly that for energy levels in $A_1$ irreps well below the $K^*$ resonance, there is dominance of the $S$--wave contribution, with much smaller contributions from the $P$--wave. We observe that the size of $\tilde{r}_n$ can vary considerably between states, but note that the energy levels in non-$A_1$ irreps lying close to the $K^*$ resonance all have approximately the same value of $\tilde{r}_n$. This is, in fact, the expected behavior when a narrow resonance is present, as we can show that in the limit of a vanishingly small width, for those states lying close to $m_R$, $\tilde{r}_n \to \sqrt{16\pi} \, \hat{c}_R$ -- a derivation is provided in Appendix~\ref{app::narrow}.

The degree to which the $\tilde{r}_n$ values are sensitive to the scattering amplitude parameterization is explored in Appendix~\ref{app::Mparams}, where we observe that the only significant difference is between the Breit-Wigner-based $P$--wave amplitudes, and the \mbox{$K$--matrix}-based $P$--wave amplitudes, and where the largest effect is at energies far from the resonance position, where the Breit-Wigner may not be accurate.

\begin{figure*}
\includegraphics[width=0.99\textwidth]{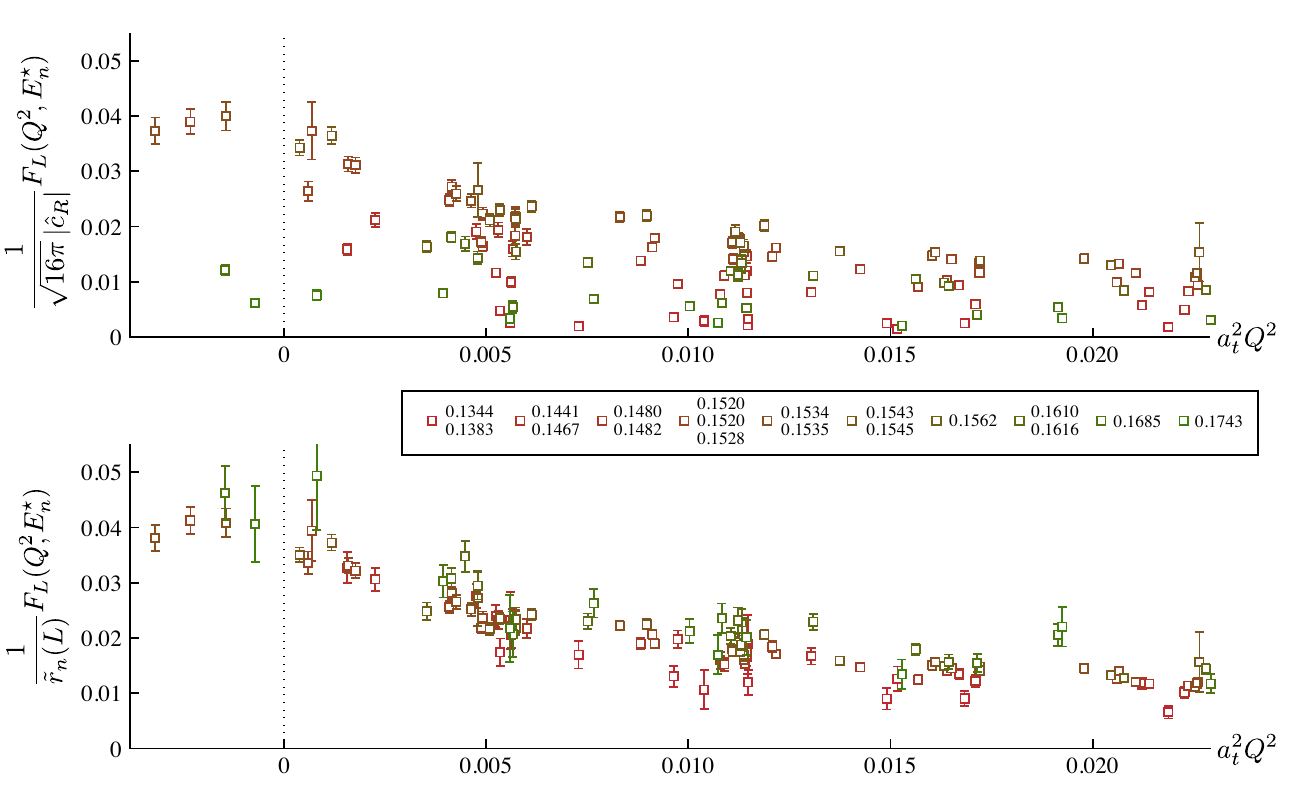}
\caption{Upper panel: Finite-volume form-factor values, $F_L$, extracted from three-point correlation functions, scaled by the constant $K^* \to K\pi$ reduced coupling, as a function of $Q^2$. Colors indicate $E_n^\star$ ranges. Lower panel: Infinite-volume form-factor values, obtained by finite-volume correcting $F_L$ according to Eqn.~\ref{FVcorr}. $\tilde{r}_n$ computed using $K\pi$ amplitude parameterization $\mathrm{KM}_\mathsf{g}$. }
\label{fv_all}
\end{figure*}

\vspace{5mm}
The upper panel of Figure~\ref{fv_all} shows the complete set of $F_L(Q^2, E_n^\star)$ values extracted from our three-point correlation functions, scaled by a common constant. It is quite clear that the data show a considerable degree of scatter, and no obvious trend with increasing energy, $E^\star_n$. The lower panel of the figure shows the same data after correction by a factor of $1/\tilde{r}_n$, where the data comes into close agreement, with only a gentle increase in magnitude with increasing $E_n^\star$ remaining.

\begin{figure*}
\includegraphics[width=0.99\textwidth]{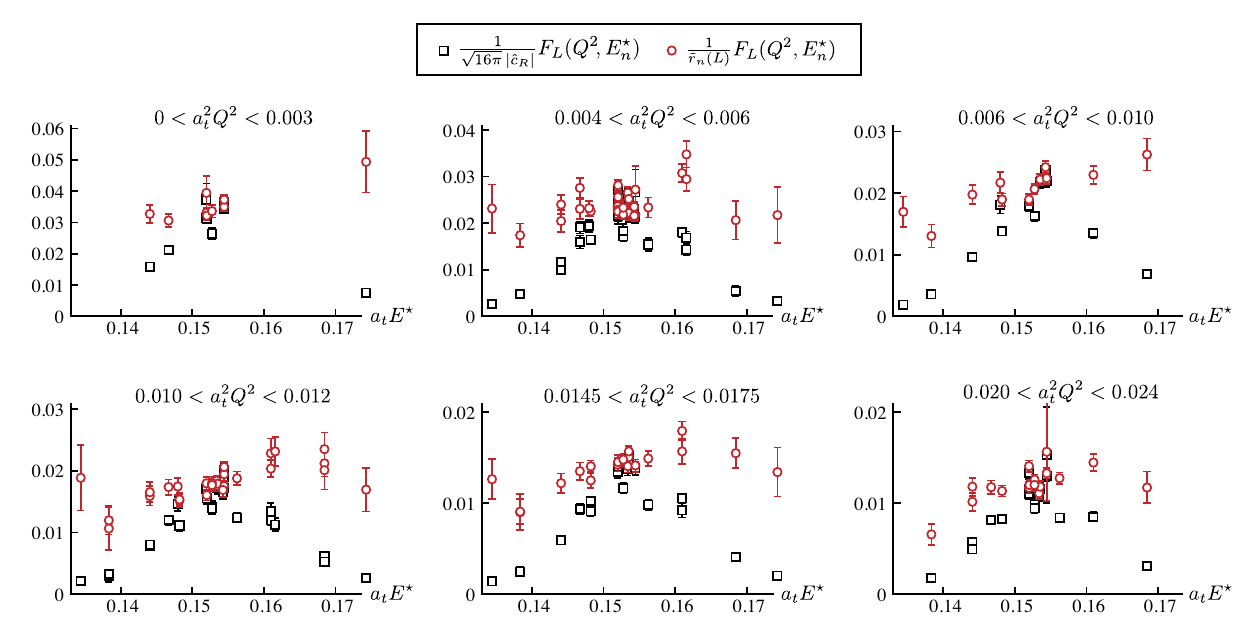}
\caption{Finite-volume (black squares) and infinite-volume (red circles) form-factor values in bins of $Q^2$.}
\label{fv_qsq_bins}
\end{figure*}

Another way to illustrate the impact of the finite-volume correction is to examine the data points binned in relatively small $Q^2$-bins, as shown in Figure~\ref{fv_qsq_bins}. The (uncorrected) black points show significant energy dependence, and scatter far outside the statistical uncertainties\footnote{Note that the ``bump-like'' structure in this data is nothing like the lineshape of the $K^*$ resonance which is a much sharper peak about $a_t E^\star \approx 0.152$.}, while the (corrected) red points show the much milder energy dependence expected of this infinite-volume quantity.

The corrected data shown in Figures~\ref{fv_all} and \ref{fv_qsq_bins} supplies the input to the next stage of the calculation, where we seek to parameterize the $Q^2$ and $E^\star$ dependence of $F(Q^2,E^\star)$ in order to obtain a description of all the data. We take two different approaches to this, motivated by the observation that after finite-volume correction, the resulting 128 data points are found to have a considerable degree of data correlation whose origin is in the high degree of correlation between the $\tilde{r}_n$ values for different energy levels.

\vspace{8mm}
The first approach, which we call ``level-by-level fitting'', circumvents the high degree of correlation by first fitting the $Q^2$ dependence of the uncorrected $F_L(Q^2, E^\star_n)$ for each energy level. The parameters in these fits are, where necessary, corrected with $\tilde{r}_n$, and then considered as functions of $E^\star$. The $E^\star$ behavior of the parameters is then fitted, yielding a parameterization of $F(Q^2,E^\star)$.

The second approach, which we call ``global fitting'', attempts to fit all of the corrected data $F(Q^2,E^\star)$ simultaneously, using an appropriate truncation of the data correlation matrix. We will find that we obtain compatible results from the two methods, utilizing within each a range of parameterization forms to assess systematic variations.

\subsection{Level-by-level fitting}

\begin{figure}
\!\!\!\includegraphics[width=1.04\columnwidth]{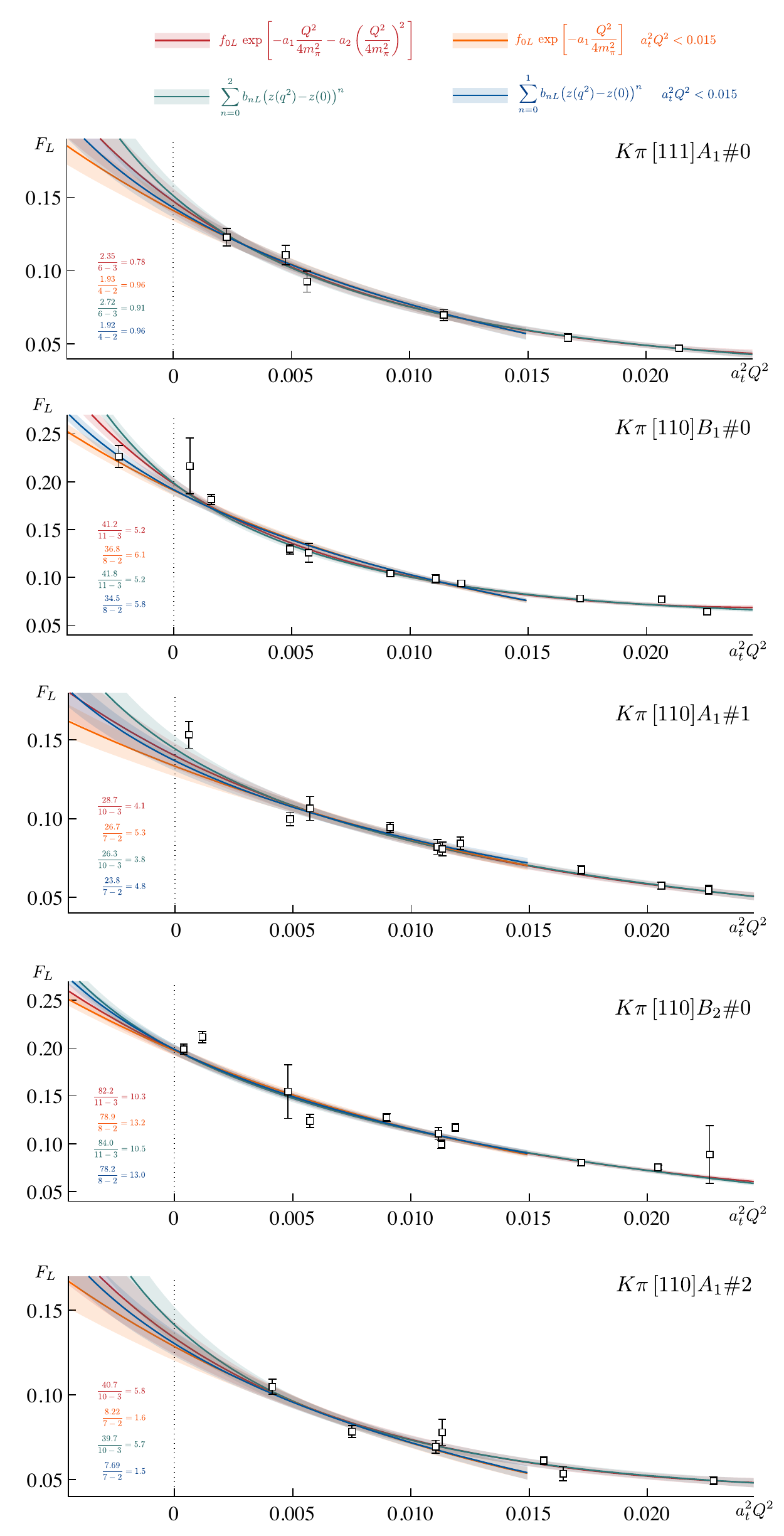}
\caption{ Fits using Eqns.~\ref{exp_poly},\ref{z_poly} to the $Q^2$ dependence of $F_L$ for five sample energy levels. Red, cyan fits to full $Q^2$ range; orange, blue to limited region, $a_t^2 Q^2 < 0.015$. }
\label{Qsq_fits}
\end{figure}

In this approach, for each $K\pi$ energy level, we consider the $Q^2$ dependence of the uncorrected $F_L$, for which we typically have a handful of discrete points in $Q^2$ (as can be seen in each vertical slice of Figure~\ref{kin}). Figure~\ref{Qsq_fits} shows a sample of this kind of fitting, for five energy levels.

The selection of parameterizations need not be sophisticated -- since no attempt will be made to analytically continue in $Q^2$, all that is really required are functions which can interpolate the data points along a limited region of the real $Q^2$ axis. In practice, we are mostly interested in $Q^2=0$ to yield the real-photon transition~\footnote{In principle, if a detailed understanding of the $Q^2$ dependence was sought, the analysis could be done at the level of the separate isospin currents, which have different singularities at timelike $Q^2$. In this first calculation only the $j_\mathrm{em}$ combination is analysed.}.

One form which proves capable of describing the data is an exponential of a polynomial in $Q^2$, 
\begin{equation} \label{exp_poly}
F_L(Q^2) = f_{0L} \cdot \exp \left[ - \sum_{n=1}^N a_n \left(\tfrac{Q^2}{4m_\pi^2}\right)^n  \right] \, .
\end{equation}
This is a phenomenological form with no particularly good physics justification, but it is at least free from unphysical nearby singularities. $4 m_\pi^2$ is introduced into a ratio with $Q^2$ as an appropriate dimensionful scale to render the parameters $\{a_n\}$ dimensionless. The subscript $L$ on $f_0$ indicates that this parameter will require finite-volume correction, while the parameters $\{a_n\}$ do not.

\vspace{5mm}
Another option with somewhat more physical motivation (see e.g. Ref.~\cite{Boyd:1994tt}), is a polynomial in a conformal mapping variable $z(Q^2)$, 
\begin{equation} \label{z_poly}
F_L(Q^2) = \sum_{n=0}^N b_{nL} \, \big( z(Q^2) - z(0) \big)^n \, ,
\end{equation}
where
\begin{equation} \label{z}
 z(Q^2) = \frac{\sqrt{Q^2 + t_\mathrm{cut}} - \sqrt{Q^2_0 + t_\mathrm{cut}} }{\sqrt{Q^2 + t_\mathrm{cut}} + \sqrt{Q^2_0 + t_\mathrm{cut}} } \, ,
\end{equation}
maps the entire complex $Q^2$ plane away from the unitarity cut (${Q^2 = -\infty \to - t_\mathrm{cut} }$) into a disk of radius $1$ in $z$. In our case where the isovector vector current features, the appropriate choice for $t_\mathrm{cut}$ is  $(2m_\pi)^2$. We may freely choose the scale $Q^2_0$ so that the $Q^2$ range of our data set (within $a_t^2 Q^2 = -0.005 \to 0.030$) lies in a region symmetrically distributed about $z=0$. The choice $a_t^2 \, Q_0^2 = 0.0035$ achieves this. Since the parameters $\{b_{nL}\}$ appear linearly in Eqn.~\ref{z_poly}, they will all require finite-volume correction by a factor $1/\tilde{r}$.

We observe in Figure~\ref{Qsq_fits} that the trend of the data can be captured by these fit-forms, over either the full $Q^2$ range, or over a limited $a_t^2 Q^2 < 0.015$ range. The $\chi^2/N_\mathrm{dof}$ values for these $Q^2$ fits can be large, and we suggest that this is not a limitation of the fit-forms but rather reflects the scatter in the data points that we put down to fluctuations due to the timeslice fitting approach, and possibly discretization effects. In those cases where there are no timelike $Q^2$ data points and there are no data points at very low spacelike $Q^2$ values, we observe that there is some model dependency in the extrapolated value of $F_L(Q^2\!=\!0)$.

\begin{figure}
\includegraphics[width=1.0\columnwidth]{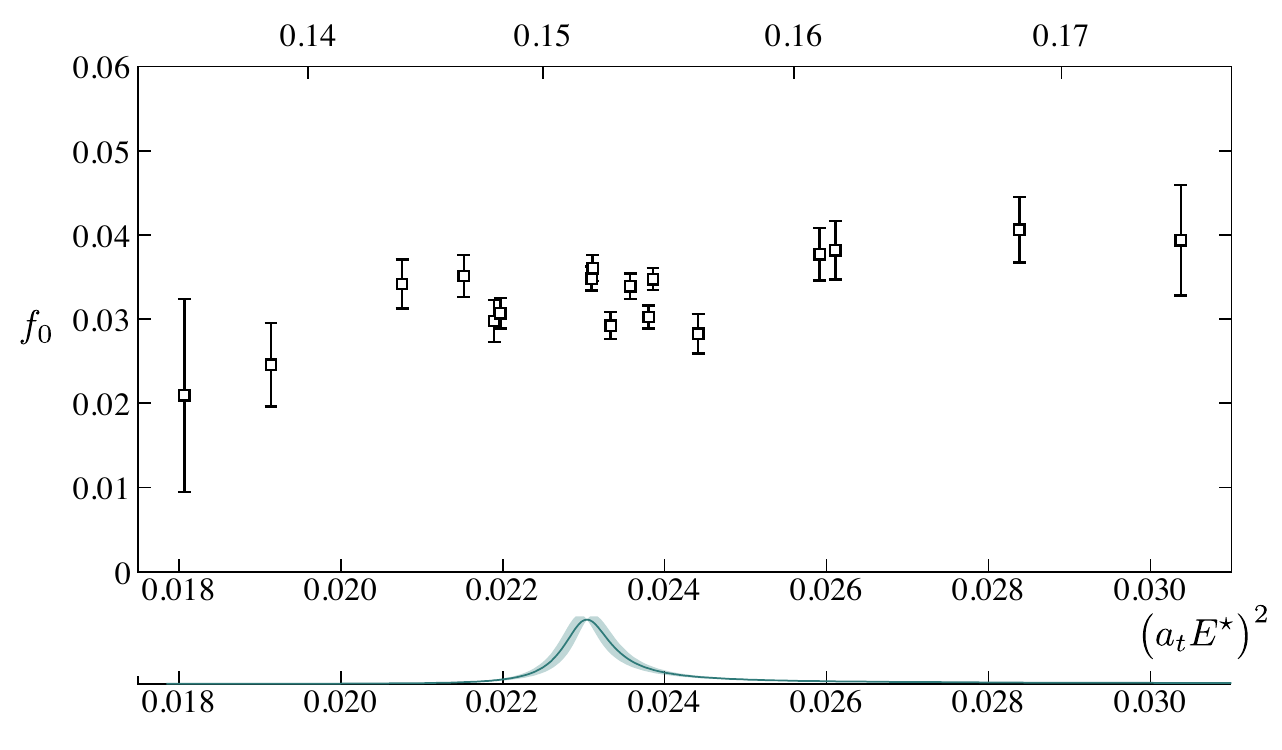}
\caption{Energy dependence of $f_0 = f_{0L}/\tilde{r}$ extracted from $Q^2$ fits using $f_{0L} \cdot \exp \left[ - a_1 \frac{Q^2}{4m_\pi^2}  \right]$ for $a_t^2 Q^2 < 0.015$, where $\mathrm{KM}_\mathsf{g}$ amplitude is used for finite-volume correction. Lower panel shows the corresponding $\mathrm{KM}_\mathsf{g}$ amplitude, indicating the position of the resonance. In this and subsequent plots the upper abscissa shows $a_t E^\star$.}
\label{f0}
\end{figure}

\begin{figure}
\includegraphics[width=1.0\columnwidth]{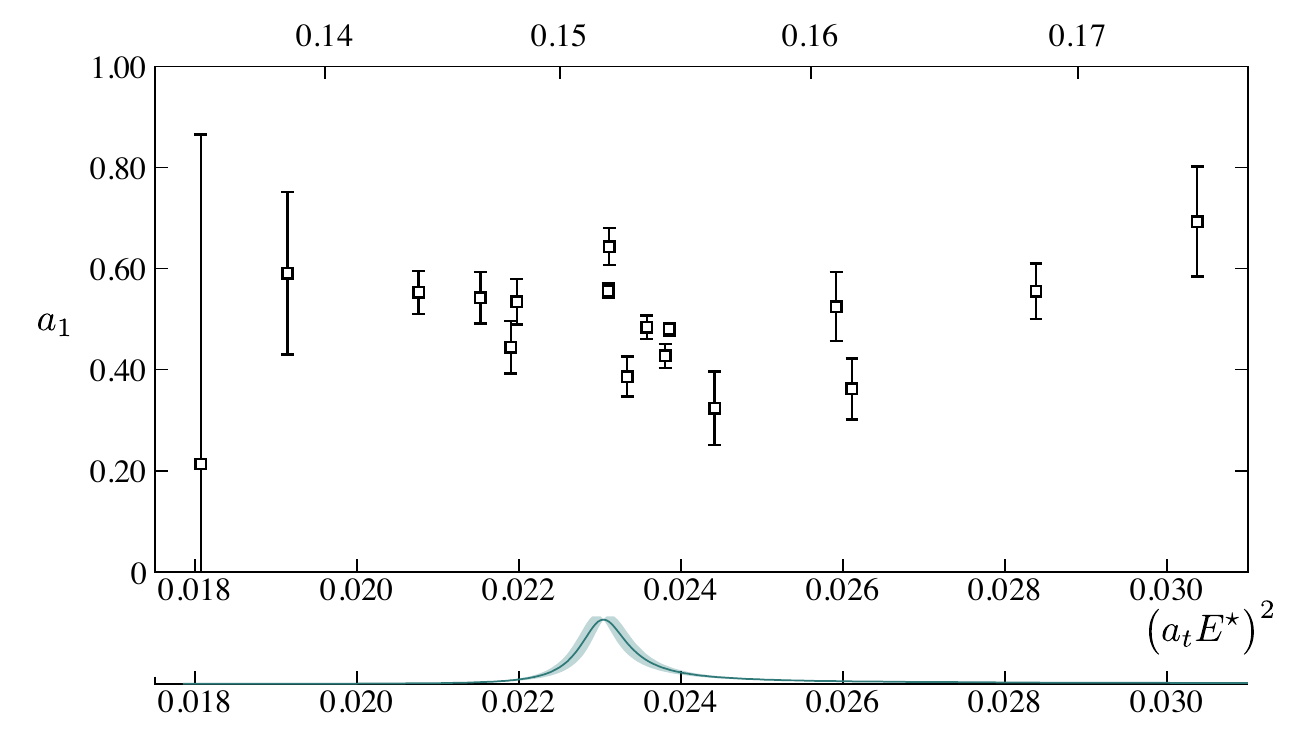}
\caption{Energy dependence of $a_1$ (which requires no finite volume correction) extracted from $Q^2$ fits using ${f_{0L} \cdot \exp \left[ - a_1 \frac{Q^2}{4m_\pi^2}  \right]}$ for $a_t^2 Q^2 < 0.015$. }
\label{a1}
\end{figure}

\vspace{5mm}
These $Q^2$ fit-forms are used to describe each of our 18 energy levels, yielding values of the parameters in Eqns.~\ref{exp_poly}, \ref{z_poly} at 18 values of $E_n^\star$. These are then appropriately finite-volume corrected with $1/\tilde{r}_n$, and we show the example case of fits using $f_{0L} \cdot \exp \left[ - a_1 \frac{Q^2}{4m_\pi^2}  \right]$ to describe data points with $a_t^2 Q^2 < 0.015$ in Figures~\ref{f0},\ref{a1}. The finite-volume corrected $f_0 = f_{0L}/ \tilde{r}$ shown in Figure~\ref{f0} is observed to have a very mild variation with $E^\star$ over a range that extends well outside the region of the resonance peak. The parameter $a_1$ which controls the fall-off with $Q^2$, and which does not require finite-volume correction, is seen in Figure~\ref{a1} to also have a rather mild dependence on $E^\star$, albeit with some scatter.

\begin{figure}
\includegraphics[width=1.0\columnwidth]{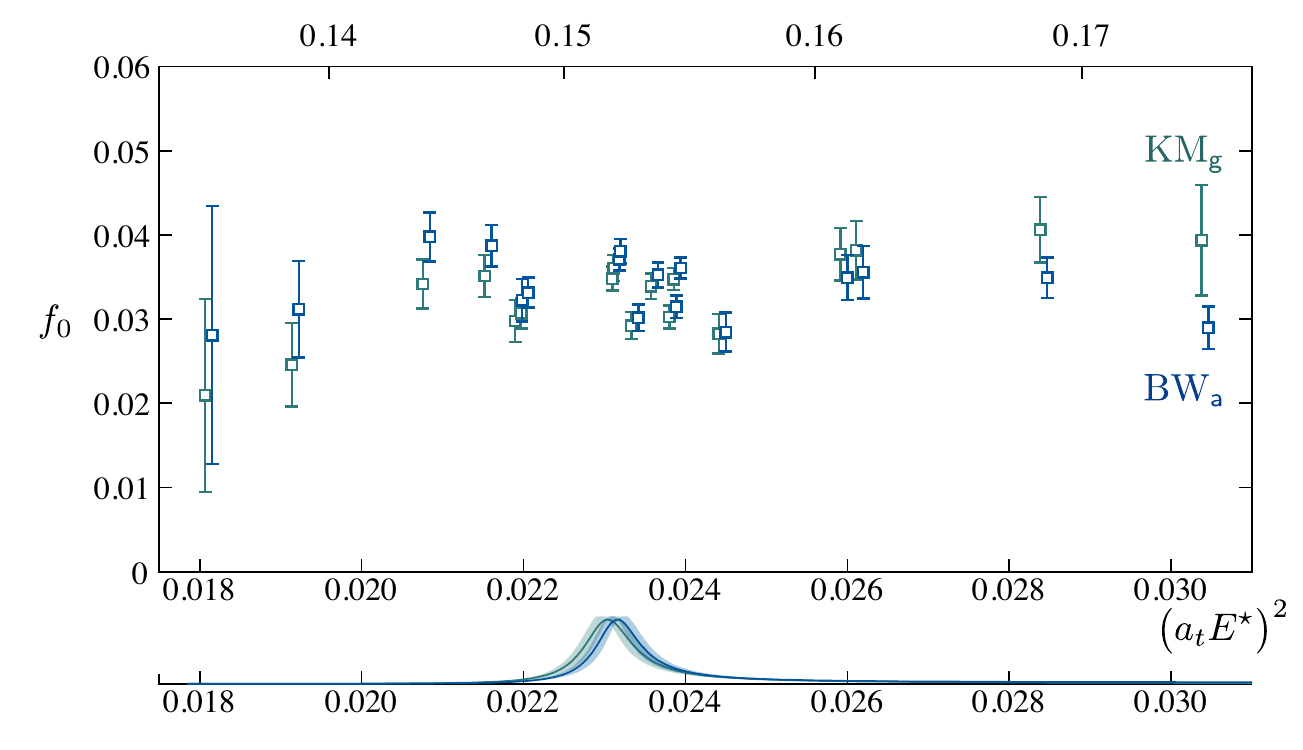}
\caption{As Figure~\ref{f0}, but for two choices of $K\pi \to K\pi$ amplitude parameterization used in the finite-volume correction. $\mathrm{BW}_\mathsf{a}$ data-points displaced slightly to the right for clarity.}
\label{f0_KM_BW}
\end{figure}

Figure~\ref{f0} illustrates $f_0$ in the case of finite-volume corrections using the elastic $K\pi \to K\pi$ amplitude parameterization $\mathrm{KM}_\mathsf{g}$. The degree of sensitivity to this choice is illustrated in Figure~\ref{f0_KM_BW} where it is compared to using $\mathrm{BW}_\mathsf{a}$, and where we observe only relatively modest change, most notable at energies away from the resonance, where we expect the Breit-Wigner amplitude to be of lesser validity.

These parameter energy dependences (and similarly those for the other parameterization choices) can be fitted with low-order polynomials in $E^{\star2}$ to ultimately yield complete parameterizations of $F(Q^2, E^\star)$. We will not show these results here as they prove to be broadly compatible with the results obtained from the more direct ``global fitting'' approach to be presented in the next subsection.

\clearpage

\subsection{Global fitting} \label{sec:global}

This second approach starts from the complete set of finite-volume-corrected $F$ data points shown in the lower panel of Figure~\ref{fv_all}. In principle these 128 data points can be directly fitted with parameterizations that are functions of both $Q^2$ and $s= E^{\star 2}$. The challenge is that the data correlation matrix for these points features many large positive off-diagonal elements that can be traced back to the high degree of correlation amongst the finite-volume correction factors (which tend to have a fractional uncertainty comparable to the fractional uncertainty on the $F_L$). 

\begin{figure*}[t]
\includegraphics[width=1.0\textwidth]{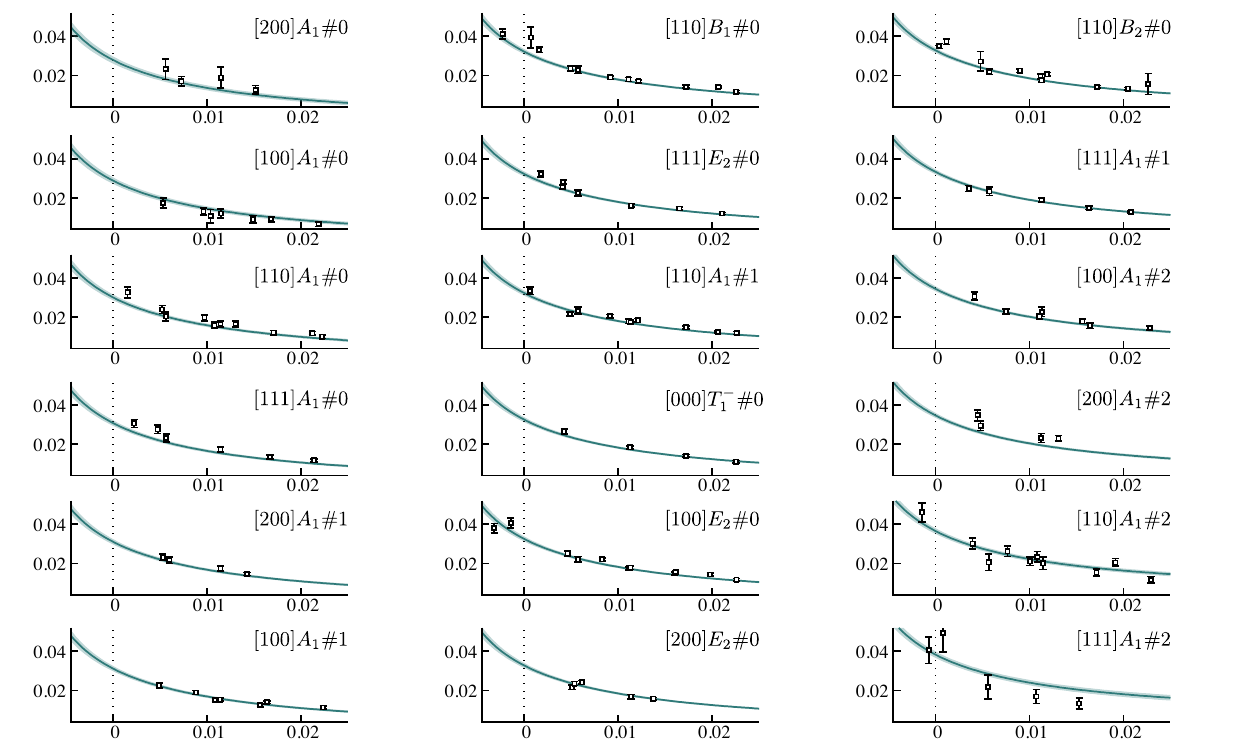}
\caption{Description of $F(Q^2, E^\star_n)$ data obtained from $F_L$ using amplitude $\mathrm{KM}_\mathsf{g}$ when fitted with Eqn.~\ref{z_poly_100}. The curves plotted are `slices' of $F(Q^2, E^\star)$ against $a_t^2\, Q^2$ when $E^\star$ is set equal to the finite-volume energies, $E_n^\star$. }
\label{global_slices}
\end{figure*}

Our calculation makes use of 348 configurations, such that the data correlation matrix is constructed from 348 outer-products, and hence has a maximal rank of 348. It is perhaps not such a surprise then that our dimension 128 data correlation matrix has a significant number of small eigenvalues. The standard procedure of inverting the data correlation matrix for use in the $\chi^2$ leads to fits which lie significantly and systematically below the data points. This motivates us to ``reset'' smaller eigenvalues, removing the corresponding eigenvectors from the matrix inverse. A discussion of how we select an eigenvalue cutoff of $\lambda_\mathrm{cut} = 0.01 \cdot \lambda_\mathrm{max}$ is presented in Appendix~\ref{app::resets}.

The parameterization forms for $F(Q^2, s\!=\! E^{\star 2})$ we use are compatible with those used in the approach followed in the previous subsection. For instance,
\begin{align}
F(Q^2, s) = \Big( &f_{0,0} + f_{0,1} \big(\tfrac{s - s_0}{s_0}\big) + \ldots  \Big) \nonumber \\
&\cdot \exp \bigg[ -
\Big( a_{1,0} + a_{1,1} \big(\tfrac{s - s_0}{s_0}\big) + \ldots  \Big) \tfrac{Q^2}{4m_\pi^2} \nonumber \\
 &\quad\quad\quad - \Big( a_{2,0} + a_{2,1} \big(\tfrac{s - s_0}{s_0}\big) + \ldots  \Big) \left(\tfrac{Q^2}{4m_\pi^2}\right)^{\!2} \nonumber \\ 
 &\quad\quad\quad+ \ldots
\bigg] \, , \label{exp_global}
\end{align}
where a convenient choice is $a_t^2 \, s_0 = (0.1520)^2$, motivated by the approximate location of the resonance bump in $K\pi \to K \pi$. Alternatively, using the conformal mapping variable,
\begin{align}
F(Q^2, s) = 
\sum_{q=0}^{n_q} \sum_{\sigma = 0}^{n_\sigma(q)} b_{q,\sigma}\cdot \left(\tfrac{s - s_0}{s_0}\right)^\sigma \cdot \big( z(Q^2) \!-\! z(0) \big)^q \, , \label{z_global}
\end{align}
so that for example we might have,
\begin{align} \label{z_poly_100}
F(Q^2, s) = 
\Big(& b_{0,0} + b_{0,1} \tfrac{s - s_0}{s_0} \Big) \nonumber \\
&+ b_{1,0} \cdot \big( z(Q^2) \!-\! z(0) \big) + 
b_{2,0} \cdot \big( z(Q^2) \!-\! z(0) \big)^2 \, ,
\end{align}
which has a universal slope in $s$ for all $Q^2$.

Figure~\ref{global_slices} illustrates the result of a global fit using the form in Eqn.~\ref{z_poly_100} when the $F_L$ data is corrected using $\tilde{r}$ computed with amplitude $\mathrm{KM}_\mathsf{g}$. The fit is able to describe the data with a $\chi^2/N_\mathrm{dof} = 81.2 / (128 - 91 -4 ) = 2.46$, where the subtracted 91 reflects the reduction in information from the resetting of data correlation eigenvalues.

\begin{figure}
\includegraphics[width=0.9\columnwidth]{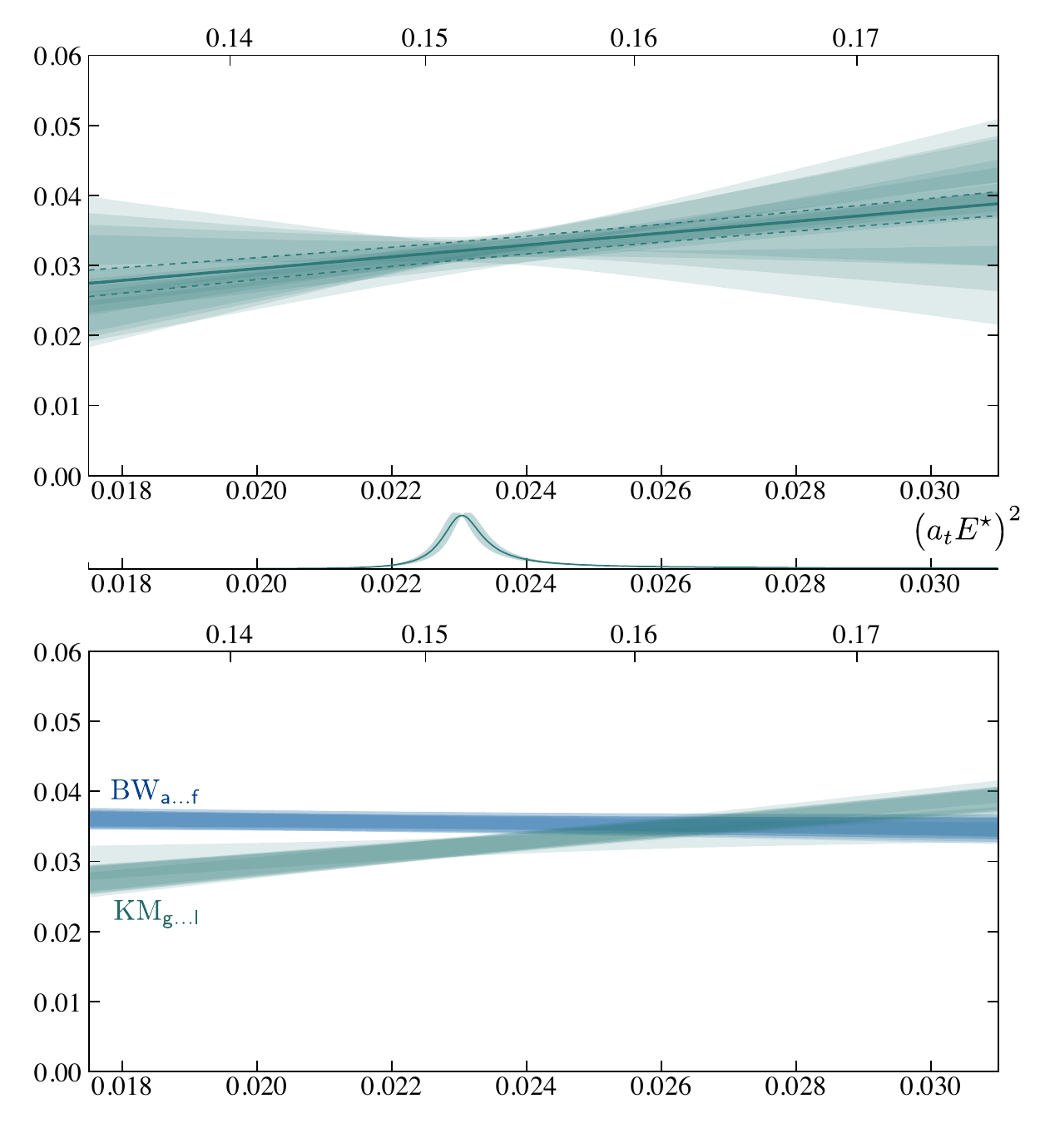}
\vspace{-7mm}
\caption{$F(Q^2\!=\!0, s)$ from ``global fitting''. Upper panel shows variation under changes of fit-form using Eqns.~\ref{exp_global}, \ref{z_global}. The band bounded by dashed lines shows the choice Eqn.~\ref{z_poly_100} as plotted in Figure~\ref{global_slices}. Lower panel shows variation under $K\pi \to K\pi$ amplitude choice (felt through the finite-volume corrections, $\tilde{r}_n$) for the fit form Eqn.~\ref{z_poly_100}.}
\label{global_F0s}
\end{figure}

The upper panel of Figure~\ref{global_F0s} shows the variation of $F(Q^2\!=\!0, s)$ under changes in parameterization for fixed choice of $K\pi \to K\pi$ amplitude $\mathrm{KM}_\mathsf{g}$. Several variations of Eqns.~\ref{exp_global},\ref{z_global} are considered, all leading to reasonable fits to the data. It is clear that the energy dependence of the form-factor at zero virtuality is mild, but the precise value of the small slope is not well determined, while the behavior around the resonance is tightly constrained. Figure~\ref{global_F0s} can be compared to the comparable quantity obtained in the ``level-by-level'' approach, and plotted in Figs~\ref{f0}, \ref{f0_KM_BW} where close agreement can be seen.

This ``global fitting'' procedure was applied to $F$ data corrected with each of the $K\pi \to K\pi$ amplitudes introduced in Section~\ref{Kpi}, $\mathrm{BW}_\mathsf{a \ldots f}$, $\mathrm{KM}_\mathsf{g \ldots l}$, and the lower panel of Figure~\ref{global_F0s} shows the resulting variation in $F(Q^2\!=\!0, s)$ for a fixed parameterization (Eqn.~\ref{z_poly_100}). There is a clear systematic difference between the $P$--wave Breit-Wigner amplitudes and those using a $K$-matrix, and the discrepancy is largest away from the resonance region. This effect is observed to be smaller in magnitude than the variation under $F$ parameterization choice shown in the upper panel.

\newpage
\section{Results}
	\label{results}

\begin{figure*}
\includegraphics[width=1.6\columnwidth]{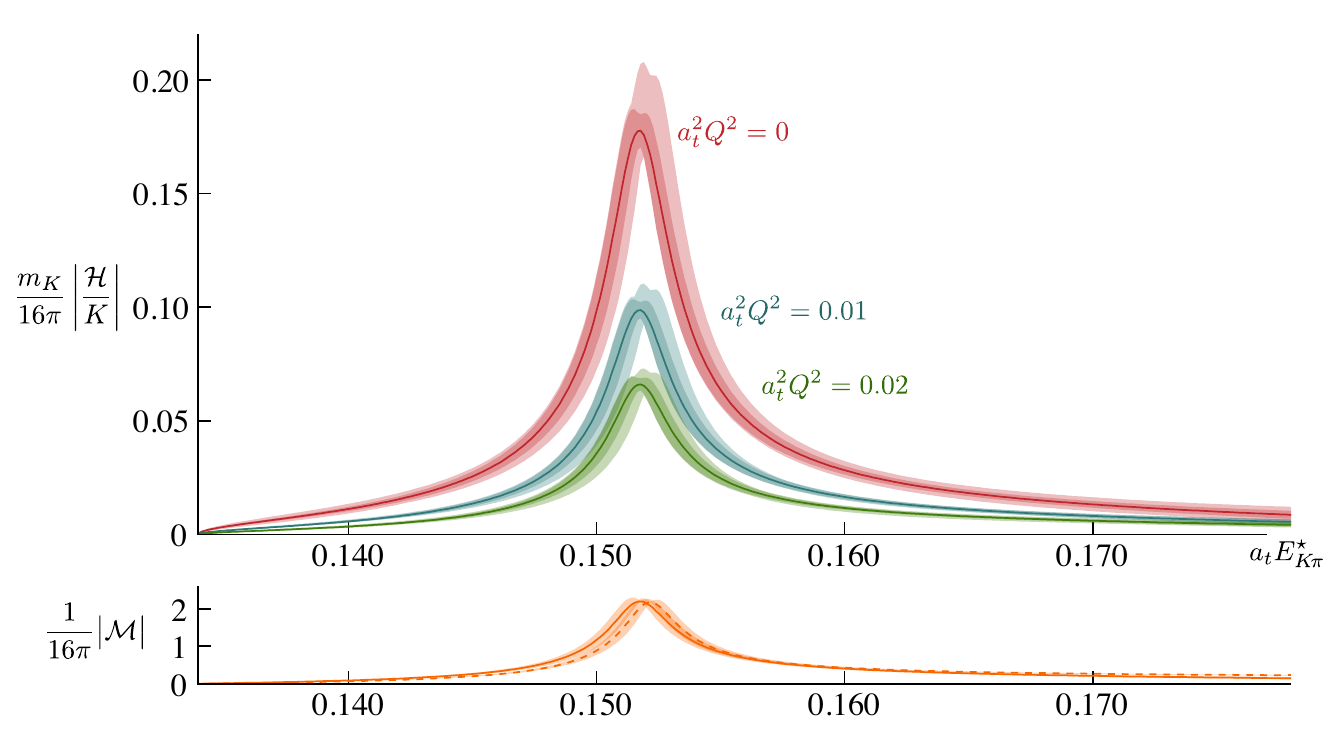}
\caption{Transition amplitude for $\gamma K^+ \to (K\pi)_{I=1/2, I_z=+1/2}$ for three values of photon virtuality. Lines and inner band correspond to the ``global fitting'' analysis using $\mathrm{KM}_\mathsf{g}$ for the $K\pi$ elastic scattering amplitude, and Eqn.~\ref{z_poly_100} as transition parameterization. The outer band shows an envelope of one-sigma variations over choices of $K\pi$ amplitude and transition amplitude parameterization form. The lower panel shows the corresponding elastic $K\pi$ $P$-wave scattering amplitude for two sample parameterizations, $\mathrm{KM}_\mathsf{g}$(solid line), $\mathrm{BW}_\mathsf{a}$(dashed line).}
\label{H_over_K}
\end{figure*}

Making use of the transition form-factor, $F(Q^2,s)$, obtained in the previous section, we can compute the transition amplitude $\mathcal{H}$, introduced in Eqn.~\ref{H} -- this is shown for three sample photon virtualities in Figure~\ref{H_over_K}, where we have divided out the kinematic factor to give an invariant quantity. As expected, the $K^*$ resonance bump present in $K\pi$ elastic scattering is also present in the transition amplitude, barely modulated in shape given the mild energy dependence of $F(Q^2,s)$.

\begin{figure}
\includegraphics[width=1.0\columnwidth]{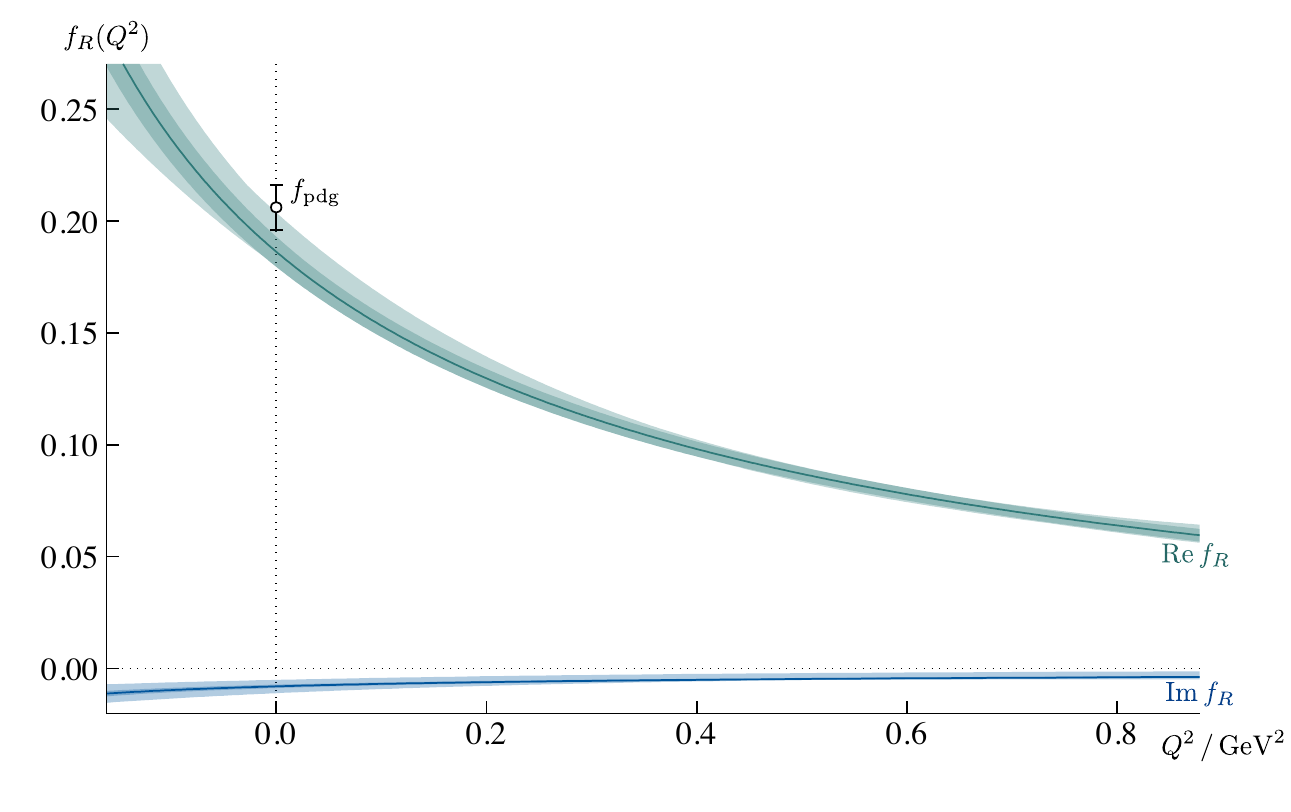}
\caption{$K^{*+} \to K^+ \gamma$ resonance transition form factor, as defined in Eqn.~\ref{fRdefn}. Lines and inner band correspond to the ``global fitting'' analysis using $\mathrm{KM}_\mathsf{g}$ for the $K\pi$ elastic scattering amplitude, and Eqn.~\ref{z_poly_100} as transition parameterization. The outer band shows an envelope of one-sigma variations over choices of $K\pi$ amplitude and transition amplitude parameterization form. Also shown an estimate for the $Q^2=0$ value extracted from the experimental radiative decay width.
}
\label{fRQsq}
\end{figure}

As indicated in Eqn.~\ref{fRdefn}, the \emph{resonance transition form-factor}, $f_R(Q^2)$, can be found by analytically continuing $F(Q^2, s)$ to the location of the resonance pole. We show the result of this in Figure~\ref{fRQsq}, where the smallness of the imaginary part can be explained by the narrow width of the $K^*$ which causes only a small departure from the real value on the real energy axis.

The value of this quantity at $Q^2=0$ is of particular interest, it being the amplitude for a real photon transition. A conservative best estimate from this calculation, accounting for the degree of fluctuation observed when varying the $K\pi$ elastic scattering amplitude, the range of $E^\star$ and $Q^2$ data considered, and the transition amplitude parameterization, is
\begin{equation} \label{fR_final}
f_R(0) = 0.185(15) - i\,  0.008(3).
\end{equation}
A somewhat comparable\footnote{Experimental analyses do not typically perform an analytic continuation to the pole, rather they assume a Breit-Wigner energy-dependence and factorize the numerator into production and decay partial widths.} quantity can be extracted from the experimental partial decay width, $\Gamma( K^{*+} \to K^+ \gamma)$. Given an amplitude for $K^{*+} \to K^+ \gamma$,
\begin{equation*}
T_{\lambda_{\!K\!\pi}, \lambda_\gamma} = e\, \epsilon^*_\mu(\lambda_\gamma) K^\mu(\lambda_{K\!\pi}) \, f \, ,
\end{equation*}
where $K^\mu$ is the same kinematic factor defined in Eqn.~\ref{A}, the decay width is given by
\begin{equation*}
 \Gamma(K^{*+} \to K^+ \gamma) = \frac{4}{3} \alpha \frac{k^{\star 3}_{K\!\gamma}}{m_K^2} \, \big| f \big|^2 \, .
\end{equation*}
The argument leading up to Eqn.~\ref{fRdefn} suggests an association between $f_R(0)$ and $f$ in the above equations that would be exact for a stable $K^*$. Using the PDG average~\cite{Workman:2022ynf} for the radiative partial decay width, and the physical values of hadron masses, we extract
\begin{equation*}
 \big|f_\mathrm{pdg}\big| = 0.206(10) \, ,
\end{equation*}
which we show in Figure~\ref{fRQsq}. We note that our ${|f_R(0)| = 0.185(15)}$, despite being computed with unphysically heavy light quark masses, is in reasonable agreement with this value\footnote{The analysis done for $\gamma \pi \to \pi \pi$ in Ref.~\cite{Niehus:2021iin}, extended simplistically to the current case, would seem to suggest that it is the quantity $f_R/m_K$ that is approximately constant with changing light quark-mass. The kaon mass in this calculation is only 5\% larger than the physical kaon mass, leading to a modest correction that worsens slightly the apparent agreement.}. \\

\vspace{5mm}
The rate of fall-off of $f_R(Q^2)$ with $Q^2$ might be used to estimate a \emph{transition radius} defined, in analogy to the charge radius of stable hadrons, via,
\begin{equation*}
\big\langle r^2 \big\rangle_{K^{*+}\!, K^+} \equiv \frac{1}{f_R(0)} \cdot \left( - 6 \frac{d}{d Q^2}f_R(Q^2) \right)\Bigg|_{Q^2=0} \, .
\end{equation*}
Our calculation suggests (again accounting for systematic variations),
\begin{equation*}
\mathrm{Re}\, \big\langle r^2 \big\rangle_{K^{*+}\!, K^+}^{1/2} = 0.69(4)\, \mathrm{fm}\, ,
\end{equation*}
and an imaginary part that is much smaller and statistically compatible with zero. This value does not differ significantly from the radius we extract from the charged kaon form-factor computed on this same lattice, $\big\langle r^2 \big\rangle_{K^+}^{1/2} = 0.55(2) \, \mathrm{fm}$.

\vspace{5mm}
We can use our transition amplitude to compute a cross-section for $\gamma K \to K \pi$. The transition amplitude \emph{for a particular final charge state} requires an isospin Clebsch-Gordan coefficient relative to the definite isospin case we have computed, e.g. 
\begin{equation*}
\Big|\mathcal{H} \big(\gamma K^+ \to K^+ \pi^0 \big) \Big| = \frac{1}{\sqrt{3}} \Big|\mathcal{H} \big(\gamma K^+ \to (K\pi)_{1/2, +1/2} \big) \Big| \, .
\end{equation*}
We will assume no contribution to this final charge-state from the non-resonant $I=\tfrac{3}{2}$ channel (which we have not computed).

\begin{figure}
\!\!\includegraphics[width=1.06\columnwidth]{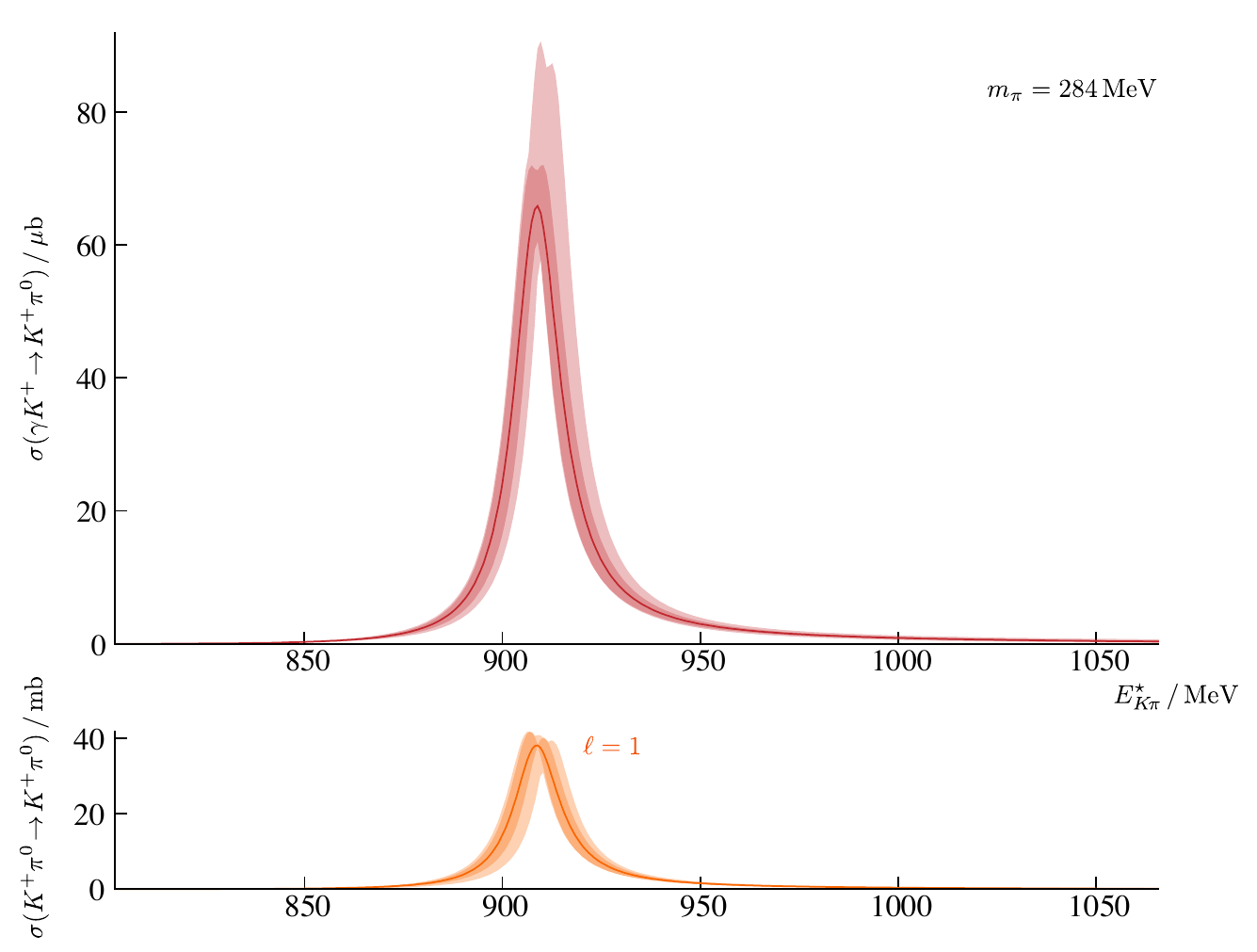}
\caption{$P$--wave contributions to cross-sections for ${\gamma K^+ \to K^+ \pi^0}$ and ${K^+ \pi^0 \to K^+ \pi^0}$ in a version of QCD with $m_\pi = $ 284 MeV.}
%
%
\label{cross-section}
\end{figure}

The differential cross-section is related to the transition matrix element in the usual way,
\begin{align*}
 \frac{d\sigma}{d\Omega}\big( \gamma K^+ &\!\!\to\! K^+ \pi^0 \big) \\
 &= \frac{1}{64 \pi^2} \frac{k^\star_{K\!\pi}}{k^\star_{K\!\gamma}} \frac{1}{s} \cdot
\, \frac{1}{2}\!\! \sum_{\lambda_{K\!\pi}, \lambda_\gamma} \bigg| \frac{1}{\sqrt{3}} \, e \, \epsilon_\mu(\lambda_\gamma) \mathcal{H}^\mu_{\lambda_{K\!\pi}} \bigg|^2 \, ,
\end{align*}
where there is an average over the two polarizations of a real photon in the initial state, and a sum over the three helicities of the $P$--wave $K\pi$ final state. The sums can be evaluated, and the integral over solid angle carried out to yield for the cross-section,
\begin{equation*}
\sigma \big( \gamma K^+ \!\!\to\! K^+ \pi^0 \big) = \frac{1}{3} \alpha \frac{k^\star_{K\!\gamma}}{k^\star_{K\!\pi}} \frac{1}{m_K^2} \big| F \, \mathcal{M} \big|^2 \, ,
\end{equation*}
which we plot for this lattice calculation\footnote{Physical units are obtained by setting the lattice scale using the mass of the $\Omega$-baryon.} in Figure~\ref{cross-section}. 

The experimentally measurable Primakoff process in which a kaon beam is scattered off a nucleus, $K A \to K \pi A$, with photon exchange isolated at small Mandelstam $t$, is proportional to the version of this cross-section for physical mass light quarks, but the data from the 1970s and 1980s was not presented in this way. A recent theoretical study, Ref.~\cite{Dax:2020dzg}, built a dispersive representation of the $\gamma K \to K \pi$ process, using as input descriptions of experimental $K\pi$ scattering, and phenomenology to constrain cross-channels, with subtraction constants left to be set by other experimental input. When the PDG values of the $K^* \to K \gamma$ radiative widths and the chiral anomaly were used, the resulting cross-section had a peak value of around 35 $\mu \mathrm{b}$, around a factor of two smaller than we have found with $m_\pi = $ 284 MeV. In Appendix~\ref{app::extrap}, we explore (in a simplistic extrapolation model) whether evolution to the physical point is likely to bring the cross-section into closer agreement with Ref.~\cite{Dax:2020dzg}.

\section{Summary}
	\label{summary}

We have presented the first lattice QCD determination of the process $\gamma K \to K \pi$, where the vector $K^*$ appears as a resonance. This required careful application of the formalism relating current matrix-elements in finite volume to those in infinite volume. In moving-frame $A_1$ irreps, which provide the broadest coverage of $E_{K\!\pi}$, the finite-volume correction factor is sensitive to the $S$--wave scattering amplitude as well as the $P$--wave scattering amplitude, and our results indicate that this correction does indeed bring the matrix-elements extracted from these irreps into good agreement with matrix elements at similar $E_{K\!\pi}$ extracted from irreps whose correction factor is not sensitive to the $S$--wave amplitude.

The result of this calculation is the $\gamma K \to K \pi$ amplitude as a function of $Q^2$ and $E_{K\!\pi}$ constrained by lattice QCD determined matrix elements at 128 discrete points in the $(Q^2, E_{K\!\pi})$ plane. The amplitude is constructed to conform to the restriction imposed by elastic unitarity, and the contribution of the $K^*$ resonance is quantified in a rigorous way by analytically continuing to the resonance pole. 

The use of a light-quark mass that is somewhat larger than the physical value, yielding a pion mass of 284 MeV, limits our ability to directly compare to related experimental measurements. In advance of a re-computation at the physical light quark mass, one might consider attempting a \emph{chiral extrapolation}, but this is not simple. As emphasized in Ref.~\cite{Niehus:2021iin}, in a process like this one, where there is a low-lying resonance, and where unitarity is an important constraint, naive applications of chiral effective field theory are unlikely to be successful. Ref.~\cite{Niehus:2021iin} considered the closely related $\gamma \pi \to \pi \pi$ process, using an approach combining dispersion relations with $SU(2)$ chiral perturbation theory, which they applied to extrapolate the lattice QCD data presented in \cite{Briceno:2016kkp, Alexandrou:2018jbt}. While the extrapolation is likely to be milder in the current case, the poorer convergence of $SU(3)$ chiral perturbation theory may limit the precision that can be obtained.

An obvious future application of the formalism explored in this paper would be to the process $B \to K^* \ell^+ \ell^-$, where experimental measurements are being used to place limits on lepton universality and more generally to limit possible extensions to the Standard Model. Doing this requires precise knowledge of the QCD form-factors in the transition $B \to K^*$ induced by the short-distance current of electroweak origin. To date, these have been evaluated in lattice QCD treating the $K^*$ as a stable particle (see for example Ref.\cite{Horgan:2013hoa}) which leaves an undetermined systematic error associated with the resonant nature of the $K^*$.


\begin{acknowledgments}
We thank our colleagues within the Hadron Spectrum Collaboration for their continued assistance, B. Kubis for providing the data of Ref.~\cite{Dax:2020dzg}, and W. Jay for explaining his model averaging approach. The authors acknowledge support from the U.S. Department of Energy contract DE-SC0018416 at William \& Mary, and contract DE-AC05-06OR23177, under which Jefferson Science Associates, LLC, manages and operates Jefferson Lab.

The software codes
{\tt Chroma}~\cite{Edwards:2004sx}, {\tt QUDA}~\cite{Clark:2009wm,Babich:2010mu}, {\tt QUDA-MG}~\cite{Clark:SC2016}, {\tt QPhiX}~\cite{ISC13Phi},
{\tt MG\_PROTO}~\cite{MGProtoDownload}, and {\tt QOPQDP}~\cite{Osborn:2010mb,Babich:2010qb} were used for the computation of the quark propagators.
The authors acknowledge support from the U.S. Department of Energy, Office of Science, Office of Advanced Scientific Computing Research and Office of Nuclear Physics, Scientific Discovery through Advanced Computing (SciDAC) program. 
Also acknowledged is support from the Exascale Computing Project (17-SC-20-SC), a collaborative effort of the U.S. Department of Energy Office of Science and the National Nuclear Security Administration.
%
%
This work was also performed on clusters at Jefferson Lab under the USQCD Collaboration and the LQCD ARRA Project.
This research was supported in part under an ALCC award, and used resources of the Oak Ridge Leadership Computing Facility at the Oak Ridge National Laboratory, which is supported by the Office of Science of the U.S. Department of Energy under Contract No. DE-AC05-00OR22725.
This research used resources of the National Energy Research Scientific Computing Center (NERSC), a DOE Office of Science User Facility supported by the Office of Science of the U.S. Department of Energy under Contract No. DE-AC02-05CH11231.
The authors acknowledge the Texas Advanced Computing Center (TACC) at The University of Texas at Austin for providing HPC resources.
Gauge configurations were generated using resources awarded from the U.S. Department of Energy INCITE program at the Oak Ridge Leadership Computing Facility, the NERSC, the NSF Teragrid at the TACC and the Pittsburgh Supercomputer Center, as well as at the Cambridge Service for Data Driven Discovery (CSD3) and Jefferson Lab.
This work was performed in part using computing facilities at William \& Mary which were provided by contributions from the National Science Foundation (MRI grant PHY-1626177), and the Commonwealth of Virginia Equipment Trust Fund.
\end{acknowledgments}

\vspace{-6mm}
\appendix

\section{Wick diagrams in $\gamma K \to K\pi$}
	\label{app::wicks}

The correlation functions we require in this calculation include cases in which the source operator is a $K\pi$-like product, and in these cases the Wick diagrams shown in the lower row of Figure~\ref{wicks} feature. When the $K\pi$-like operator has $(I,I_z)=\left(\tfrac{1}{2}, +\tfrac{1}{2}\right)$, the three isospin-basis current insertions correspond to the following linear combinations of Wick diagrams (up to an overall factor):  
\begin{align*}
 \langle j_\rho \rangle   &= -\tfrac{3}{2} \mathsf{c}_l + \tfrac{1}{2} \mathsf{a} + \mathsf{p} \nonumber \\
 \langle j_{\omega_l} \rangle &= -\tfrac{3}{2} \mathsf{c}_l - \tfrac{3}{2} \mathsf{a} + 3 \mathsf{d}_l \nonumber \\
 \langle j_{\omega_s} \rangle &= -\tfrac{3}{\sqrt{2}} \mathsf{c}_s  + \tfrac{3}{\sqrt{2}} \mathsf{d}_s \, .
\end{align*}
In the particular combination of these needed for $j_\mathrm{em}$, the diagram $\mathsf{a}$ actually cancels, and the disconnected pieces enter proportional to $Z_V^l \mathsf{d}_l - Z_V^s \mathsf{d}_s$ such that in the $SU(3)$ flavor limit, the net disconnected contribution is zero. 

We might anticipate that the contribution of diagram $\mathsf{p}$ will be small: one way to view it is that the current $j_\rho$ behaves like an isovector, vector meson which transitions to a pion through $t$-channel exchange of color-singlet isoscalar $C=-$ objects, and diagram $\mathsf{p}$ corresponds to the \emph{disconnected} contribution to these. Considered in the $t$-channel, the relevant objects coupling to $\rho^0 \pi^0$ are the $h_J$ and $\omega_J$ mesons, and these are not believed to have large disconnected contributions~\cite{Dudek:2013yja}. Diagram $\mathsf{p}$ is nevertheless computed without further approximation in our calculation.%

For comparison, when the $K\pi$-like operator has $(I,I_z)= \left(\tfrac{3}{2}, +\tfrac{1}{2} \right)$, the three isospin-basis current insertions correspond to the following linear combinations of Wick diagrams (up to an overall factor):  
\begin{equation*}
 \langle j_\rho \rangle       =  -\sqrt{2} \mathsf{a} + \sqrt{2} \mathsf{p}, \, \langle j_{\omega_l} \rangle = 0 ,\, 
 \langle j_{\omega_s} \rangle = 0 \, .
\end{equation*}

\section{Finite-volume normalization for a narrow resonance}
	\label{app::narrow}

When a partial-wave contains a narrow resonance, the finite-volume normalization factors for discrete energy eigenstates close to the resonance mass take a particularly simple and illustrative form. For simplicity, consider a single elastic partial-wave of angular momentum $\ell$, with a single resonance pole lying close to the real energy axis,
\begin{equation}\label{Mres}
 \mathcal{M}(E^\star) = 16\pi \frac{ (c_R)^2}{\big(m_R - i \tfrac{1}{2}\Gamma_R \big)^2 - E^{\star 2}   } + \mathrm{reg} \, ,
\end{equation} 
where the pole dominates over the part regular in $E^{\star 2}$ for energies $E^{\star} \approx m_R$. 

For an amplitude like this with $\Gamma_R$ small, solution of the quantization condition, $0 = \det\big[ F^{-1} + \mathcal{M} \big]$, in any irrep containing $\ell$, for any volume, will yield a discrete energy eigenvalue $E^\star_R(L)$ parametrically close to $m_R$. The corresponding finite-volume state normalization can be found starting from
\begin{equation*}
\widetilde{R}_n = 2 E_n \cdot \lim_{E\to E_n} \big( E - E_n \big) \frac{1}{F^{-1} + \mathcal{M}} \,,
\end{equation*}
by noting that there is no reason for the slope of the finite-volume function $F^{-1}$ to be large at the resonance mass, while the slope of $\mathcal{M}$ \emph{is} large there so that,
\begin{equation*}
\frac{dF^{-1} }{dE} \bigg|_{E=E_R} \ll \frac{d\mathcal{M}}{dE}  \bigg|_{E=E_R} \, ,
\end{equation*}
and hence for the energy-level near the resonance mass,
\begin{equation*}
\widetilde{R}_R \to \frac{2 E_R}{\frac{d\mathcal{M}}{dE}  \big|_{E_R}} = \frac{2 E^\star_R}{\frac{d\mathcal{M}}{dE^\star}  \big|_{E^\star_R}} \, .
\end{equation*}

The factor suggested in Ref.~\cite{Briceno:2021xlc}, and applied as finite-volume correction in this paper, $\sqrt{- \frac{2E^\star_n}{\mu_0^{\star\prime}}} \frac{1}{k^{\star \ell}}$, in the elastic, single partial-wave case considered here is just ${\sqrt{\widetilde{R}_n} \cdot \mathcal{M} \cdot \frac{1}{k^{\star \ell}} }$. Insertion of Eqn.~\ref{Mres} yields
\begin{equation*}
\sqrt{- \frac{2E^\star_R}{\mu_0^{\star\prime}}} \frac{1}{k^{\star \ell}} = \sqrt{16\pi} \, \frac{c_R}{k^{\star \ell}} + \ldots \, ,
\end{equation*}
where the corrections vanish as $\Gamma_R \to 0$. Because the \mbox{$\ell$-wave} has a threshold behavior $\mathcal{M} \sim (k^\star)^{2\ell}$, it is convenient to define reduced couplings $\hat{c}_R = c_R / k_R^{\star \ell}$, so that for a narrow resonance
\begin{equation} \label{ll-narrow}
\sqrt{- \frac{2E^\star_R}{\mu_0^{\star\prime}}} \frac{1}{k^{\star \ell}} = \sqrt{16\pi}\,  \hat{c}_R  \, .
\end{equation}
This volume-independent result is in accord with our decomposition of the infinite-volume transition amplitude, $\mathcal{H} = K F \tfrac{1}{k^{\star \ell}} \mathcal{M}$, since for the narrow resonance,
\begin{equation*} 
\mathcal{H} = 
\Big( K F \sqrt{16\pi} \, \hat{c}_R \Big) 
\cdot \frac{1}{\big(m_R - i \tfrac{1}{2}\Gamma_R \big)^2 - E^{\star 2}   }
\cdot \Big( \sqrt{16\pi} \, \hat{c}_R k^{\star \ell} \Big) \, ,
\end{equation*}
where the three factors can be interpreted as the ${\gamma K \to K^*}$ vertex, the $K^*$ propagator, and the ${K^* \to K \pi}$ vertex. As shown in Eqn.~\ref{FVcorr}, 
\begin{equation*}
  F_L = \sqrt{- \frac{2E^\star_R}{\mu_0^{\star\prime}}} \frac{1}{k^{\star \ell}} \,\cdot F \, ,
\end{equation*}
so for a resonance with a vanishing width,
\begin{equation*}
  F_L =  F \, \sqrt{16\pi} \, \hat{c}_R \, ,
\end{equation*}
and the finite-volume computed quantity is indeed the volume-independent (stable) $K^* \to K \gamma$ form-factor.

\section{$K \pi$ amplitude parameterizations}
	\label{app::Mparams}

The $K\pi$ elastic scattering amplitude parameterizations we use are a subset of those investigated in Ref.~\cite{Wilson:2019wfr}, and more details can be found in that paper and in references therein.

For the choices $\mathrm{BW}_\mathsf{a \ldots f}$ the $P$--wave amplitude is a Breit-Wigner,
\begin{align*}
 \mathcal{M}^{\ell=1}(s) = \frac{16\pi}{\rho(s)}& \frac{\sqrt{s} \, \Gamma(s)}{m_{\mathrm{BW}}^2 - s - i \sqrt{s} \, \Gamma(s) } \, , \quad \Gamma(s) = g_\mathrm{BW}^2 \frac{k^{\star3}}{s} \, ,
\end{align*}
where $m_\mathrm{BW}$, $g_\mathrm{BW}$ are free parameters. The $S$-wave amplitudes are
\begin{align*}
\mathcal{M}^{\ell = 0}_\mathsf{a}(s) &= \frac{16\pi}{ \left( \gamma_0 + \gamma_1 \big( \tfrac{s - s_\mathrm{thr}}{s_\mathrm{thr}} \big) \right)^{-1} + I_\mathrm{thr}(s) } \, , \nonumber \\[1ex]
\mathcal{M}^{\ell = 0}_\mathsf{b}(s) &= \frac{16\pi}{ \left( \gamma_0 + \gamma_1 \big( \tfrac{s - s_\mathrm{thr}}{s_\mathrm{thr}} \big) + \gamma_2 \big( \tfrac{s - s_\mathrm{thr}}{s_\mathrm{thr}} \big)^2 \right)^{-1} + I_\mathrm{thr}(s) } \, ,\nonumber \\[1ex]
\mathcal{M}^{\ell = 0}_\mathsf{c}(s) &= \frac{16\pi \, (s-s_A)}{ \left( \gamma_0 + \gamma_1 \big( \tfrac{s - s_\mathrm{thr}}{s_\mathrm{thr}} \big) \right)^{-1} - i \rho(s) \, (s-s_A) } \, ,\nonumber \\[1ex]
\mathcal{M}^{\ell = 0}_\mathsf{d}(s) &= \frac{16\pi }{ \left( \gamma_0 + \gamma_1 \big( \tfrac{s - s_\mathrm{thr}}{s_\mathrm{thr}} \big) \right)^{-1} - i \rho(s)  }\, , \nonumber \\[1ex]
\mathcal{M}^{\ell = 0}_\mathsf{e}(s) &= \frac{16\pi \, (s-s_A)}{ \gamma_0 + \gamma_1 \big( \tfrac{s - s_\mathrm{thr}}{s_\mathrm{thr}} \big) + I_\mathrm{thr}(s) \, (s-s_A) } \, ,\nonumber \\[1ex]
\mathcal{M}^{\ell = 0}_\mathsf{f}(s) &= \frac{16\pi }{\rho(s)} \frac{k^\star}{ a^{-1} + \tfrac{1}{2} r k^{\star2} - i k^\star  } \, ,\nonumber \\[1ex]
\end{align*}
where $I_\mathrm{thr}(s)$ is the Chew-Mandelstam phase-space subtracted at threshold, $s = s_\mathrm{thr}$. Amplitudes $\mathsf{c}$ and $\mathsf{e}$ include Adler zeros with $s_A$ fixed at the tree-level location.

For the choices $\mathrm{KM}_\mathsf{g \ldots j}$, the $S$--wave amplitude is the same as in amplitude $\mathsf{a}$, while for $\mathrm{KM}_\mathsf{k}$ the $S$--wave is the same as $\mathsf{d}$, and for $\mathrm{KM}_\mathsf{l}$ it is the same as $\mathsf{e}$. The $P$--wave amplitudes are
\begin{align*}
\mathcal{M}^{\ell = 1}_\mathsf{g}(s) &= \frac{16\pi}{ \tfrac{1}{ 4 k^{\star2} }\left( \tfrac{g^2}{m^2 - s} + \gamma_0 \right)^{-1} + I_\mathrm{pole}(s) } \, \\
\mathcal{M}^{\ell = 1}_\mathsf{h}(s) &= \frac{16\pi}{ \tfrac{1}{ 4 k^{\star2} }\left( \tfrac{g^2}{m^2 - s} + \gamma_0 + \gamma_1 \big( \tfrac{s - s_\mathrm{thr}}{s_\mathrm{thr}} \big) \right)^{-1} + I_\mathrm{pole}(s) } \, \\
\mathcal{M}^{\ell = 1}_\mathsf{i}(s) &= \frac{16\pi}{ \tfrac{1}{ 4 k^{\star2} }\left( \frac{ \big( g_0 + g_1\tfrac{s - s_\mathrm{thr}}{s_\mathrm{thr}} \big)^2  }{m^2 - s} \right)^{-1} + I_\mathrm{pole}(s) } \, \\
\mathcal{M}^{\ell = 1}_\mathsf{j}(s) &= \frac{16\pi}{ \tfrac{1}{ 4 k^{\star2} }\left( \tfrac{ \big( g_0 + g_1\tfrac{s - s_\mathrm{thr}}{s_\mathrm{thr}} \big)^2 }{m^2 - s} + \gamma_0  \right)^{-1} + I_\mathrm{pole}(s) } \, \\
\mathcal{M}^{\ell = 1}_\mathsf{k}(s) &= \frac{16\pi}{ \tfrac{1}{ 4 k^{\star2} }\left( \tfrac{g^2}{m^2 - s} + \gamma_0 \right)^{-1} - i \rho(s)} \, \\
\mathcal{M}^{\ell = 1}_\mathsf{l}(s) &= \frac{16\pi}{ \tfrac{1}{ 4 k^{\star2} }\left( \tfrac{g^2}{m^2 - s} + \gamma_0 \right)^{-1} + I_\mathrm{pole}(s) } \, \\
\end{align*}
where $I_\mathrm{pole}(s)$ is the Chew-Mandelstam phase-space subtracted at the location of the $K$-matrix pole, $s=m^2$.

\begin{figure}[b]
\!\!\includegraphics[width=1.04\columnwidth]{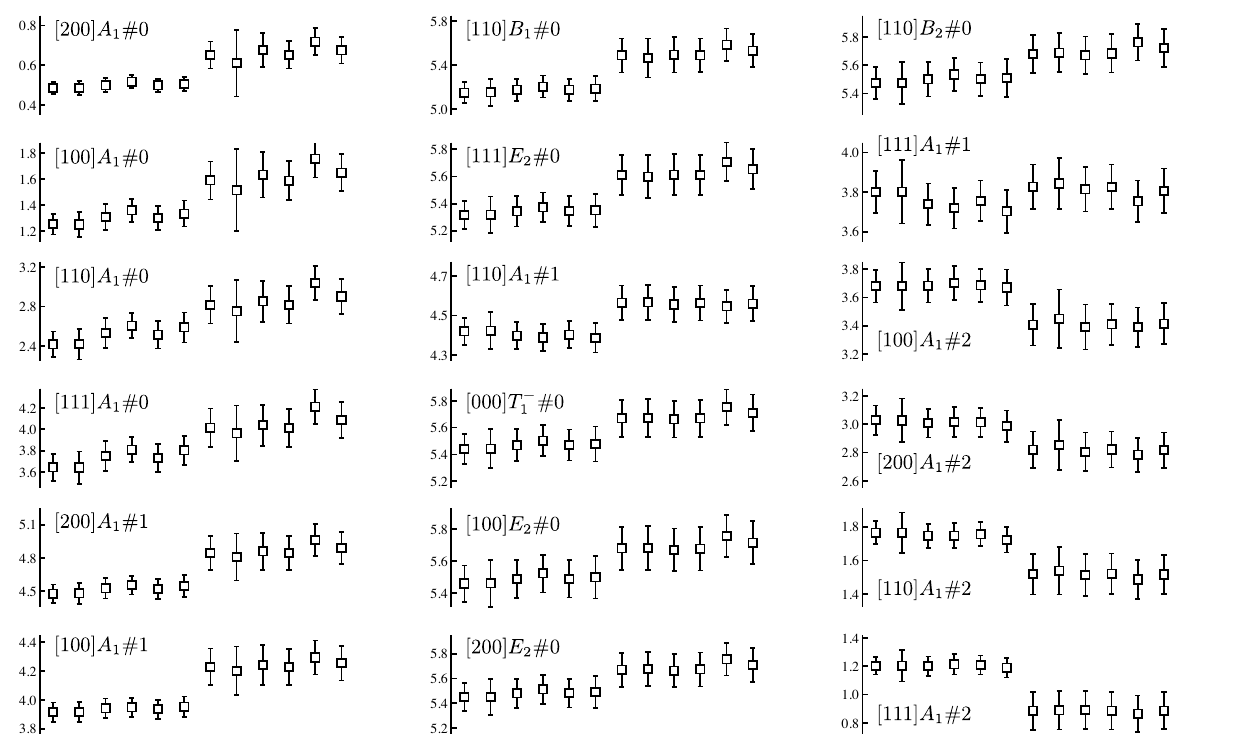}
\caption{$\tilde{r}_n$ values by energy level for twelve $K\pi \to K\pi$ scattering amplitude parameterizations $\mathsf{a} \ldots \mathsf{l}$ as described in the text.}
\label{ll_amp_var}
\end{figure}

Each parameterization pair is used to describe the finite-volume spectrum, which sets the values of the free parameters. From the amplitudes we can compute the finite-volume factors $\tilde{r}_n$ for each state in the spectrum, and these are shown in Figure~\ref{ll_amp_var}. The main variation is observed to be between those amplitudes which use a Breit-Wigner to describe the $P$--wave and those which use a $K$-matrix. We note also the apparently statistically precise factors for the Breit-Wigner choices well above and well below the resonance -- this is an artifact of the lack of freedom within the Breit-Wigner amplitude to vary away from the resonance, a freedom that \emph{is} present in the $K$-matrix cases.

\section{Timeslice fitting approach}
	\label{app::threept}

We perform correlated fits to the time dependence of each $F_L(t)$ using four fit functions,
\begin{align*}
	&F_L \\
	&F_L + a_\mathrm{src}\, e^{ - \delta E_\mathrm{src} t} \\
	&F_L + a_\mathrm{snk}\, e^{ - \delta E_\mathrm{snk} (\Delta t - t)} \\
	&F_L + a_\mathrm{src}\, e^{ - \delta E_\mathrm{src} t} + a_\mathrm{snk}\, e^{ - \delta E_\mathrm{snk} (\Delta t - t)} \, ,
\end{align*}
for large numbers of fit-windows $[t_\mathrm{min},t_\mathrm{max}]$, always with $t_\mathrm{max} - t_\mathrm{min} \ge 9$. For the second, third and fourth fit forms, a broad Bayesian prior is placed on the energy-shift parameter(s), $a_t \, \delta E = 0.15 \pm 0.15$. For each fit, a weight $w = \exp\big[-\tfrac{1}{2}( \chi^2 - 2N_\mathrm{dof}) \big]$ is computed, and the thirty fits with largest weights are retained. The weights are normalized to sum to one, and the resulting quantities are assigned the meaning of ``model probabilities''~\cite{Jay:2020jkz}. The ensemble values of the determined constants $F_{L,i}$ are then averaged using these probabilities, to yield a ``model average'' estimate of $F_L$,
\begin{equation*}
	F_L^{\mathrm{mod. avg.}} = \frac{\sum_{i=1}^{30} w_i \, F_{L,i}}{ \sum_{i=1}^{30} w_i} \, .
\end{equation*}

\begin{figure*}
\includegraphics[width=0.76\textwidth]{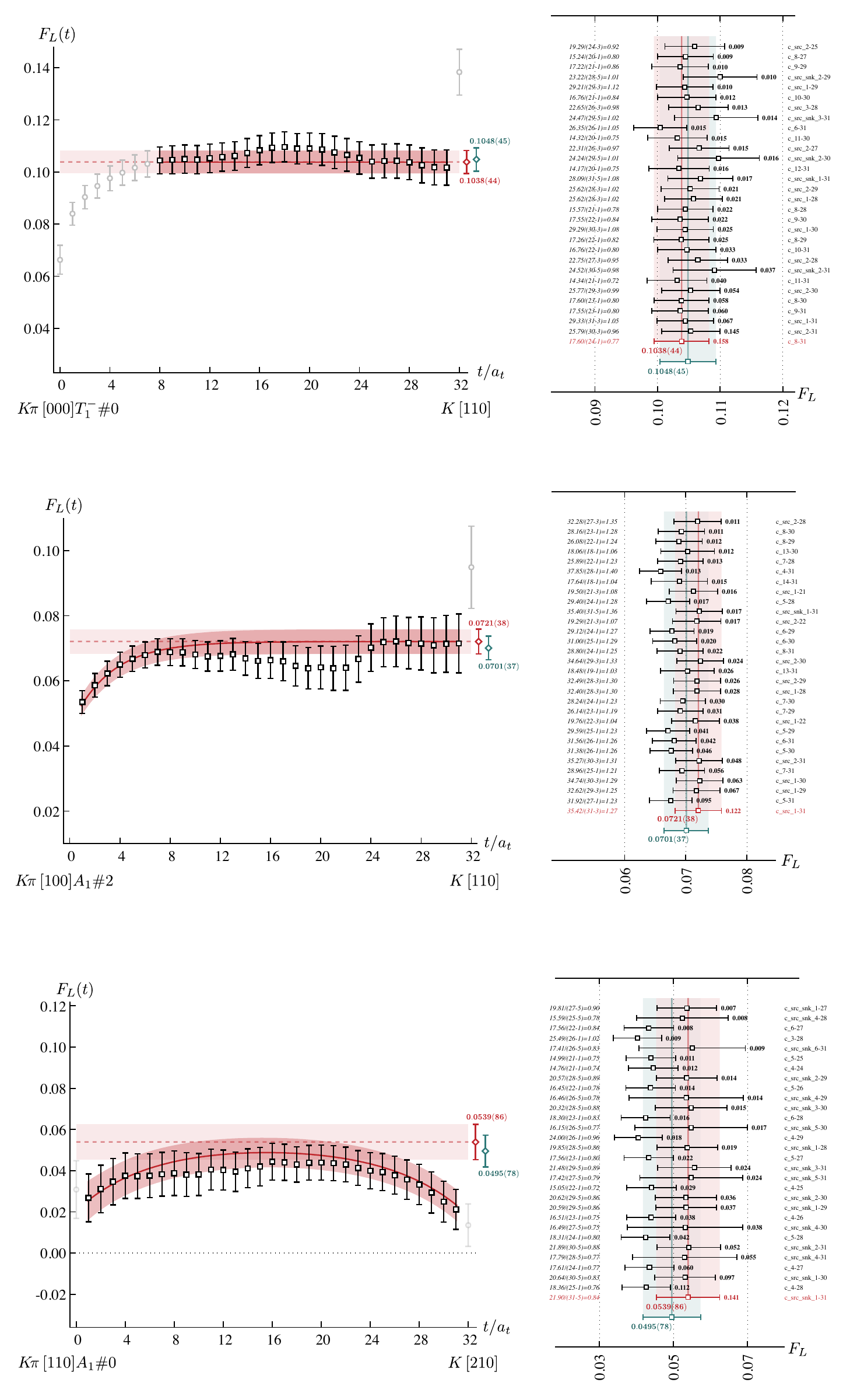}
\caption{Fits to timeslice dependence of three sample $F_L(t)$. Left side of each plot shows best single description over varying fit-windows, as quantified by the $w_i$ value. Right side of each plot shows variation over different fit-windows with, for each fit, the $\chi^2/N_\mathrm{dof}$ (italic), the fit probability, $\frac{w_i}{\sum_i w_i}$ (bold) and a description of the fit: ``c\_tmin-tmax'' indicates a constant fit to the fit-window $[t_\mathrm{min}, t_\mathrm{max}]$, ``c\_src'' a constant plus an exponential at the source, ``c\_snk'' a constant plus an exponential at the sink, and ``c\_src\_snk'' a constant plus an exponential at each of source and sink. The $w$-weighted average of the thirty highest probability fits is given by the blue band. }
\label{timeslice}
\end{figure*}

Examples of fits to three sample $F_L(t)$ are shown in Figure~\ref{timeslice}, where the left panel in each case shows the single fit with the largest value of $w$, whose $F_L$ value is indicated by the red band in the right panel. The right panel shows the $F_{L,i}$ value for the thirty largest $w$ fits, and the cyan band shows the ``model average''. We observe that in these cases the difference between the model average and the single best fit value is not large, but that the model average reflects the fact that some fits of comparable $w$ value to the best fit, do differ systematically from the best fit. 

\begin{figure*}
\includegraphics[width=0.76\textwidth]{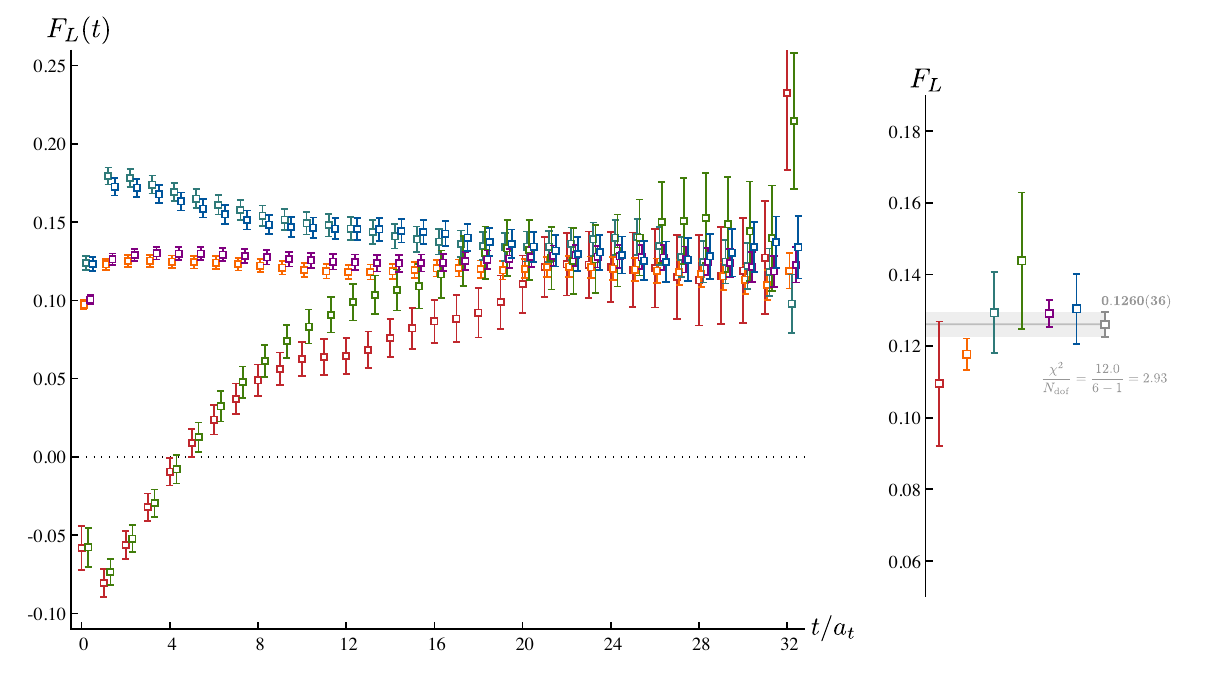}
\caption{An example of six $F_L(t)$ at the same $(Q^2, E^\star_n)$ corresponding to different momentum directions, irrep rows and current subductions. The left panel shows that they all likely plateau to the same constant value, but experience different amounts of excited-state pollution and have different signal/noise. The right panel shows the fitted $F_L$ values for each (using the model averaging procedure) which are then fitted to a constant to yield a single $F_L$ value to be used in later stages of the analysis.
}
\label{averaging}
\end{figure*}

In many cases there are multiple $F_L(t)$ at the same kinematic point $(Q^2, E^\star_n)$, and we choose to combine these by fitting each one independently, as described above, and then averaging the results with a correlated fit to a constant. The reason for this can be seen in Figure~\ref{averaging}, where we observe that the different $F_L(t)$, whilst likely plateauing to the same value, have very different excited state pollutions, and quite different signal/noise ratios. The right panel of Figure~\ref{averaging} shows the result of the averaging of the fitted $F_L$ to yield a single estimate that is used in subsequent computations.

\newpage
\section{Global fitting}
	\label{app::resets}

\begin{figure*}
\includegraphics[width=0.8\textwidth]{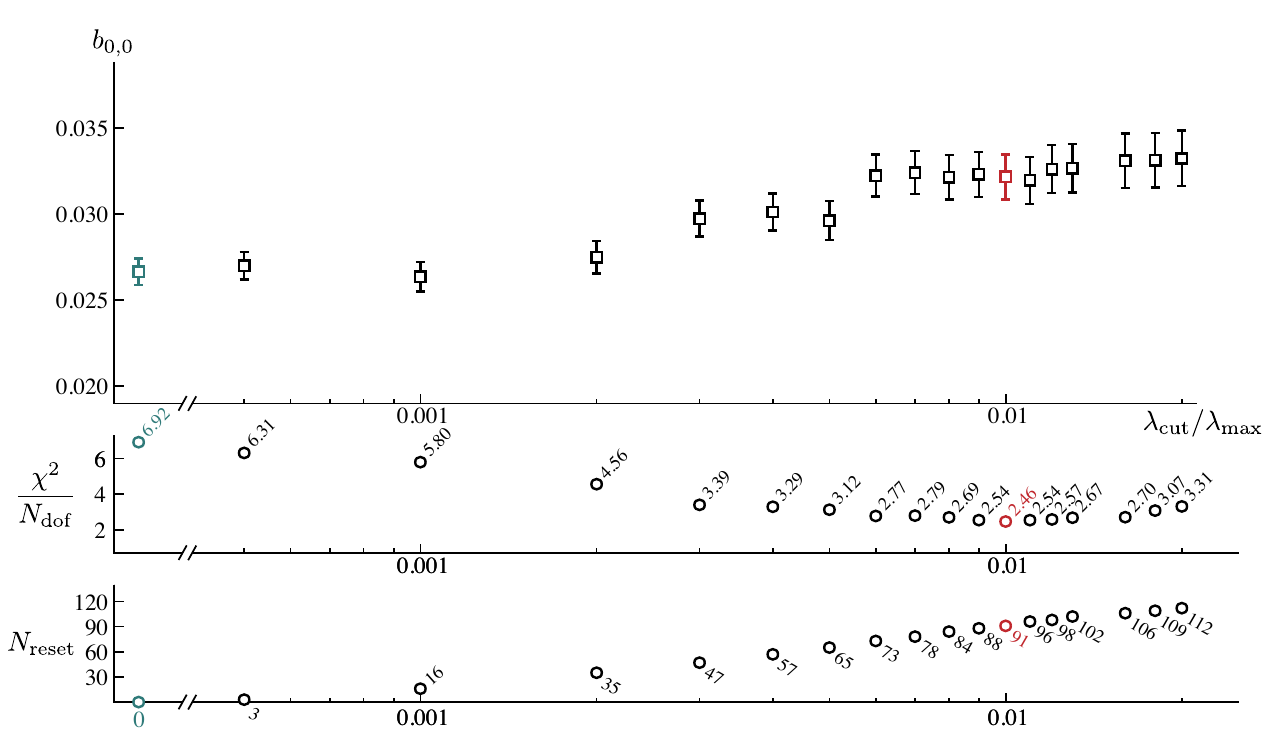}
\caption{The top panel shows the value of parameter $b_{0,0}$ in a ``global fit'' to 128 $F$ points using Eqn.~\ref{z_poly_100}, as a function of the cut used to reset eigenvalues of the data correlation. The middle and lower panels show the corresponding $\chi^2/N_\mathrm{dof}$ and the number of reset eigenvalues. The blue points indicate the result of not resetting any eigenvalues, and the red points indicate the choice made for the global fits presented in Section~\ref{sec:global}. }
\label{sv}
\end{figure*}

As discussed in the body of the paper, attempts to fit all 128 $(Q^2, E^\star_n)$ points using the simple inverse of the data correlation matrix in the $\chi^2$ lead to solutions which lie significantly and systematically below the data. We propose that with only 348 configurations, we are not producing reliable estimates of the entire data correlation matrix, and opt to reduce the impact of this by eliminating the poorly-estimated smaller eigenvalues. We reduce the size of the data space by eliminating the linear combinations of data corresponding to those eigenvectors of the data correlation matrix with eigenvalue smaller than $\lambda_\mathrm{cut}$. When computing the $\chi^2/N_\mathrm{dof}$ we reduce the number of degrees-of-freedom by the number of ``reset'' eigenvalues. Figure~\ref{sv} shows the result of this procedure as a function of $\lambda_\mathrm{cut}$, where we observe that for $\lambda_\mathrm{cut} \gtrsim 0.006 \, \lambda_\mathrm{max}$, the value of the fit parameter $b_{0,0}$ plateaus, and the $\chi^2/N_\mathrm{dof}$ varies rather little. We choose to use $\lambda_\mathrm{cut} = 0.01 \, \lambda_\mathrm{max}$ as our default in the analysis presented in the paper, which happens to be the point at which the $\chi^2/N_\mathrm{dof}$ takes its minimum value.

Using the data correlation matrix eigenvalue cut found above, we explore the variation in fits under changes in $F(Q^2, E^\star)$ parameterization, and $K\pi$ elastic scattering amplitude parameterization.

Table III
 ~illustrates a sample of $F(Q^2, E^\star)$ parameterization variations when the $F_L$ data is corrected with $\mathrm{KM}_\mathsf{g}$. The amplitudes are given labels as follows: e.g. ``exp\_poly\_110'' indicates Eqn.~\ref{exp_global} where $f_{0,0}, f_{0,1}$, $a_{1,0}, a_{1,1}$ and $a_{2,0}$ are free parameters, and ``z\_poly\_110'' indicates Eqn.~\ref{z_global} where $b_{0,0}, b_{0,1}$, $b_{1,0}, b_{1,1}$ and $b_{2,0}$ are free parameters. The data ranges are: ``all'', indicating all 128 data points, ``small--$Q^2$'' indicating only data with $a_t^2 Q^2 < 0.015$, and ``res. region'' indicating only data in the window $0.15 < a_t E^\star < 0.16$. 

\begin{table*} \label{tab::amp2g}
\renewcommand{\arraystretch}{1.3}
\begin{tabular}{rr @{\hskip 5mm} r @{\hskip 5mm} l}
\multicolumn{1}{c}{fit form} & \multicolumn{1}{c}{data} & \multicolumn{1}{c}{$\chi^2/N_\mathrm{dof}$} & \multicolumn{1}{c}{$f_R(0)$}\\[0.6ex]
\hline\\[-2.4ex]
exp\_poly\_100 & all   & $83/(128-91-4) = 2.5$ & $0.1868(68) - i \, 0.0095(9)$ \\
z\_poly\_100   & all   & $81/(128-91-4) = 2.5$ & $0.1862(67) - i \, 0.0079(8)$ \\
exp\_poly\_110 & all   & $82/(128-91-5) = 2.6$ & $0.1887(75) - i \, 0.0083(27)$ \\
z\_poly\_110   & all   & $81/(128-91-5) = 2.5$ & $0.1869(67) - i \, 0.0070(13)$ \\
exp\_poly\_100 & small--$Q^2$   & $115/(91-50-4) = 3.1$ & $0.1852(67) - i \, 0.0096(9)$ \\
exp\_poly\_10  & small--$Q^2$   & $136/(91-50-3) = 3.6$ & $0.1752(65) - i \, 0.0089(8)$ \\
z\_poly\_100 & small--$Q^2$   & $117/(91-50-4) = 3.2$ & $0.1840(66) - i \, 0.0082(7)$ \\
z\_poly\_11 & small--$Q^2$   & $120/(91-50-4) = 3.2$ & $0.1842(62) - i \, 0.0049(13)$ \\
exp\_poly\_000 & res. region   & $75/(64-37-3) = 3.1$ & $0.1869(63) - i \, 0.0062(8)$ \\
z\_poly\_000 & res. region   & $75/(64-37-3) = 3.1$ & $0.1852(62) - i \, 0.0061(8)$ \\
\end{tabular}
\caption{Variation with $F(Q^2, E^\star)$ parameterization and data range.}
\end{table*}

Table IV
~illustrates variation using changes in $K\pi$ elastic scattering amplitude when the parameterization ``z\_poly\_100'' (Eqn.~\ref{z_poly_100}) is used. These variations, as well as the variations observed in ``level-by-level'' analysis are used to come to the conservative estimate presented in Eqn.~\ref{fR_final}.

\begin{table*} \label{tab::Kpi}
\renewcommand{\arraystretch}{1.3}
\begin{tabular}{lr @{\hskip 5mm} l}
amplitude & \multicolumn{1}{c}{$\chi^2/N_\mathrm{dof}$} & \multicolumn{1}{c}{$f_R(0)$} \\[0.6ex]
$\mathrm{BW}_\mathsf{a}$ & $102/(128-84 - 4) = 2.6$ & $0.1918(63) - i \, 0.0083(5)$ \\
$\mathrm{BW}_\mathsf{b}$ & $85/(128-90 - 4) = 2.5$ & $0.1924(69) - i \, 0.0084(6)$ \\
$\mathrm{BW}_\mathsf{c}$ & $93/(128-86 - 4) = 2.5$ & $0.1937(65) - i \, 0.0086(5)$ \\
$\mathrm{BW}_\mathsf{d}$ & $92/(128-85 - 4) = 2.4$ & $0.1975(64) - i \, 0.0089(5)$ \\
$\mathrm{BW}_\mathsf{e}$ & $91/(128-84 - 4) = 2.4$ & $0.1940(64) - i \, 0.0086(5)$ \\
$\mathrm{BW}_\mathsf{f}$ & $82/(128-89 - 4) = 2.3$ & $0.1963(67) - i \, 0.0087(6)$ \\
$\mathrm{KM}_\mathsf{g}$ & $81/(128-91 - 4) = 2.5$ & $0.1862(67) - i \, 0.0079(8)$ \\
$\mathrm{KM}_\mathsf{h}$ & $77/(128-96 - 4) = 2.7$ & $0.1849(71) - i \, 0.0075(19)$ \\
$\mathrm{KM}_\mathsf{i}$ & $84/(128-93 - 4) = 2.7$ & $0.1857(67) - i \, 0.0075(9)$ \\
$\mathrm{KM}_\mathsf{j}$ & $82/(128-91 - 4) = 2.5$ & $0.1860(67) - i \, 0.0080(7)$ \\
$\mathrm{KM}_\mathsf{k}$ & $83/(128-90 - 4) = 2.4$ & $0.1895(66) - i \, 0.0082(7)$ \\
$\mathrm{KM}_\mathsf{l}$ & $80/(128-91 - 4) = 2.4$ & $0.1883(66) - i \, 0.0080(8)$ \\
\end{tabular}
\caption{Variation with $K\pi$ elastic scattering amplitude parameterization.}
\end{table*}

\section{Simplistic extrapolation to physical kinematics}
	\label{app::extrap}

A phenomenologically motivated extrapolation to physical kinematics can be justified by the observation made in Figure 4 of Ref.~\cite{Wilson:2019wfr} that the reduced coupling appears to be largely quark mass independent. Taking this literally, the total hadronic width of the $K^*$ would be
\begin{equation*}
 \Gamma_R = 3 \cdot \Gamma(K^+ \pi^0) = 3 \cdot \frac{2}{3} \frac{k^{\star 3}_{K\!\pi} }{m_R^2} \,|\hat{c}_R|^2 = 42(3) \, \mathrm{MeV} \, ,
\end{equation*}
where the physical mass $m_R = 892$ MeV is used for the mass of the $K^*$ in the kinematic quantities. This is in reasonable agreement with the PDG width~\cite{Workman:2022ynf}.

It is not obvious how $f_R(0)$ should evolve with changing quark mass, but if we assume that it does not change, we can make the following, rather crude, estimate of the radiative decay width,
\begin{equation*}
	\Gamma(K^{*+} \!\!\to\!\! K^+ \gamma) = \tfrac{4}{3} \alpha \tfrac{ k^{\star 3}_{K\!\gamma} }{m_K^2} \big| f_R(0) \big|^2 
	= 40(6)\, \mathrm{keV} \, ,
\end{equation*}
which is also in reasonable agreement with the PDG averaged partial width~\cite{Workman:2022ynf}. The agreement is slightly worse if $f_R(0)/m_K$ is assumed to be constant (as suggested by a simplistic extension of the argument presented in Ref.~\cite{Niehus:2021iin}), as then $\Gamma(K^{*+} \!\to\! K^+ \gamma) = 37(5)\, \mathrm{keV}$.

\begin{figure}
\includegraphics[width=\columnwidth]{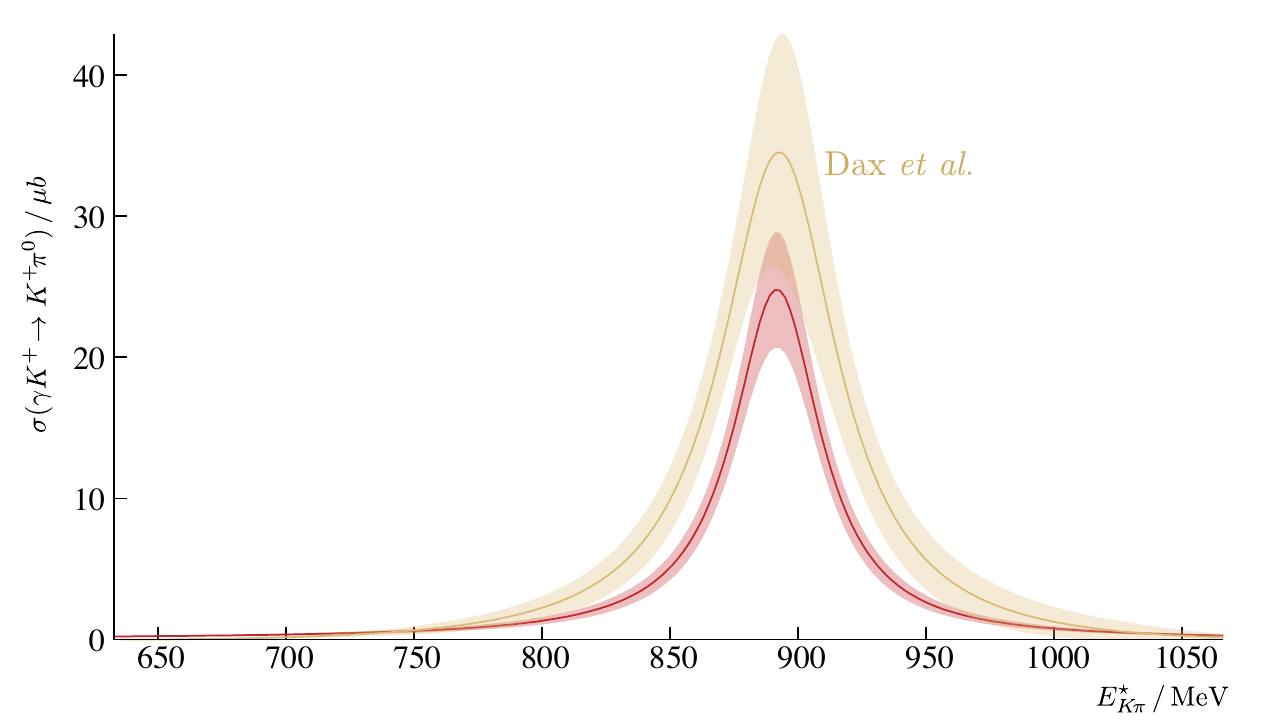}
\caption{Simplistic extrapolation of cross-section to physical light quark mass as described in the text. Compared to twice subtracted dispersion results of Dax \emph{et al.}~\cite{Dax:2020dzg}, with subtraction constants fixed by experimental radiative transition width and chiral anomaly (their Figure 8). }
\label{kubis}
\end{figure}

The extrapolated hadronic width, and the (first) extrapolated radiative width can be used in a simplistic pole-only form for the cross-section for physical kinematics,
\begin{equation*}
	\sigma( \gamma K^+ \to K^+ \pi^0 ) = \frac{2\pi}{k^{\star 2}_{K\!\gamma}} 
	\frac{ m_R^2 \, \Gamma_R \, \Gamma(K^{*+} \!\!\to\!\! K^+ \gamma) } {\left| \big(m_R - i \Gamma_R/2\big)^2 - s \right|^2 } \, ,\end{equation*}
using again the physical value of $m_R$ in all kinematic quantities. Figure~\ref{kubis} shows this cross-section estimate plotted along with the estimate given in Figure 8 of Ref.\cite{Dax:2020dzg} coming from a dispersive approach making use of the PDG radiative decay partial widths and the chiral anomaly to set the subtraction constants.

\bibliographystyle{apsrev4-1}
\bibliography{bib}

\begin{thebibliography}{37}%
\makeatletter
\providecommand \@ifxundefined [1]{%
 \@ifx{#1\undefined}
}%
\providecommand \@ifnum [1]{%
 \ifnum #1\expandafter \@firstoftwo
 \else \expandafter \@secondoftwo
 \fi
}%
\providecommand \@ifx [1]{%
 \ifx #1\expandafter \@firstoftwo
 \else \expandafter \@secondoftwo
 \fi
}%
\providecommand \natexlab [1]{#1}%
\providecommand \enquote  [1]{``#1''}%
\providecommand \bibnamefont  [1]{#1}%
\providecommand \bibfnamefont [1]{#1}%
\providecommand \citenamefont [1]{#1}%
\providecommand \href@noop [0]{\@secondoftwo}%
\providecommand \href [0]{\begingroup \@sanitize@url \@href}%
\providecommand \@href[1]{\@@startlink{#1}\@@href}%
\providecommand \@@href[1]{\endgroup#1\@@endlink}%
\providecommand \@sanitize@url [0]{\catcode `\\12\catcode `\$12\catcode
  `\&12\catcode `\#12\catcode `\^12\catcode `\_12\catcode `\%12\relax}%
\providecommand \@@startlink[1]{}%
\providecommand \@@endlink[0]{}%
\providecommand \url  [0]{\begingroup\@sanitize@url \@url }%
\providecommand \@url [1]{\endgroup\@href {#1}{\urlprefix }}%
\providecommand \urlprefix  [0]{URL }%
\providecommand \Eprint [0]{\href }%
\providecommand \doibase [0]{http://dx.doi.org/}%
\providecommand \selectlanguage [0]{\@gobble}%
\providecommand \bibinfo  [0]{\@secondoftwo}%
\providecommand \bibfield  [0]{\@secondoftwo}%
\providecommand \translation [1]{[#1]}%
\providecommand \BibitemOpen [0]{}%
\providecommand \bibitemStop [0]{}%
\providecommand \bibitemNoStop [0]{.\EOS\space}%
\providecommand \EOS [0]{\spacefactor3000\relax}%
\providecommand \BibitemShut  [1]{\csname bibitem#1\endcsname}%
\let\auto@bib@innerbib\@empty
\bibitem [{\citenamefont {Carithers}\ \emph {et~al.}(1975)\citenamefont
  {Carithers}, \citenamefont {Muhlemann}, \citenamefont {Underwood},\ and\
  \citenamefont {Ryan}}]{Carithers:1975cg}%
  \BibitemOpen
  \bibfield  {author} {\bibinfo {author} {\bibfnamefont {W.~C.}\ \bibnamefont
  {Carithers}}, \bibinfo {author} {\bibfnamefont {P.}~\bibnamefont
  {Muhlemann}}, \bibinfo {author} {\bibfnamefont {D.}~\bibnamefont
  {Underwood}}, \ and\ \bibinfo {author} {\bibfnamefont {D.~G.}\ \bibnamefont
  {Ryan}},\ }\href {\doibase 10.1103/PhysRevLett.35.349} {\bibfield  {journal}
  {\bibinfo  {journal} {Phys. Rev. Lett.}\ }\textbf {\bibinfo {volume} {35}},\
  \bibinfo {pages} {349} (\bibinfo {year} {1975})}\BibitemShut {NoStop}%
\bibitem [{\citenamefont {Berg}\ \emph {et~al.}(1981)\citenamefont {Berg} \emph
  {et~al.}}]{Berg:1980cp}%
  \BibitemOpen
  \bibfield  {author} {\bibinfo {author} {\bibfnamefont {D.}~\bibnamefont
  {Berg}} \emph {et~al.},\ }\href {\doibase 10.1016/0370-2693(81)90380-4}
  {\bibfield  {journal} {\bibinfo  {journal} {Phys. Lett. B}\ }\textbf
  {\bibinfo {volume} {98}},\ \bibinfo {pages} {119} (\bibinfo {year}
  {1981})}\BibitemShut {NoStop}%
\bibitem [{\citenamefont {Chandlee}\ \emph {et~al.}(1983)\citenamefont
  {Chandlee} \emph {et~al.}}]{Chandlee:1983hf}%
  \BibitemOpen
  \bibfield  {author} {\bibinfo {author} {\bibfnamefont {C.}~\bibnamefont
  {Chandlee}} \emph {et~al.},\ }\href {\doibase 10.1103/PhysRevLett.51.168}
  {\bibfield  {journal} {\bibinfo  {journal} {Phys. Rev. Lett.}\ }\textbf
  {\bibinfo {volume} {51}},\ \bibinfo {pages} {168} (\bibinfo {year}
  {1983})}\BibitemShut {NoStop}%
\bibitem [{\citenamefont {Carlsmith}\ \emph {et~al.}(1986)\citenamefont
  {Carlsmith} \emph {et~al.}}]{Carlsmith:1985ep}%
  \BibitemOpen
  \bibfield  {author} {\bibinfo {author} {\bibfnamefont {D.}~\bibnamefont
  {Carlsmith}} \emph {et~al.},\ }\href {\doibase 10.1103/PhysRevLett.56.18}
  {\bibfield  {journal} {\bibinfo  {journal} {Phys. Rev. Lett.}\ }\textbf
  {\bibinfo {volume} {56}},\ \bibinfo {pages} {18} (\bibinfo {year}
  {1986})}\BibitemShut {NoStop}%
\bibitem [{\citenamefont {Dax}\ \emph {et~al.}(2021)\citenamefont {Dax},
  \citenamefont {Stamen},\ and\ \citenamefont {Kubis}}]{Dax:2020dzg}%
  \BibitemOpen
  \bibfield  {author} {\bibinfo {author} {\bibfnamefont {M.}~\bibnamefont
  {Dax}}, \bibinfo {author} {\bibfnamefont {D.}~\bibnamefont {Stamen}}, \ and\
  \bibinfo {author} {\bibfnamefont {B.}~\bibnamefont {Kubis}},\ }\href
  {\doibase 10.1140/epjc/s10052-021-08951-x} {\bibfield  {journal} {\bibinfo
  {journal} {Eur. Phys. J. C}\ }\textbf {\bibinfo {volume} {81}},\ \bibinfo
  {pages} {221} (\bibinfo {year} {2021})},\ \Eprint
  {http://arxiv.org/abs/2012.04655} {arXiv:2012.04655 [hep-ph]} \BibitemShut
  {NoStop}%
\bibitem [{\citenamefont {Briceno}\ \emph {et~al.}(2018)\citenamefont
  {Briceno}, \citenamefont {Dudek},\ and\ \citenamefont
  {Young}}]{Briceno:2017max}%
  \BibitemOpen
  \bibfield  {author} {\bibinfo {author} {\bibfnamefont {R.~A.}\ \bibnamefont
  {Briceno}}, \bibinfo {author} {\bibfnamefont {J.~J.}\ \bibnamefont {Dudek}},
  \ and\ \bibinfo {author} {\bibfnamefont {R.~D.}\ \bibnamefont {Young}},\
  }\href {\doibase 10.1103/RevModPhys.90.025001} {\bibfield  {journal}
  {\bibinfo  {journal} {Rev. Mod. Phys.}\ }\textbf {\bibinfo {volume} {90}},\
  \bibinfo {pages} {025001} (\bibinfo {year} {2018})},\ \Eprint
  {http://arxiv.org/abs/1706.06223} {arXiv:1706.06223 [hep-lat]} \BibitemShut
  {NoStop}%
\bibitem [{\citenamefont {Lellouch}\ and\ \citenamefont
  {Luscher}(2001)}]{Lellouch:2000pv}%
  \BibitemOpen
  \bibfield  {author} {\bibinfo {author} {\bibfnamefont {L.}~\bibnamefont
  {Lellouch}}\ and\ \bibinfo {author} {\bibfnamefont {M.}~\bibnamefont
  {Luscher}},\ }\href {\doibase 10.1007/s002200100410} {\bibfield  {journal}
  {\bibinfo  {journal} {Commun. Math. Phys.}\ }\textbf {\bibinfo {volume}
  {219}},\ \bibinfo {pages} {31} (\bibinfo {year} {2001})},\ \Eprint
  {http://arxiv.org/abs/hep-lat/0003023} {arXiv:hep-lat/0003023} \BibitemShut
  {NoStop}%
\bibitem [{\citenamefont {Brice\~no}\ \emph {et~al.}(2015)\citenamefont
  {Brice\~no}, \citenamefont {Hansen},\ and\ \citenamefont
  {Walker-Loud}}]{Briceno:2014uqa}%
  \BibitemOpen
  \bibfield  {author} {\bibinfo {author} {\bibfnamefont {R.~A.}\ \bibnamefont
  {Brice\~no}}, \bibinfo {author} {\bibfnamefont {M.~T.}\ \bibnamefont
  {Hansen}}, \ and\ \bibinfo {author} {\bibfnamefont {A.}~\bibnamefont
  {Walker-Loud}},\ }\href {\doibase 10.1103/PhysRevD.91.034501} {\bibfield
  {journal} {\bibinfo  {journal} {Phys. Rev. D}\ }\textbf {\bibinfo {volume}
  {91}},\ \bibinfo {pages} {034501} (\bibinfo {year} {2015})},\ \Eprint
  {http://arxiv.org/abs/1406.5965} {arXiv:1406.5965 [hep-lat]} \BibitemShut
  {NoStop}%
\bibitem [{\citenamefont {Briceño}\ and\ \citenamefont
  {Hansen}(2015)}]{Briceno:2015csa}%
  \BibitemOpen
  \bibfield  {author} {\bibinfo {author} {\bibfnamefont {R.~A.}\ \bibnamefont
  {Briceño}}\ and\ \bibinfo {author} {\bibfnamefont {M.~T.}\ \bibnamefont
  {Hansen}},\ }\href {\doibase 10.1103/PhysRevD.92.074509} {\bibfield
  {journal} {\bibinfo  {journal} {Phys. Rev.}\ }\textbf {\bibinfo {volume}
  {D92}},\ \bibinfo {pages} {074509} (\bibinfo {year} {2015})},\ \Eprint
  {http://arxiv.org/abs/1502.04314} {arXiv:1502.04314 [hep-lat]} \BibitemShut
  {NoStop}%
\bibitem [{\citenamefont {Briceno}\ \emph {et~al.}(2015)\citenamefont
  {Briceno}, \citenamefont {Dudek}, \citenamefont {Edwards}, \citenamefont
  {Shultz}, \citenamefont {Thomas},\ and\ \citenamefont
  {Wilson}}]{Briceno:2015dca}%
  \BibitemOpen
  \bibfield  {author} {\bibinfo {author} {\bibfnamefont {R.~A.}\ \bibnamefont
  {Briceno}}, \bibinfo {author} {\bibfnamefont {J.~J.}\ \bibnamefont {Dudek}},
  \bibinfo {author} {\bibfnamefont {R.~G.}\ \bibnamefont {Edwards}}, \bibinfo
  {author} {\bibfnamefont {C.~J.}\ \bibnamefont {Shultz}}, \bibinfo {author}
  {\bibfnamefont {C.~E.}\ \bibnamefont {Thomas}}, \ and\ \bibinfo {author}
  {\bibfnamefont {D.~J.}\ \bibnamefont {Wilson}},\ }\href {\doibase
  10.1103/PhysRevLett.115.242001} {\bibfield  {journal} {\bibinfo  {journal}
  {Phys. Rev. Lett.}\ }\textbf {\bibinfo {volume} {115}},\ \bibinfo {pages}
  {242001} (\bibinfo {year} {2015})},\ \Eprint
  {http://arxiv.org/abs/1507.06622} {arXiv:1507.06622 [hep-ph]} \BibitemShut
  {NoStop}%
\bibitem [{\citenamefont {Brice\~no}\ \emph {et~al.}(2016)\citenamefont
  {Brice\~no}, \citenamefont {Dudek}, \citenamefont {Edwards}, \citenamefont
  {Shultz}, \citenamefont {Thomas},\ and\ \citenamefont
  {Wilson}}]{Briceno:2016kkp}%
  \BibitemOpen
  \bibfield  {author} {\bibinfo {author} {\bibfnamefont {R.~A.}\ \bibnamefont
  {Brice\~no}}, \bibinfo {author} {\bibfnamefont {J.~J.}\ \bibnamefont
  {Dudek}}, \bibinfo {author} {\bibfnamefont {R.~G.}\ \bibnamefont {Edwards}},
  \bibinfo {author} {\bibfnamefont {C.~J.}\ \bibnamefont {Shultz}}, \bibinfo
  {author} {\bibfnamefont {C.~E.}\ \bibnamefont {Thomas}}, \ and\ \bibinfo
  {author} {\bibfnamefont {D.~J.}\ \bibnamefont {Wilson}},\ }\href {\doibase
  10.1103/PhysRevD.93.114508} {\bibfield  {journal} {\bibinfo  {journal} {Phys.
  Rev. D}\ }\textbf {\bibinfo {volume} {93}},\ \bibinfo {pages} {114508}
  (\bibinfo {year} {2016})},\ \Eprint {http://arxiv.org/abs/1604.03530}
  {arXiv:1604.03530 [hep-ph]} \BibitemShut {NoStop}%
\bibitem [{\citenamefont {Alexandrou}\ \emph {et~al.}(2018)\citenamefont
  {Alexandrou}, \citenamefont {Leskovec}, \citenamefont {Meinel}, \citenamefont
  {Negele}, \citenamefont {Paul}, \citenamefont {Petschlies}, \citenamefont
  {Pochinsky}, \citenamefont {Rendon},\ and\ \citenamefont
  {Syritsyn}}]{Alexandrou:2018jbt}%
  \BibitemOpen
  \bibfield  {author} {\bibinfo {author} {\bibfnamefont {C.}~\bibnamefont
  {Alexandrou}}, \bibinfo {author} {\bibfnamefont {L.}~\bibnamefont
  {Leskovec}}, \bibinfo {author} {\bibfnamefont {S.}~\bibnamefont {Meinel}},
  \bibinfo {author} {\bibfnamefont {J.}~\bibnamefont {Negele}}, \bibinfo
  {author} {\bibfnamefont {S.}~\bibnamefont {Paul}}, \bibinfo {author}
  {\bibfnamefont {M.}~\bibnamefont {Petschlies}}, \bibinfo {author}
  {\bibfnamefont {A.}~\bibnamefont {Pochinsky}}, \bibinfo {author}
  {\bibfnamefont {G.}~\bibnamefont {Rendon}}, \ and\ \bibinfo {author}
  {\bibfnamefont {S.}~\bibnamefont {Syritsyn}},\ }\href {\doibase
  10.1103/PhysRevD.98.074502} {\bibfield  {journal} {\bibinfo  {journal} {Phys.
  Rev. D}\ }\textbf {\bibinfo {volume} {98}},\ \bibinfo {pages} {074502}
  (\bibinfo {year} {2018})},\ \Eprint {http://arxiv.org/abs/1807.08357}
  {arXiv:1807.08357 [hep-lat]} \BibitemShut {NoStop}%
\bibitem [{\citenamefont {Leskovec}\ and\ \citenamefont
  {Prelovsek}(2012)}]{Leskovec:2012gb}%
  \BibitemOpen
  \bibfield  {author} {\bibinfo {author} {\bibfnamefont {L.}~\bibnamefont
  {Leskovec}}\ and\ \bibinfo {author} {\bibfnamefont {S.}~\bibnamefont
  {Prelovsek}},\ }\href {\doibase 10.1103/PhysRevD.85.114507} {\bibfield
  {journal} {\bibinfo  {journal} {Phys. Rev.}\ }\textbf {\bibinfo {volume}
  {D85}},\ \bibinfo {pages} {114507} (\bibinfo {year} {2012})},\ \Eprint
  {http://arxiv.org/abs/1202.2145} {arXiv:1202.2145 [hep-lat]} \BibitemShut
  {NoStop}%
\bibitem [{\citenamefont {Shultz}\ \emph {et~al.}(2015)\citenamefont {Shultz},
  \citenamefont {Dudek},\ and\ \citenamefont {Edwards}}]{Shultz:2015pfa}%
  \BibitemOpen
  \bibfield  {author} {\bibinfo {author} {\bibfnamefont {C.~J.}\ \bibnamefont
  {Shultz}}, \bibinfo {author} {\bibfnamefont {J.~J.}\ \bibnamefont {Dudek}}, \
  and\ \bibinfo {author} {\bibfnamefont {R.~G.}\ \bibnamefont {Edwards}},\
  }\href {\doibase 10.1103/PhysRevD.91.114501} {\bibfield  {journal} {\bibinfo
  {journal} {Phys. Rev.}\ }\textbf {\bibinfo {volume} {D91}},\ \bibinfo {pages}
  {114501} (\bibinfo {year} {2015})},\ \Eprint
  {http://arxiv.org/abs/1501.07457} {arXiv:1501.07457 [hep-lat]} \BibitemShut
  {NoStop}%
\bibitem [{\citenamefont {Edwards}\ \emph {et~al.}(2008)\citenamefont
  {Edwards}, \citenamefont {Joo},\ and\ \citenamefont {Lin}}]{Edwards:2008ja}%
  \BibitemOpen
  \bibfield  {author} {\bibinfo {author} {\bibfnamefont {R.~G.}\ \bibnamefont
  {Edwards}}, \bibinfo {author} {\bibfnamefont {B.}~\bibnamefont {Joo}}, \ and\
  \bibinfo {author} {\bibfnamefont {H.-W.}\ \bibnamefont {Lin}},\ }\href
  {\doibase 10.1103/PhysRevD.78.054501} {\bibfield  {journal} {\bibinfo
  {journal} {Phys. Rev.}\ }\textbf {\bibinfo {volume} {D78}},\ \bibinfo {pages}
  {054501} (\bibinfo {year} {2008})},\ \Eprint {http://arxiv.org/abs/0803.3960}
  {arXiv:0803.3960 [hep-lat]} \BibitemShut {NoStop}%
\bibitem [{\citenamefont {Lin}\ \emph {et~al.}(2009)\citenamefont {Lin} \emph
  {et~al.}}]{HadronSpectrum:2008xlg}%
  \BibitemOpen
  \bibfield  {author} {\bibinfo {author} {\bibfnamefont {H.-W.}\ \bibnamefont
  {Lin}} \emph {et~al.} (\bibinfo {collaboration} {Hadron Spectrum}),\ }\href
  {\doibase 10.1103/PhysRevD.79.034502} {\bibfield  {journal} {\bibinfo
  {journal} {Phys. Rev. D}\ }\textbf {\bibinfo {volume} {79}},\ \bibinfo
  {pages} {034502} (\bibinfo {year} {2009})},\ \Eprint
  {http://arxiv.org/abs/0810.3588} {arXiv:0810.3588 [hep-lat]} \BibitemShut
  {NoStop}%
\bibitem [{\citenamefont {Wilson}\ \emph {et~al.}(2019)\citenamefont {Wilson},
  \citenamefont {Briceno}, \citenamefont {Dudek}, \citenamefont {Edwards},\
  and\ \citenamefont {Thomas}}]{Wilson:2019wfr}%
  \BibitemOpen
  \bibfield  {author} {\bibinfo {author} {\bibfnamefont {D.~J.}\ \bibnamefont
  {Wilson}}, \bibinfo {author} {\bibfnamefont {R.~A.}\ \bibnamefont {Briceno}},
  \bibinfo {author} {\bibfnamefont {J.~J.}\ \bibnamefont {Dudek}}, \bibinfo
  {author} {\bibfnamefont {R.~G.}\ \bibnamefont {Edwards}}, \ and\ \bibinfo
  {author} {\bibfnamefont {C.~E.}\ \bibnamefont {Thomas}},\ }\href {\doibase
  10.1103/PhysRevLett.123.042002} {\bibfield  {journal} {\bibinfo  {journal}
  {Phys. Rev. Lett.}\ }\textbf {\bibinfo {volume} {123}},\ \bibinfo {pages}
  {042002} (\bibinfo {year} {2019})},\ \Eprint
  {http://arxiv.org/abs/1904.03188} {arXiv:1904.03188 [hep-lat]} \BibitemShut
  {NoStop}%
\bibitem [{\citenamefont {Peardon}\ \emph {et~al.}(2009)\citenamefont
  {Peardon}, \citenamefont {Bulava}, \citenamefont {Foley}, \citenamefont
  {Morningstar}, \citenamefont {Dudek}, \citenamefont {Edwards}, \citenamefont
  {Joo}, \citenamefont {Lin}, \citenamefont {Richards},\ and\ \citenamefont
  {Juge}}]{HadronSpectrum:2009krc}%
  \BibitemOpen
  \bibfield  {author} {\bibinfo {author} {\bibfnamefont {M.}~\bibnamefont
  {Peardon}}, \bibinfo {author} {\bibfnamefont {J.}~\bibnamefont {Bulava}},
  \bibinfo {author} {\bibfnamefont {J.}~\bibnamefont {Foley}}, \bibinfo
  {author} {\bibfnamefont {C.}~\bibnamefont {Morningstar}}, \bibinfo {author}
  {\bibfnamefont {J.}~\bibnamefont {Dudek}}, \bibinfo {author} {\bibfnamefont
  {R.~G.}\ \bibnamefont {Edwards}}, \bibinfo {author} {\bibfnamefont
  {B.}~\bibnamefont {Joo}}, \bibinfo {author} {\bibfnamefont {H.-W.}\
  \bibnamefont {Lin}}, \bibinfo {author} {\bibfnamefont {D.~G.}\ \bibnamefont
  {Richards}}, \ and\ \bibinfo {author} {\bibfnamefont {K.~J.}\ \bibnamefont
  {Juge}} (\bibinfo {collaboration} {Hadron Spectrum}),\ }\href {\doibase
  10.1103/PhysRevD.80.054506} {\bibfield  {journal} {\bibinfo  {journal} {Phys.
  Rev. D}\ }\textbf {\bibinfo {volume} {80}},\ \bibinfo {pages} {054506}
  (\bibinfo {year} {2009})},\ \Eprint {http://arxiv.org/abs/0905.2160}
  {arXiv:0905.2160 [hep-lat]} \BibitemShut {NoStop}%
\bibitem [{\citenamefont {Omnes}(1958)}]{Omnes:1958hv}%
  \BibitemOpen
  \bibfield  {author} {\bibinfo {author} {\bibfnamefont {R.}~\bibnamefont
  {Omnes}},\ }\href {\doibase 10.1007/BF02747746} {\bibfield  {journal}
  {\bibinfo  {journal} {Nuovo Cim.}\ }\textbf {\bibinfo {volume} {8}},\
  \bibinfo {pages} {316} (\bibinfo {year} {1958})}\BibitemShut {NoStop}%
\bibitem [{\citenamefont {Thomas}\ \emph {et~al.}(2012)\citenamefont {Thomas},
  \citenamefont {Edwards},\ and\ \citenamefont {Dudek}}]{Thomas:2011rh}%
  \BibitemOpen
  \bibfield  {author} {\bibinfo {author} {\bibfnamefont {C.~E.}\ \bibnamefont
  {Thomas}}, \bibinfo {author} {\bibfnamefont {R.~G.}\ \bibnamefont {Edwards}},
  \ and\ \bibinfo {author} {\bibfnamefont {J.~J.}\ \bibnamefont {Dudek}},\
  }\href {\doibase 10.1103/PhysRevD.85.014507, 10.1103/PhysRevD.85.039901}
  {\bibfield  {journal} {\bibinfo  {journal} {Phys. Rev.}\ }\textbf {\bibinfo
  {volume} {D85}},\ \bibinfo {pages} {014507} (\bibinfo {year} {2012})},\
  \Eprint {http://arxiv.org/abs/1107.1930} {arXiv:1107.1930 [hep-lat]}
  \BibitemShut {NoStop}%
\bibitem [{\citenamefont {Brice\~no}\ \emph {et~al.}(2021)\citenamefont
  {Brice\~no}, \citenamefont {Dudek},\ and\ \citenamefont
  {Leskovec}}]{Briceno:2021xlc}%
  \BibitemOpen
  \bibfield  {author} {\bibinfo {author} {\bibfnamefont {R.~A.}\ \bibnamefont
  {Brice\~no}}, \bibinfo {author} {\bibfnamefont {J.~J.}\ \bibnamefont
  {Dudek}}, \ and\ \bibinfo {author} {\bibfnamefont {L.}~\bibnamefont
  {Leskovec}},\ }\href {\doibase 10.1103/PhysRevD.104.054509} {\bibfield
  {journal} {\bibinfo  {journal} {Phys. Rev. D}\ }\textbf {\bibinfo {volume}
  {104}},\ \bibinfo {pages} {054509} (\bibinfo {year} {2021})},\ \Eprint
  {http://arxiv.org/abs/2105.02017} {arXiv:2105.02017 [hep-lat]} \BibitemShut
  {NoStop}%
\bibitem [{\citenamefont {Dudek}\ \emph
  {et~al.}(2013{\natexlab{a}})\citenamefont {Dudek}, \citenamefont {Edwards},\
  and\ \citenamefont {Thomas}}]{Dudek:2012xn}%
  \BibitemOpen
  \bibfield  {author} {\bibinfo {author} {\bibfnamefont {J.~J.}\ \bibnamefont
  {Dudek}}, \bibinfo {author} {\bibfnamefont {R.~G.}\ \bibnamefont {Edwards}},
  \ and\ \bibinfo {author} {\bibfnamefont {C.~E.}\ \bibnamefont {Thomas}}
  (\bibinfo {collaboration} {Hadron Spectrum}),\ }\href {\doibase
  10.1103/PhysRevD.87.034505, 10.1103/PhysRevD.90.099902} {\bibfield  {journal}
  {\bibinfo  {journal} {Phys. Rev.}\ }\textbf {\bibinfo {volume} {D87}},\
  \bibinfo {pages} {034505} (\bibinfo {year} {2013}{\natexlab{a}})},\ \bibinfo
  {note} {[Erratum: Phys. Rev.D90,no.9,099902(2014)]},\ \Eprint
  {http://arxiv.org/abs/1212.0830} {arXiv:1212.0830 [hep-ph]} \BibitemShut
  {NoStop}%
\bibitem [{\citenamefont {Dudek}\ \emph {et~al.}(2012)\citenamefont {Dudek},
  \citenamefont {Edwards},\ and\ \citenamefont {Thomas}}]{Dudek:2012gj}%
  \BibitemOpen
  \bibfield  {author} {\bibinfo {author} {\bibfnamefont {J.~J.}\ \bibnamefont
  {Dudek}}, \bibinfo {author} {\bibfnamefont {R.~G.}\ \bibnamefont {Edwards}},
  \ and\ \bibinfo {author} {\bibfnamefont {C.~E.}\ \bibnamefont {Thomas}},\
  }\href {\doibase 10.1103/PhysRevD.86.034031} {\bibfield  {journal} {\bibinfo
  {journal} {Phys. Rev.}\ }\textbf {\bibinfo {volume} {D86}},\ \bibinfo {pages}
  {034031} (\bibinfo {year} {2012})},\ \Eprint {http://arxiv.org/abs/1203.6041}
  {arXiv:1203.6041 [hep-ph]} \BibitemShut {NoStop}%
\bibitem [{\citenamefont {Jay}\ and\ \citenamefont {Neil}(2021)}]{Jay:2020jkz}%
  \BibitemOpen
  \bibfield  {author} {\bibinfo {author} {\bibfnamefont {W.~I.}\ \bibnamefont
  {Jay}}\ and\ \bibinfo {author} {\bibfnamefont {E.~T.}\ \bibnamefont {Neil}},\
  }\href {\doibase 10.1103/PhysRevD.103.114502} {\bibfield  {journal} {\bibinfo
   {journal} {Phys. Rev. D}\ }\textbf {\bibinfo {volume} {103}},\ \bibinfo
  {pages} {114502} (\bibinfo {year} {2021})},\ \Eprint
  {http://arxiv.org/abs/2008.01069} {arXiv:2008.01069 [stat.ME]} \BibitemShut
  {NoStop}%
\bibitem [{\citenamefont {Boyd}\ \emph {et~al.}(1995)\citenamefont {Boyd},
  \citenamefont {Grinstein},\ and\ \citenamefont {Lebed}}]{Boyd:1994tt}%
  \BibitemOpen
  \bibfield  {author} {\bibinfo {author} {\bibfnamefont {C.~G.}\ \bibnamefont
  {Boyd}}, \bibinfo {author} {\bibfnamefont {B.}~\bibnamefont {Grinstein}}, \
  and\ \bibinfo {author} {\bibfnamefont {R.~F.}\ \bibnamefont {Lebed}},\ }\href
  {\doibase 10.1103/PhysRevLett.74.4603} {\bibfield  {journal} {\bibinfo
  {journal} {Phys. Rev. Lett.}\ }\textbf {\bibinfo {volume} {74}},\ \bibinfo
  {pages} {4603} (\bibinfo {year} {1995})},\ \Eprint
  {http://arxiv.org/abs/hep-ph/9412324} {arXiv:hep-ph/9412324} \BibitemShut
  {NoStop}%
\bibitem [{\citenamefont {Workman}(2022)}]{Workman:2022ynf}%
  \BibitemOpen
  \bibfield  {author} {\bibinfo {author} {\bibfnamefont {R.~L.}\ \bibnamefont
  {Workman}} (\bibinfo {collaboration} {Particle Data Group}),\ }\href@noop {}
  {\bibfield  {journal} {\bibinfo  {journal} {PTEP}\ }\textbf {\bibinfo
  {volume} {2022}},\ \bibinfo {pages} {083C01} (\bibinfo {year}
  {2022})}\BibitemShut {NoStop}%
\bibitem [{\citenamefont {Niehus}\ \emph {et~al.}(2021)\citenamefont {Niehus},
  \citenamefont {Hoferichter},\ and\ \citenamefont {Kubis}}]{Niehus:2021iin}%
  \BibitemOpen
  \bibfield  {author} {\bibinfo {author} {\bibfnamefont {M.}~\bibnamefont
  {Niehus}}, \bibinfo {author} {\bibfnamefont {M.}~\bibnamefont {Hoferichter}},
  \ and\ \bibinfo {author} {\bibfnamefont {B.}~\bibnamefont {Kubis}},\ }\href
  {\doibase 10.1007/JHEP12(2021)038} {\bibfield  {journal} {\bibinfo  {journal}
  {JHEP}\ }\textbf {\bibinfo {volume} {12}},\ \bibinfo {pages} {038} (\bibinfo
  {year} {2021})},\ \Eprint {http://arxiv.org/abs/2110.11372} {arXiv:2110.11372
  [hep-ph]} \BibitemShut {NoStop}%
\bibitem [{\citenamefont {Horgan}\ \emph {et~al.}(2014)\citenamefont {Horgan},
  \citenamefont {Liu}, \citenamefont {Meinel},\ and\ \citenamefont
  {Wingate}}]{Horgan:2013hoa}%
  \BibitemOpen
  \bibfield  {author} {\bibinfo {author} {\bibfnamefont {R.~R.}\ \bibnamefont
  {Horgan}}, \bibinfo {author} {\bibfnamefont {Z.}~\bibnamefont {Liu}},
  \bibinfo {author} {\bibfnamefont {S.}~\bibnamefont {Meinel}}, \ and\ \bibinfo
  {author} {\bibfnamefont {M.}~\bibnamefont {Wingate}},\ }\href {\doibase
  10.1103/PhysRevD.89.094501} {\bibfield  {journal} {\bibinfo  {journal} {Phys.
  Rev. D}\ }\textbf {\bibinfo {volume} {89}},\ \bibinfo {pages} {094501}
  (\bibinfo {year} {2014})},\ \Eprint {http://arxiv.org/abs/1310.3722}
  {arXiv:1310.3722 [hep-lat]} \BibitemShut {NoStop}%
\bibitem [{\citenamefont {Edwards}\ and\ \citenamefont
  {Joo}(2005)}]{Edwards:2004sx}%
  \BibitemOpen
  \bibfield  {author} {\bibinfo {author} {\bibfnamefont {R.~G.}\ \bibnamefont
  {Edwards}}\ and\ \bibinfo {author} {\bibfnamefont {B.}~\bibnamefont {Joo}}
  (\bibinfo {collaboration} {SciDAC, LHPC, UKQCD}),\ }\bibfield  {booktitle}
  {\emph {\bibinfo {booktitle} {{Lattice field theory. Proceedings, 22nd
  International Symposium, Lattice 2004, Batavia, USA, June 21-26, 2004}}},\
  }\href {\doibase 10.1016/j.nuclphysbps.2004.11.254} {\bibfield  {journal}
  {\bibinfo  {journal} {Nucl. Phys. Proc. Suppl.}\ }\textbf {\bibinfo {volume}
  {140}},\ \bibinfo {pages} {832} (\bibinfo {year} {2005})},\ \Eprint
  {http://arxiv.org/abs/hep-lat/0409003} {arXiv:hep-lat/0409003 [hep-lat]}
  \BibitemShut {NoStop}%
\bibitem [{\citenamefont {Clark}\ \emph {et~al.}(2010)\citenamefont {Clark},
  \citenamefont {Babich}, \citenamefont {Barros}, \citenamefont {Brower},\ and\
  \citenamefont {Rebbi}}]{Clark:2009wm}%
  \BibitemOpen
  \bibfield  {author} {\bibinfo {author} {\bibfnamefont {M.~A.}\ \bibnamefont
  {Clark}}, \bibinfo {author} {\bibfnamefont {R.}~\bibnamefont {Babich}},
  \bibinfo {author} {\bibfnamefont {K.}~\bibnamefont {Barros}}, \bibinfo
  {author} {\bibfnamefont {R.~C.}\ \bibnamefont {Brower}}, \ and\ \bibinfo
  {author} {\bibfnamefont {C.}~\bibnamefont {Rebbi}},\ }\href {\doibase
  10.1016/j.cpc.2010.05.002} {\bibfield  {journal} {\bibinfo  {journal}
  {Comput. Phys. Commun.}\ }\textbf {\bibinfo {volume} {181}},\ \bibinfo
  {pages} {1517} (\bibinfo {year} {2010})},\ \Eprint
  {http://arxiv.org/abs/0911.3191} {arXiv:0911.3191 [hep-lat]} \BibitemShut
  {NoStop}%
\bibitem [{\citenamefont {Babich}\ \emph
  {et~al.}(2010{\natexlab{a}})\citenamefont {Babich}, \citenamefont {Clark},\
  and\ \citenamefont {Joo}}]{Babich:2010mu}%
  \BibitemOpen
  \bibfield  {author} {\bibinfo {author} {\bibfnamefont {R.}~\bibnamefont
  {Babich}}, \bibinfo {author} {\bibfnamefont {M.~A.}\ \bibnamefont {Clark}}, \
  and\ \bibinfo {author} {\bibfnamefont {B.}~\bibnamefont {Joo}},\ }in\ \href
  {http://www1.jlab.org/Ul/publications/view_pub.cfm?pub_id=10186} {\emph
  {\bibinfo {booktitle} {{SC 10 (Supercomputing 2010) New Orleans, Louisiana,
  November 13-19, 2010}}}}\ (\bibinfo {year} {2010})\ \Eprint
  {http://arxiv.org/abs/1011.0024} {arXiv:1011.0024 [hep-lat]} \BibitemShut
  {NoStop}%
\bibitem [{\citenamefont {Clark}\ \emph {et~al.}(2016)\citenamefont {Clark},
  \citenamefont {Joo}, \citenamefont {Strelchenko}, \citenamefont {Cheng},
  \citenamefont {Gambhir},\ and\ \citenamefont {Brower}}]{Clark:SC2016}%
  \BibitemOpen
  \bibfield  {author} {\bibinfo {author} {\bibfnamefont {K.}~\bibnamefont
  {Clark}}, \bibinfo {author} {\bibfnamefont {B.}~\bibnamefont {Joo}}, \bibinfo
  {author} {\bibfnamefont {A.}~\bibnamefont {Strelchenko}}, \bibinfo {author}
  {\bibfnamefont {M.}~\bibnamefont {Cheng}}, \bibinfo {author} {\bibfnamefont
  {A.}~\bibnamefont {Gambhir}}, \ and\ \bibinfo {author} {\bibfnamefont
  {R.}~\bibnamefont {Brower}},\ }in\ \href@noop {} {\emph {\bibinfo {booktitle}
  {{Proceedings of SC 16 (Supercomputing 2016) Salt Lake City, Utah, November
  2016}}}}\ (\bibinfo {year} {2016})\BibitemShut {NoStop}%
\bibitem [{\citenamefont {Jo\'o}\ \emph {et~al.}(2013)\citenamefont {Jo\'o},
  \citenamefont {Kalamkar}, \citenamefont {Vaidyanathan}, \citenamefont
  {Smelyanskiy}, \citenamefont {Pamnany}, \citenamefont {Lee}, \citenamefont
  {Dubey},\ and\ \citenamefont {Watson}}]{ISC13Phi}%
  \BibitemOpen
  \bibfield  {author} {\bibinfo {author} {\bibfnamefont {B.}~\bibnamefont
  {Jo\'o}}, \bibinfo {author} {\bibfnamefont {D.}~\bibnamefont {Kalamkar}},
  \bibinfo {author} {\bibfnamefont {K.}~\bibnamefont {Vaidyanathan}}, \bibinfo
  {author} {\bibfnamefont {M.}~\bibnamefont {Smelyanskiy}}, \bibinfo {author}
  {\bibfnamefont {K.}~\bibnamefont {Pamnany}}, \bibinfo {author} {\bibfnamefont
  {V.}~\bibnamefont {Lee}}, \bibinfo {author} {\bibfnamefont {P.}~\bibnamefont
  {Dubey}}, \ and\ \bibinfo {author} {\bibfnamefont {W.}~\bibnamefont
  {Watson}},\ }in\ \href {\doibase 10.1007/978-3-642-38750-0_4} {\emph
  {\bibinfo {booktitle} {Supercomputing}}},\ \bibinfo {series} {Lecture Notes
  in Computer Science}, Vol.\ \bibinfo {volume} {7905},\ \bibinfo {editor}
  {edited by\ \bibinfo {editor} {\bibfnamefont {J.}~\bibnamefont {Kunkel}},
  \bibinfo {editor} {\bibfnamefont {T.}~\bibnamefont {Ludwig}}, \ and\ \bibinfo
  {editor} {\bibfnamefont {H.}~\bibnamefont {Meuer}}}\ (\bibinfo  {publisher}
  {Springer Berlin Heidelberg},\ \bibinfo {year} {2013})\ pp.\ \bibinfo {pages}
  {40--54}\BibitemShut {NoStop}%
\bibitem [{\citenamefont {Jo\'o}()}]{MGProtoDownload}%
  \BibitemOpen
  \bibfield  {author} {\bibinfo {author} {\bibfnamefont {B.}~\bibnamefont
  {Jo\'o}},\ }\href@noop {} {\enquote {\bibinfo {title} {{MG\_PROTO: A
  Multigrid Library for QCD}},}\ }\bibinfo {howpublished}
  {\url{https://github.com/JeffersonLab/mg_proto/}}\BibitemShut {NoStop}%
\bibitem [{\citenamefont {Osborn}\ \emph {et~al.}(2010)\citenamefont {Osborn},
  \citenamefont {Babich}, \citenamefont {Brannick}, \citenamefont {Brower},
  \citenamefont {Clark}, \citenamefont {Cohen},\ and\ \citenamefont
  {Rebbi}}]{Osborn:2010mb}%
  \BibitemOpen
  \bibfield  {author} {\bibinfo {author} {\bibfnamefont {J.~C.}\ \bibnamefont
  {Osborn}}, \bibinfo {author} {\bibfnamefont {R.}~\bibnamefont {Babich}},
  \bibinfo {author} {\bibfnamefont {J.}~\bibnamefont {Brannick}}, \bibinfo
  {author} {\bibfnamefont {R.~C.}\ \bibnamefont {Brower}}, \bibinfo {author}
  {\bibfnamefont {M.~A.}\ \bibnamefont {Clark}}, \bibinfo {author}
  {\bibfnamefont {S.~D.}\ \bibnamefont {Cohen}}, \ and\ \bibinfo {author}
  {\bibfnamefont {C.}~\bibnamefont {Rebbi}},\ }\bibfield  {booktitle} {\emph
  {\bibinfo {booktitle} {{Proceedings, 28th International Symposium on Lattice
  field theory (Lattice 2010): Villasimius, Italy, June 14-19, 2010}}},\
  }\href@noop {} {\bibfield  {journal} {\bibinfo  {journal} {PoS}\ }\textbf
  {\bibinfo {volume} {LATTICE2010}},\ \bibinfo {pages} {037} (\bibinfo {year}
  {2010})},\ \Eprint {http://arxiv.org/abs/1011.2775} {arXiv:1011.2775
  [hep-lat]} \BibitemShut {NoStop}%
\bibitem [{\citenamefont {Babich}\ \emph
  {et~al.}(2010{\natexlab{b}})\citenamefont {Babich}, \citenamefont {Brannick},
  \citenamefont {Brower}, \citenamefont {Clark}, \citenamefont {Manteuffel},
  \citenamefont {McCormick}, \citenamefont {Osborn},\ and\ \citenamefont
  {Rebbi}}]{Babich:2010qb}%
  \BibitemOpen
  \bibfield  {author} {\bibinfo {author} {\bibfnamefont {R.}~\bibnamefont
  {Babich}}, \bibinfo {author} {\bibfnamefont {J.}~\bibnamefont {Brannick}},
  \bibinfo {author} {\bibfnamefont {R.~C.}\ \bibnamefont {Brower}}, \bibinfo
  {author} {\bibfnamefont {M.~A.}\ \bibnamefont {Clark}}, \bibinfo {author}
  {\bibfnamefont {T.~A.}\ \bibnamefont {Manteuffel}}, \bibinfo {author}
  {\bibfnamefont {S.~F.}\ \bibnamefont {McCormick}}, \bibinfo {author}
  {\bibfnamefont {J.~C.}\ \bibnamefont {Osborn}}, \ and\ \bibinfo {author}
  {\bibfnamefont {C.}~\bibnamefont {Rebbi}},\ }\href {\doibase
  10.1103/PhysRevLett.105.201602} {\bibfield  {journal} {\bibinfo  {journal}
  {Phys. Rev. Lett.}\ }\textbf {\bibinfo {volume} {105}},\ \bibinfo {pages}
  {201602} (\bibinfo {year} {2010}{\natexlab{b}})},\ \Eprint
  {http://arxiv.org/abs/1005.3043} {arXiv:1005.3043 [hep-lat]} \BibitemShut
  {NoStop}%
\bibitem [{\citenamefont {Dudek}\ \emph
  {et~al.}(2013{\natexlab{b}})\citenamefont {Dudek}, \citenamefont {Edwards},
  \citenamefont {Guo},\ and\ \citenamefont {Thomas}}]{Dudek:2013yja}%
  \BibitemOpen
  \bibfield  {author} {\bibinfo {author} {\bibfnamefont {J.~J.}\ \bibnamefont
  {Dudek}}, \bibinfo {author} {\bibfnamefont {R.~G.}\ \bibnamefont {Edwards}},
  \bibinfo {author} {\bibfnamefont {P.}~\bibnamefont {Guo}}, \ and\ \bibinfo
  {author} {\bibfnamefont {C.~E.}\ \bibnamefont {Thomas}} (\bibinfo
  {collaboration} {Hadron Spectrum}),\ }\href {\doibase
  10.1103/PhysRevD.88.094505} {\bibfield  {journal} {\bibinfo  {journal} {Phys.
  Rev.}\ }\textbf {\bibinfo {volume} {D88}},\ \bibinfo {pages} {094505}
  (\bibinfo {year} {2013}{\natexlab{b}})},\ \Eprint
  {http://arxiv.org/abs/1309.2608} {arXiv:1309.2608 [hep-lat]} \BibitemShut
  {NoStop}%
\end{thebibliography}%


\end{document}